\documentclass[english,reprint,aps,prl,amsmath,twocolumn,superscriptaddress,showpacs,notitlepage,nofootinbib,longbibliography]{revtex4-2}

\usepackage{microtype}
\usepackage{newtxtext,newtxmath,xspace}
\usepackage{amsbsy,bm,bbold}
\usepackage{siunitx}
\usepackage{graphicx,color,xcolor,epsfig,rotate}
\usepackage{fancyhdr}
\usepackage{gensymb}
\usepackage[colorlinks=true, 
            linkcolor=blue, 
            urlcolor=blue,
            citecolor=blue]{hyperref}
\usepackage{soul}
\usepackage{orcidlink}
\setstcolor{red}
\allowdisplaybreaks[2]
\newcommand{\iu}{{i\mkern1mu}}

\newcommand{\CoP}{Na$_{2}$BaCo(PO$_4$)$_{2}$}
\newcommand{\NiP}{Na$_{2}$BaNi(PO$_4$)$_{2}$}

\usepackage{tikz}
\definecolor{mygreen}{RGB}{50,200,50}
\definecolor{myblue}{RGB}{100,149,237}
\definecolor{myred}{RGB}{211,56,28}

\newcommand{\Tensor}[4]{
\filldraw[draw=mygreen!100,thick,fill=white!20] (#1,#2) circle (#3) ;
\draw [mygreen, thick] (#1-#3*0.7071,#2+#3*0.7071) to (#1-#3*0.7071-#4,#2+#3*0.7071+#4) ; % left
}

\newcommand{\CenterB}[4]{
\filldraw[draw=myred!100,thick,fill=myred!100] (#1,#2) circle (#3) ;
\draw (#1+#3*2.8,#2) node [above] {$\bm{r}$} ;
\draw (#1+#3*2.6,#2) node [below] {$B$} ;
\draw [myred, thick] (#1-#3*0.7071,#2+#3*0.7071) to (#1-#3*0.7071-#4,#2+#3*0.7071+#4) ; % left
}

\newcommand{\CenterA}[4]{
\filldraw[draw=mygreen!100,thick,fill=white!20] (#1,#2) circle (#3) ;
\draw (#1+#3*2.,#2) node [below] {$A$} ;
\draw [mygreen, thick] (#1-#3*0.7071,#2+#3*0.7071) to (#1-#3*0.7071-#4,#2+#3*0.7071+#4) ; % left
}

\newcommand{\triangular}[3]{
\filldraw[draw=myblue!100,thick,fill=myblue!100] (#1-#3/2,#2-#3*0.2887) -- (#1+#3/2,#2-#3*0.2887) -- (#1,#2+#3*0.5774) ;
}

\newcommand{\trianglelattice}[1]{
\draw[dashed] (-2*#1,-#1*0.2887) -- (3.8*#1,-#1*0.2887);
\draw[dashed] (-2*#1,-#1*0.2887-#1*0.8660) -- (3.8*#1,-#1*0.2887-#1*0.8660);
\draw[dashed] (-2*#1,-#1*0.2887-#1*0.8660*2) -- (3.8*#1,-#1*0.2887-#1*0.8660*2);
\draw[dashed] (-2*#1,-#1*0.2887+#1*0.8660*2) -- (3.8*#1,-#1*0.2887+#1*0.8660*2);
\draw[dashed] (-2*#1,-#1*0.2887+#1*0.8660*1) -- (3.8*#1,-#1*0.2887+#1*0.8660*1);

\Dottedlinel{-1.5*#1}{-#1*0.2887}{1.2}{2.3}
\Dottedlinel{-0.5*#1}{-#1*0.2887}{2.3}{2.3}
\Dottedlinel{0.5*#1}{-#1*0.2887}{2.3}{2.3}
\Dottedlinel{1.5*#1}{-#1*0.2887}{2.3}{2.3}
\Dottedlinel{2.5*#1}{-#1*0.2887}{2.3}{2.3}
\Dottedlinel{3.5*#1}{-#1*0.2887}{2.3}{0.5}
\Dottedlinel{3.5*#1}{-#1*7/6*1.7321}{0.3}{0.3}
\Dottedliner{-1.5*#1}{-#1*0.2887}{2.3}{1.1}
\Dottedliner{-0.5*#1}{-#1*0.2887}{2.3}{2.3}
\Dottedliner{0.5*#1}{-#1*0.2887}{2.3}{2.3}
\Dottedliner{1.5*#1}{-#1*0.2887}{2.3}{2.3}
\Dottedliner{2.5*#1}{-#1*0.2887}{2.3}{2.3}
\Dottedliner{3.5*#1}{-#1*0.2887}{0.5}{2.3}
}

\newcommand{\blackline}[4]{
\draw[gray,thick] (#1-#3*1.7321*0.5,#2-#3*0.5)--(#1+#4*1.7321*0.5,#2+#4*0.5);
}

\newcommand{\Dottedlinel}[4]{
\draw[dashed] (#1-#3*0.5,#2-#3*0.5*1.7321)--(#1+#4*0.5,#2+#4*0.5*1.7321);
}

\newcommand{\Dottedliner}[4]{
\draw[dashed] (#1+#3*0.5,#2-#3*0.5*1.7321)--(#1-#4*0.5,#2+#4*0.5*1.7321);
}

\makeatother

\usepackage{babel}
\begin{document}

\title{Possible Observation of Quadrupole Waves in Spin Nematics}

\author{Jieming~Sheng}
\thanks{These authors contributed equally to this work}
\affiliation{Department of Physics, Southern University of Science and Technology, Shenzhen 518055, China}
\affiliation{School of Physical Sciences, Great Bay University and Great Bay Institute for Advanced Study, Dongguan 523000, China}

\author{Jiahang~Hu}
\thanks{These authors contributed equally to this work}
\affiliation{Beijing National Laboratory for Condensed Matter Physics and Institute of Physics,
Chinese Academy of Sciences, Beijing 100190, China}
\affiliation{School of Physical Sciences, University of Chinese Academy of Sciences, Beijing 100049, China}

\author{Lei~Xu}
\affiliation{School of Physics, Zhejiang University, Hangzhou 310058, China}

\author{Le~Wang}
\affiliation{Department of Physics, Southern University of Science and Technology, Shenzhen 518055, China}
\affiliation{Shenzhen Institute for Quantum Science and Engineering, Shenzhen 518055, China}

\author{Xiaojian~Shi}
\affiliation{Center for Correlated Matter, Zhejiang University, Hangzhou 310058, China}
\affiliation{School of Physics, Zhejiang University, Hangzhou 310058, China}

\author{Runze~Chi}
\affiliation{Beijing National Laboratory for Condensed Matter Physics and Institute of Physics,
Chinese Academy of Sciences, Beijing 100190, China}
\affiliation{School of Physical Sciences, University of Chinese Academy of Sciences, Beijing 100049, China}
\affiliation{Division of Chemistry and Chemical Engineering, California Institute of Technology, Pasadena, California 91125, USA}

\author{Dehong~Yu}
\affiliation{Australian Nuclear Science and Technology Organisation, Lucas Heights, New South Wales 2234, Australia}

\author{Andrey~Podlesnyak}
\affiliation{Neutron Scattering Division, Oak Ridge National Laboratory, Oak Ridge, Tennessee 37831, USA}

\author{Pharit~Piyawongwatthana}
\affiliation{Materials and Life Science Division, J-PARC Center, Japan Atomic Energy Agency, Tokai, Naka, Ibaraki 319-1195, Japan}

\author{Naoki~Murai}
\affiliation{Materials and Life Science Division, J-PARC Center, Japan Atomic Energy Agency, Tokai, Naka, Ibaraki 319-1195, Japan}

\author{Seiko~Ohira-Kawamura}
\affiliation{Materials and Life Science Division, J-PARC Center, Japan Atomic Energy Agency, Tokai, Naka, Ibaraki 319-1195, Japan}

\author{Huiqiu~Yuan}
\affiliation{Center for Correlated Matter, Zhejiang University, Hangzhou 310058, China}
\affiliation{School of Physics, Zhejiang University, Hangzhou 310058, China}

\author{Ling~Wang}
\affiliation{School of Physics, Zhejiang University, Hangzhou 310058, China}

\author{Jia-Wei Mei}
\email[Corresponding author: ]{meijw@sustech.edu.cn}
\affiliation{Department of Physics, Southern University of Science and Technology, Shenzhen 518055, China}
\affiliation{Shenzhen Institute for Quantum Science and Engineering, Shenzhen 518055, China}
\affiliation{Shenzhen Key Laboratory of Advanced Quantum Functional Materials and Devices,  Southern University of Science and Technology, Shenzhen 518055, China}

\author{Hai-Jun~Liao}
\email[Corresponding author: ]{navyphysics@iphy.ac.cn}
\affiliation{Beijing National Laboratory for Condensed Matter Physics and Institute of Physics,
Chinese Academy of Sciences, Beijing 100190, China}
\affiliation{Songshan Lake Materials Laboratory, Dongguan, Guangdong 523808, China}

\author{Tao~Xiang}
\email[Corresponding author: ]{txiang@iphy.ac.cn}
\affiliation{Beijing National Laboratory for Condensed Matter Physics and Institute of Physics,
Chinese Academy of Sciences, Beijing 100190, China}
\affiliation{School of Physical Sciences, University of Chinese Academy of Sciences, Beijing 100049, China}
\affiliation{Beijing Academy of Quantum Information Sciences, Beijing, 100190, China}

\author{Liusuo~Wu}
\email[Corresponding author: ]{wuls@sustech.edu.cn}
\affiliation{Department of Physics, Southern University of Science and Technology, Shenzhen 518055, China}
\affiliation{Quantum Science Center of Guangdong-Hong Kong-Macao Greater Bay Area (Guangdong), Shenzhen 518045, China}

\author{Zhentao~Wang}
\email[Corresponding author: ]{ztwang@zju.edu.cn}
\affiliation{Center for Correlated Matter, Zhejiang University, Hangzhou 310058, China}
\affiliation{School of Physics, Zhejiang University, Hangzhou 310058, China}

\begin{abstract}
Discovery of new states of matter is a key objective in modern condensed matter physics, which often leads to revolutionary technological advancements such as superconductivity.
Quantum spin nematic, a ``hidden order'' that evades conventional magnetic probes, is one such state. \NiP{} is a potential spin nematic material, suggested by the observation of a two-magnon Bose-Einstein condensation from above the saturation field. However, direct confirmation of the spin nematicity remains elusive. 
This Letter presents inelastic neutron scattering spectra from the putative spin nematic phases of \NiP{}, revealing low-energy quadrupole waves that are absent in the neighboring conventional magnetic phases. A spin-one model quantitatively captures the full details of the spin excitation spectra across all low-temperature phases, providing direct evidence of the spin nematic orders. 
Additionally, we show evidence of the three-magnon continuum and two-magnon bound states in the $1/3$-magnetization plateau, revealing condensation of the two-magnon bound state as the origin of the low-field spin nematic supersolid phase.

\end{abstract}

\date{\today}

\maketitle

\begin{figure}[tbp!]
\includegraphics[width=0.99\columnwidth]{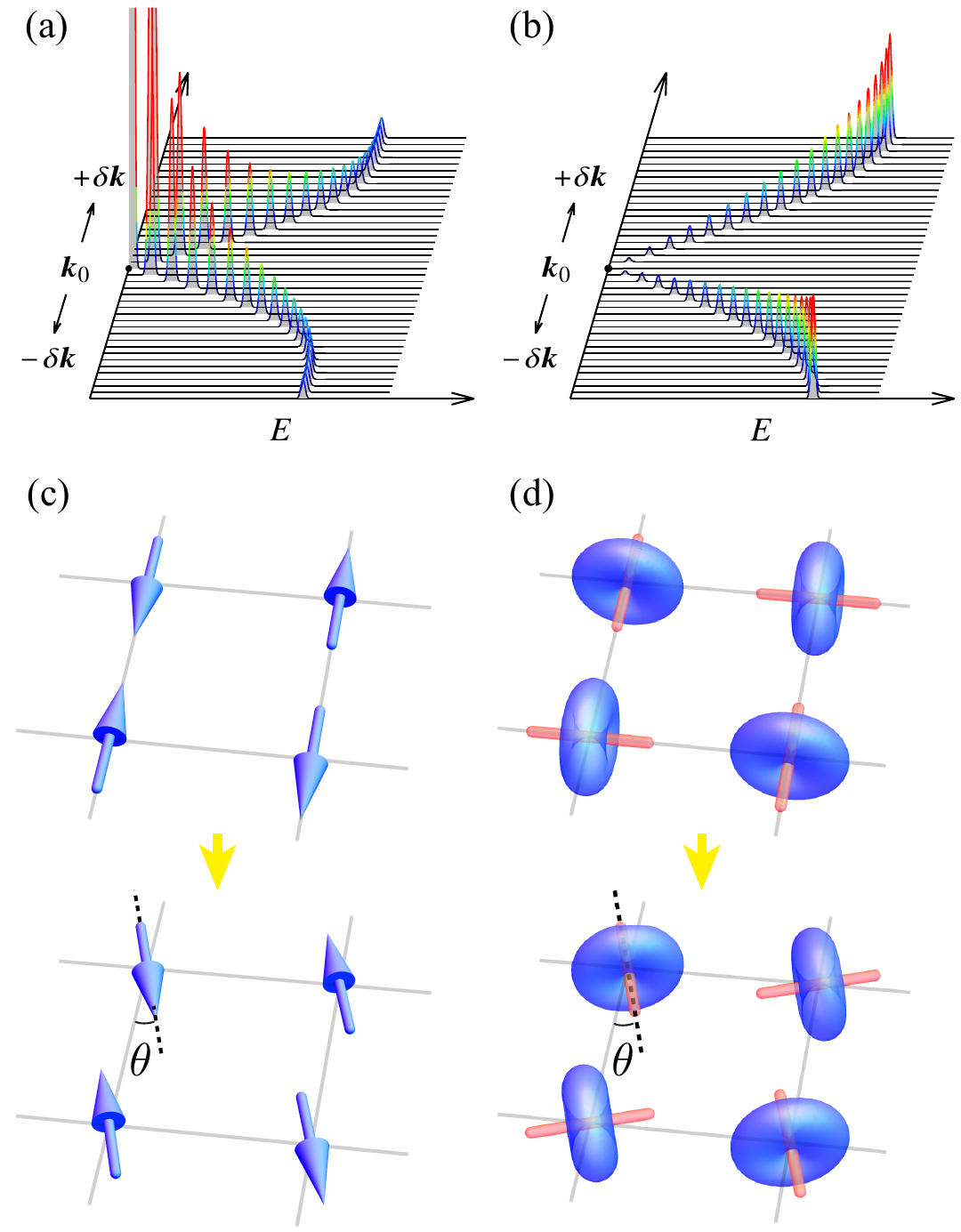}
\caption{{Low-energy excitations of conventional magnetic order and SN order.} (a) Goldstone mode of dipole order where the DSSF diverges at the ordering wave vector $\bm{k}_0$. (b) Goldstone mode of SN order where the DSSF is zero at $\bm{k}_0$ and scales linearly with the distance $\delta k$ away from $\bm{k}_0$ at low energy. (c) Schematics of the Goldstone mode for conventional magnetic order, where a global rotation with $\theta \rightarrow 0$ of dipole moments costs zero energy. (d) Schematics of the Goldstone mode for SN order, where a global rotation with $\theta \rightarrow 0$ of quadrupole moments costs zero energy. }
\label{fig:schematic}
\end{figure}

Quantum spin nematic (SN) order is a type of unconventional magnetic state in which spontaneous spin rotational symmetry breaking occurs in the quadrupolar instead of the dipolar sector, and as a result time-reversal symmetry is preserved~\cite{BlumeM1969,MatveevVM1973,AndreevAF1984,PapanicolaouN1988,ChubukovAV1990}. While there are a few promising SN candidates~\cite{SvistovLE2011,OrlovaA2017,PovarovKY2019,SkoulatosM2019,BhartiyaVK2019,IshikawaH2015,JansonO2016,YoshidaM2017,KohamaY2019,KimH2024,WangZ2018_SSM,FoghE2024,LiuR2025,SantiniP2009_RMP}, obtaining {\it direct} evidence is challenging due to zero dipolar moments in such a hidden order. 
In theory, the spin excitations of the SN should differ markedly from those of conventional magnetic order~\cite{TsunetsuguH2006,LauchliA2006,ShindouR2013_nematic_DSF,SmeraldA2013}. In inelastic neutron scattering (INS) experiments, the Goldstone mode of conventional magnetic order exhibits a low-energy spin wave with diverging intensity at zero energy. In contrast, in SN phase the INS cross section is zero at the ordering wave vector $\bm{k}_0$; due to dynamic mixing of dipole-quadrupole moments, a low-energy quadrupole wave shows up with an intensity proportional to the distance $\delta k$ away from $\bm{k}_0$ (see Fig.~\ref{fig:schematic}). Thus, identification of the quadrupole wave emerging from the SN order serves as a strong evidence of its existence.

The triangular lattice (TL) compound \NiP{}~\cite{LiN2021,DingF2021} has recently emerged as a promising candidate for SN order arising from Bose-Einstein condensation (BEC)~\cite{ZapfV2014_RMP} of a two-magnon bound state at the saturation field~\cite{ShengJ2025_NiP}. While indirect evidence for SN has been observed, including the two-magnon BEC transition and $T^2$ behavior in specific heat within the proposed SN phase~\cite{ShengJ2025_NiP}, direct confirmation is still lacking.
To address this, we conducted INS measurements in various low-temperature phases of \NiP{}, revealing the existence of low-energy quadrupole waves in the proposed SN phases.
We achieved quantitative agreement with the INS results across all fields using
infinite 
projected entangled-pair states (iPEPS) calculations on a TL spin-one model,  further supporting the SN nature of the material. 
To gain further insight on the low-field SN phase, we examined the multimagnon spectra in the $1/3$-magnetization plateau, demonstrating that the BEC transition of the two-magnon bound state is the common origin of both SN phases in \NiP{}.

%Through a comprehensive comparison between the INS data and theoretical calculations, we reveal the existence of low-energy quadrupole waves in two distinct phases of \NiP{}.
%The existence of the quadrupole waves, together with the  indeed proves the SN nature of the two phases as predicted in Ref.~\cite{ShengJ2025_NiP}. Previously, the SN phase below the saturation field $B_\text{s}$ in \NiP{} was shown to be induced by BEC of two-magnon bound states inside the fully polarized (FP) phase. In this work, we will demonstrate that two-magnon bound states are also stable in the up-up-down (UUD) phase that has $1/3$ magnetization plateau, and the low-field SN-supersolid phase is also induced by condensation of the two-magnon bound state.

The Hamiltonian of \NiP{} has been firmly established in Ref.~\cite{ShengJ2025_NiP}, where the spin excitations in the fully polarized (FP) phase were shown to be accurately described by a spin-one {\it XXZ} model with moderate easy-axis single-ion anisotropy on a TL,
\begin{equation}\label{eq:Ham}
\begin{split}
\mathcal{H} &= J \sum_{\langle ij\rangle} \left( S_i^x S_j^x + S_i^y S_j^y + \Delta S_i^z S_j^z \right) \\
&\quad -D \sum_i \left(S_i^z \right)^2 - g_c \mu_B B \sum_i S_i^z,
\end{split}
\end{equation}
where $J$ is the antiferromagnetic spin exchange on nearest neighbor bonds, $\Delta$ quantifies the {\it XXZ} anisotropy, $D$ is the strength of the single-ion anisotropy, and $B$ is the magnitude of the Zeeman field applied along the $c$ axis. The parameters are $J=\qty{0.032}{meV}$, $\Delta=1.13$, $D/J=3.97$, and $g_c=2.24$ according to Ref.~\cite{ShengJ2025_NiP}, leaving no adjustable model parameters in this Letter.

Three quantum critical points (QCPs) \{$B_\text{c1}$, $B_\text{c2}$, $B_\text{s}$\} can be reached by varying the magnetic field $B$, separating four phases at $T=0$. Two of them were clearly identified as conventional magnetic phases, namely the up-up-down (UUD) phase with a $1/3$-magnetization plateau and the FP phase.
These QCPs were confirmed by density matrix renormalization group (DMRG) calculations on model \eqref{eq:Ham}. 
Moreover, the order parameters computed by DMRG indeed suggest that the phases neighboring the $1/3$-plateau are SNs~\cite{ShengJ2025_NiP}. 

\begin{figure}[tbp!]
\includegraphics[width=0.99\columnwidth]{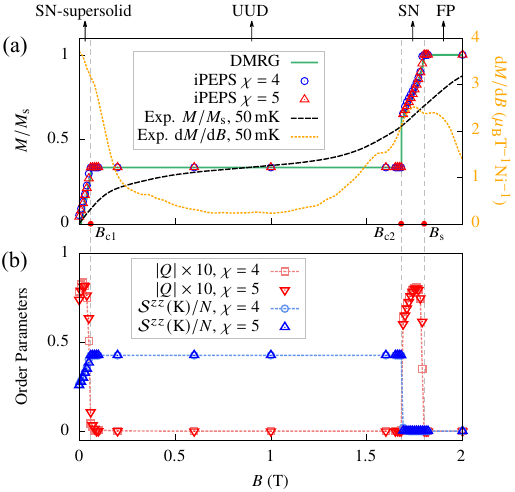}
\caption{{Field dependence of the model \eqref{eq:Ham} at $T=0$}. (a) Magnetization from DMRG (reproduced from Ref.~\cite{ShengJ2025_NiP}) and iPEPS calculations with bond dimension $\chi$. The experimentally measured $M/M_\text{s}$ and $\mathrm{d}M/\mathrm{d}B$ at \qty{50}{mK} from Ref.~\cite{ShengJ2025_NiP} are overlaid for comparison. (b) The SN order parameter $|Q|$ and the solid order parameter $\mathcal{S}^{zz}(\text{K})$ calculated by iPEPS. }
\label{fig:magnetization}
\end{figure}

As the DMRG calculations were performed on a finite lattice and the magnitudes of the SN order parameters are small, extrapolation to the thermodynamic limit is not straightforward. 
To cross-check this point, we performed iPEPS calculations on model \eqref{eq:Ham},  which revealed a magnetization profile that agrees quantitatively with the DMRG results [Fig.~\ref{fig:magnetization}(a)]. 
The order parameters that characterize the SN and ``solid'' phases are
\begin{subequations}
\begin{align}
|Q| & \equiv \sqrt{\langle Q^{x^2-y^2} \rangle^2 + \langle Q^{xy} \rangle^2},\\
\mathcal{S}^{zz}(\bm{k}) & \equiv \frac{1}{N}\sum_{i,j}e^{-\iu \bm{k} \cdot \left( \bm{r}_i - \bm{r}_j \right)} \langle S_i^z  S_j^z\rangle,
\end{align}
\end{subequations}
where $N\rightarrow \infty$ is the number of lattice sites, $Q^{x^2-y^2}\equiv \frac{1}{2N}\sum_i \left( S_i^+ S_i^+ + S_i^-S_i^- \right)$, $Q^{xy}\equiv \frac{-\iu}{2N}\sum_i \left( S_i^+ S_i^+ - S_i^-S_i^- \right)$, and we fixed the momentum $\bm{k}$ at the $K$ point to capture the three-sublattice order. 
Figure~\ref{fig:magnetization}(b) shows that both phases neighboring the $1/3$ plateau are indeed SN. Additionally, the phase at $B<B_\text{c1}$ breaks both the U(1) and translational symmetries, forming an SN-supersolid.

\begin{figure*}[tbp!]
\includegraphics[width=0.95\textwidth]{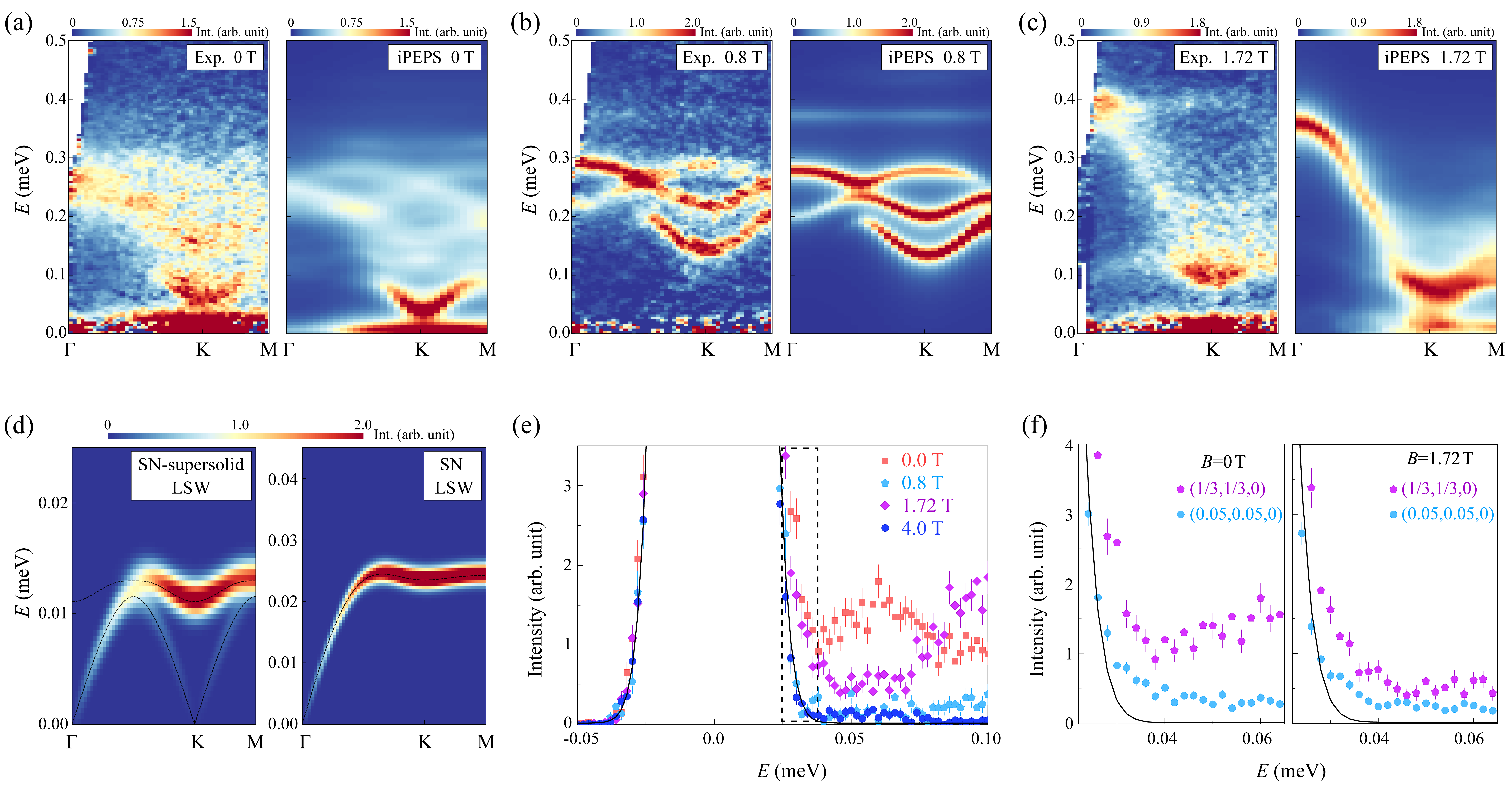}
\caption{{Spin excitation spectra of \NiP{} in the SN-supersolid, UUD, and SN phases.} (a)--(c) Background-subtracted INS intensities at \qty{60}{mK} compared to the DSSF $\mathcal{S}(\bm{k},E)$ calculated by iPEPS at $T=0$ with bond dimension $\chi=4$. INS data collected at \qty{60}{mK} and \qty{4}{T} were used as the background for subtraction. (d) DSSF $\mathcal{S}(\bm{k},E)$ by LSW calculations of the effective model \eqref{eq:Heff} in the SN-supersolid phase ($B=\qty{0.01}{T}$) and the SN phase ($B=\qty{1.87}{T}$), where a Gaussian broadening factor $\sigma=\qty{0.001}{meV}$ is used. (e) Raw INS intensities at the $K$ point measured at $T=\qty{60}{mK}$. The dashed rectangle highlights the low-energy quadrupole waves exclusive to the SN phases. (f) Raw INS intensities measured at $T=\qty{60}{mK}$ with $B=0$ and \qty{1.72}{T}. The solid lines in (e) and (f) are Gaussian fits to the diffusive background. }
\label{fig:neutron}
\end{figure*}

%While there is already compelling indirect evidence supporting the existence of SN's in \NiP{}, an experimental probe of the quadrupolar degree of freedom directly inside the presumed SN-supersolid and SN phases is still lacking. 
To gain further insight into the nature of the $T=0$ phases, we performed INS measurements under various magnetic fields (see Supplemental Material~\cite{SM}). Figures~\ref{fig:neutron}(a)--\ref{fig:neutron}(c) show the $T=\qty{60}{mK}$ INS results at $B=0$, $0.8$, and \qty{1.72}{T}, relevant to the SN-supersolid, UUD, and SN phases, respectively. 
Among phases below saturation field, the UUD state has a collinear spin structure and relatively weak quantum fluctuations. As a result, three sharp spin wave branches are clearly visible [Fig.~\ref{fig:neutron}(b)].
In contrast, the phases at $B=\qty{0}{T}$ [Fig.~\ref{fig:neutron}(a)] and \qty{1.72}{T} [Fig.~\ref{fig:neutron}(c)] show qualitatively distinct features compared to the UUD state, suggesting their different natures of the underlying ground states.
Notably, at intermediate to high energies, the spin excitation spectra at both $B=0$ and \qty{1.72}{T} show a broadened feature for extended regions in the \{$\bm{k}$, $E$\} phase space, likely due to the multimagnon decay process~\cite{ZhitomirskyME2013_RMP}. In fact, such continuumlike spin excitations were also recently reported in the noncollinear ordered phases of a similar compound \CoP{}~\cite{ShengJ2025_CoP,ChiR2024,GaoY2024_CoP}.

A key finding of this study is the presence of distinct narrow modes near zero energy in Figs.~\ref{fig:neutron}(a) and \ref{fig:neutron}(c). This feature is unique to the SN phases of \NiP{} and is not observed in any phase of the sibling compound \CoP{}. 
To single out these narrow low-energy modes more clearly, the raw INS data with different magnetic fields at the $K$ point are shown in Fig.~\ref{fig:neutron}(e). On the negative-energy-transfer side ($E<0$), the INS intensity at different magnetic fields perfectly overlap onto a single curve that characterizes the diffusive background. 
In contrast, the low-energy INS intensities at $B=0$ and \qty{1.72}{T} are clearly beyond the quasielastic scattering background on the positive-energy-transfer side ($E>0$), confirming their existence unambiguously [highlighted by the dashed rectangle in Fig.~\ref{fig:neutron}(e)]. For both $B=0$ and \qty{1.72}{T}, this mode is more clearly seen at the $K$ point, and near the $\Gamma$ point it is practically indistinguishable from the background [Fig.~\ref{fig:neutron}(f)].

To make a quantitative comparison, we utilized a recently developed tensor network method~\cite{VanderstraetenL2015,VanderstraetenL2019,PonsioenB2020,PonsioenB2022,ChiR2022} to compute the dynamic spin structure factor (DSSF) $\mathcal{S}(\bm{k},E)$ at zero temperature ($T=0$). This approach leverages the single-mode excitation ansatz~\cite{OstlundS1995} of iPEPS combined with automatic differentiation~\cite{LiaoHJ2019,SM} and has been effectively applied to several TL antiferromagnets, including $\rm Ba_3CoSb_2O_9$~\cite{ChiR2022}, $\rm Na_2BaCo(PO_4)_2$~\cite{ChiR2024}, and $\rm K_2Co(SeO_3)_2$~\cite{XuY2025}.
%Further methodological details are provided in the Supplemental Material (SM).

Figures~\ref{fig:neutron}(a)--\ref{fig:neutron}(c) show a remarkable agreement between INS and iPEPS results across all phases, which is uncommon in typical frustrated quantum magnets. 
Moreover, the narrow low-energy modes are clearly present in phases with a finite quadrupolar order parameter $|Q|$ [Fig.~\ref{fig:magnetization}(b)], suggesting that these low-energy modes are unique to the SN phases of \NiP{}.

%This low-energy mode is present in both the INS data and the iPEPS calculations, suggesting that it is not an artifact of the INS background subtraction. 

%Surprisingly, the same set of parameters as determined in Ref.~\cite{ShengJ2025_NiP} leads to quantitative agreements between experimental data and iPEPS calculations in all phases. Particularly, the quantum fluctuations are suppressed in the FP phase, where iPEPS agrees perfectly with the published INS data (see Fig.?? in supplementary materials). 
%Among phases below saturation field, the UUD state has a collinear spin structure and relatively weak quantum fluctuations. As a result, three sharp spin wave branches are clearly visible in both experimental and iPEPS results (Fig.~\ref{fig:neutron}{\bf B}). 
%The phases at $B=\qty{0}{T}$ and \qty{1.72}{T} show qualitatively different features compared to the sharp spin waves inside the FP and UUD phases, suggesting their noncollinear nature.

%These observations of the low-energy modes at $B=0$ and \qty{1.72}{T} are in qualitative agreement with iPEPS calculations (see Figs.~\ref{fig:neutron}{\bf A}-{\bf C}). Since the accuracy of the DSSF by iPEPS is often inadequate to capture the Goldstone modes from continuous symmetry breaking~\cite{ChiR2022,ChiR2024}, we resort to semi-classical spin wave calculations that reveal the qualitative behavior of these low-energy modes {\color{red} [xx Question to Haijun: are these statements accurate? xx]}.

For modes near zero energy, iPEPS with finite $\chi$ lacks quantitative accuracy due to slower convergence~\cite{ChiR2022}. It would be helpful if an intuitive quasiparticle picture could be used to understand the origin of these narrow low-energy modes. We note that the typical linear spin wave (LSW) theory based on Holstein-Primakoff bosons~\cite{HolsteinT1940} does not directly apply to the Hamiltonian \eqref{eq:Ham}, since the semiclassical SU(2) coherent states completely miss the quadrupolar moments. An alternative generalized linear spin wave (GLSW) approach typically adopted for $S=1$ is based on SU(3) coherent states~\cite{MatveevVM1973,PapanicolaouN1988,BatistaCD2004_SUN,TsunetsuguH2006,LauchliA2006}. However, this approach also does not capture the SN phases of the model~\cite{SeifertUFP2022}, as the site-factorized wave function underestimates the quantum entanglement in these phases.

To overcome these difficulties, recall that for \NiP{} we have $D/J=3.97$, where the $|S_i^z = \pm 1 \rangle$ doublet is relatively lower in energy compared to the $|S_i^z=0\rangle$ state, leading to an effective $s=1/2$ model at low temperature~\cite{ShengJ2025_NiP},
\begin{equation}\label{eq:Heff}
\mathcal{H}_\text{eff} = -\frac{J^2}{D} \sum_{\langle ij \rangle } \left[ s_i^x s_j^x + s_i^y s_j^y +\tilde{\Delta} s_i^z s_j^z  \right]  -\tilde{H} \sum_i s_i^z,
\end{equation}
where $\tilde{\Delta}\equiv -2 \left( 1 + 2 \Delta D /J\right) $, $\tilde{H}\equiv 2 g_c \mu_B B$, $s_i^+ = P_0 (S_i^+ S_i^+ /2) P_0$, $s_i^- = P_0 (S_i^- S_i^- /2) P_0$, and $s_i^z = P_0 (S_i^z/2) P_0$. Here $P_0$ is the projection operator to the doublet space.

The semiclassical approximation of the effective model \eqref{eq:Heff} using SU(2) coherent states faithfully reproduces all phases observed in DMRG and iPEPS results~\cite{YamamotoD2014,SM}. While this approach slightly shifts the critical fields to $B_\text{c1}^\text{(eff)}\approx 0.05$, $B_\text{c2}^\text{(eff)}\approx 1.86$, and $B_\text{s}^\text{(eff)}\approx \qty{1.95}{T}$, the $T=0$ phase diagram remains qualitatively unchanged.
Thus, we are allowed to use Eq.~\eqref{eq:Heff} to extract the low-energy part of the DSSF for the original model~\eqref{eq:Ham}. Note that, since the spin-one operators $S_i^+$ and $S_i^-$ do not directly connect the low-energy $|\pm 1\rangle$ states of the effective model, the mapping of the DSSF between the two models is
\begin{equation}\label{eq:DSSF}
\mathcal{S}(\bm{k},E) \equiv \sum_{\alpha=x,y,z} \mathcal{S}^{\alpha \alpha}(\bm{k},E) \approx 4 \mathcal{S}_\text{eff}^{zz}(\bm{k},E),
\end{equation}
whereas the transverse modes are suppressed. 

Figure~\ref{fig:neutron}(d) shows the DSSF $\mathcal{S}(\bm{k},E)$ of the low-energy modes in both the SN-supersolid ($B=\qty{0.01}{T}$) and the SN ($B=\qty{1.87}{T}$) phases, calculated by LSW theory for the effective model \eqref{eq:Heff} at $T=0$. The gapless behavior of these modes is indeed consistent with spontaneous U(1) symmetry breaking responsible for the finite in-plane quadrupolar moments [Fig.~\ref{fig:magnetization}(b)]. Moreover, the intensity of $\mathcal{S}(\bm{k},E)$ at the ordering wave vector $\bm{k}_0$ ($\bm{k}_0=K$ for SN-supersolid and $\bm{k}_0=\Gamma$ for SN) is zero: $\mathcal{S}(\bm{k}_0,0) = 0$; as we move away from $\bm{k}_0$, the intensity scales linearly as a function of $\delta q$. These behaviors reveal the quadrupole wave nature of the narrow low-energy modes~\cite{TsunetsuguH2006,LauchliA2006,ShindouR2013_nematic_DSF,SmeraldA2013}, providing strong evidence of the SN phases surrounding the 1/3 plateau.

A key question arises regarding the microscopic origins of SN formation in both phases.  Reference \cite{ShengJ2025_NiP} partially addressed this by demonstrating that the SN phase between $B_\text{c2} < B < B_\text{s}$ results from the BEC transition of a two-magnon bound state formed in the FP state.
This insight prompted us to further investigate the UUD state adjacent to the SN-supersolid phase, 
which actually contains rich magnetic excitations despite its simple structure.
%In \NiP{}, the two-magnon bound state in the FP phase was shown to go through a BEC transition that induces the SN phase below the saturation field~\cite{ShengJ2025_NiP}. A natural question is: Could the two-magnon bound state in the UUD phase also be responsible for the SN-supersolid phase below $B_\text{c1}$?
%In the following, we will discuss further details on the spin excitation spectra inside the UUD phase. 

In addition to the three sharp spin waves, a weak flat excitation slightly below \qty{0.4}{meV} is visible in both the INS and iPEPS data at $B=\qty{0.8}{T}$ [Fig.~\ref{fig:neutron}(b)]. According to iPEPS calculations, this mode arises exclusively from the $\mathcal{S}^{xx+yy}(\bm{k},E)$ components, representing the three-magnon continuum of transverse fluctuations.
In fact, this three-magnon continuum is already captured at the semiclassical level by GLSW theory (see Supplemental Material~\cite{SM}). 
Figure~\ref{fig:multi-magnon}(a) shows the evolution of both the one-boson excitations and the free two-boson continuum in $\mathcal{S}(\Gamma,E)$. Note that the elementary bosonic excitations calculated by the GLSW approach are not exactly one magnons in nature, because some of the bosonic operators in the GLSW method connect $\Delta S^z=2$ states in the Hilbert space. 
As the slopes of the two-boson continuum are identical to the three sharp one-magnon excitations [Fig.~\ref{fig:multi-magnon}(a)], it is clear that they are fluctuations of three single spin flips. In other words, the ``two-boson'' continuum shown in Fig.~\ref{fig:multi-magnon}(a) is indeed the three-magnon continuum revealed by INS and iPEPS in Fig.~\ref{fig:neutron}(b).

Importantly, the GLSW method identified two modes with twice the slopes of the one-magnon branches [dashed lines in Fig.~\ref{fig:multi-magnon}(a)], which are not visible in the INS intensities [Fig.~\ref{fig:neutron}(b)]. 
Similar to the case of the FP state, these two modes are the two-magnon bound states induced by the easy-axis single-ion anisotropy~\cite{SilberglittR1970,OguchiT1971}. The zero INS intensity of the two-magnon bound states in both the UUD and the FP phases is rooted in the U(1) symmetry, which forbids finite matrix elements between states differing by $\Delta S^z = \pm 2$ with operator $S_{\bm{k}}^\alpha$.
In systems with U(1) symmetry-breaking anisotropies, two-magnon bound states may become visible in INS due to their hybridization with one-magnon bands~\cite{BaiX2021}.

%As indicated by the slopes (positive/negative slope for $\mathcal{S}^{+-}$/$\mathcal{S}^{-+}$),

%the solid lines in Fig.~\ref{fig:multi-magnon}{\bf A} are the 1-magnon exciations, while the dashed lines represent two-magnon bound states induced by the easy-axis single-ion anisotropy~\cite{SilberglittR1970,OguchiT1971}. Similarly, the slopes of the ``2-boson continuum'' are the same as the 1-magon excitations, revealing their nature as the three-magnon continuum (fluctuations of 3 single-spin-flips).

\begin{figure}[tbp!]
\includegraphics[width=0.99\columnwidth]{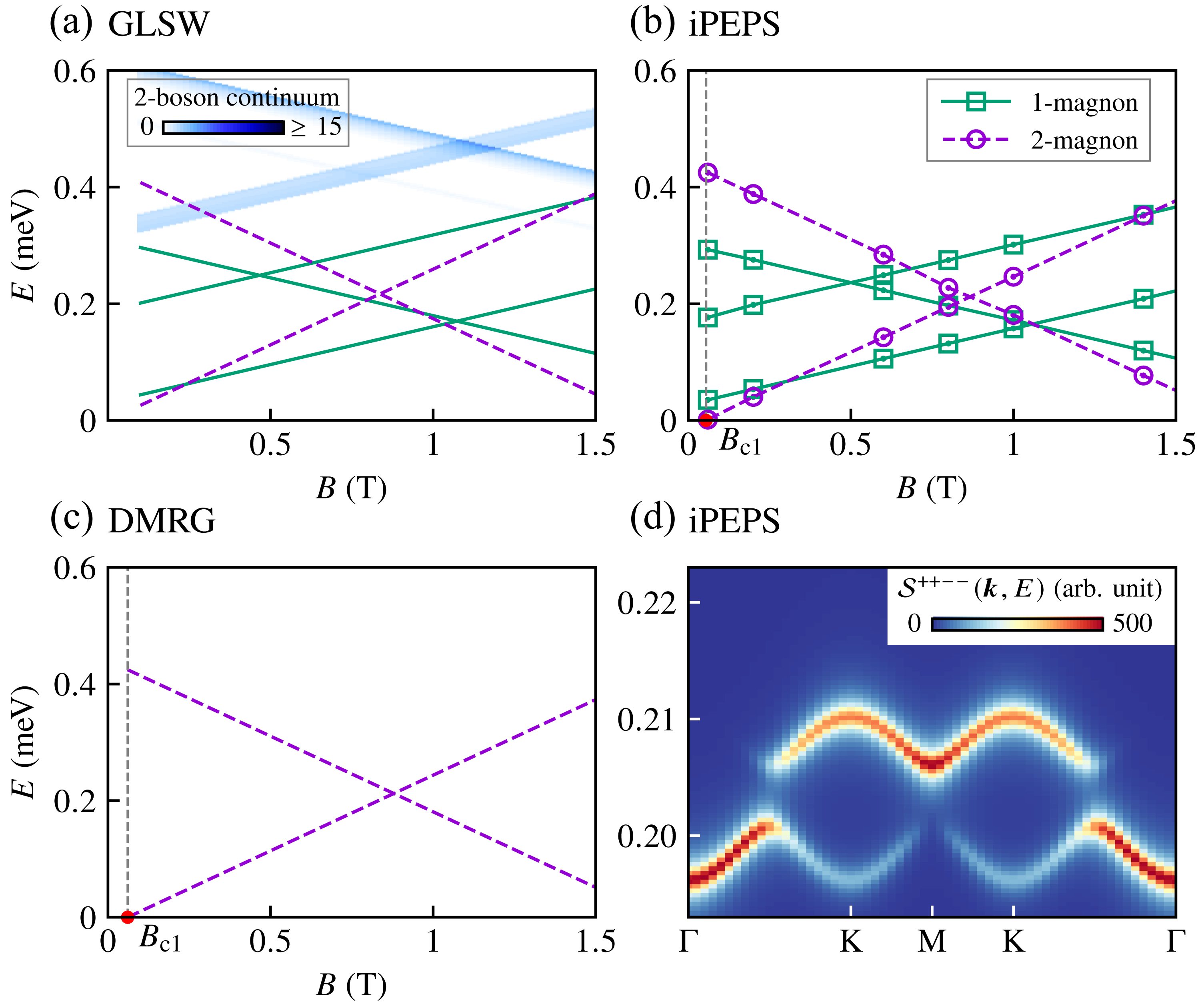}
\caption{{Single- and multimagnon excitations.}
(a) $T=0$ GLSW results of the one-magnon excitations (solid lines) and two-magnon bound states (dashed lines) at the $\Gamma$ point, and the two-boson continuum contribution to the transverse fluctuations $\mathcal{S}^{xx+yy}(\Gamma,E)$ (heat map). 
(b) $T=0$ iPEPS results of the one-magnon (squares) and two-magnon bound states (circles) at the $\Gamma$ point calculated with bond dimension $\chi=4$, where the lines are guides to the eye. 
(c) $T=0$ DMRG results of the lowest-energy two-magnon bound states calculated on a $9\times 6$ TL torus with bond dimension 3500. The two-magnon BEC critical point is indicated by the red dot at $B_\text{c1}$ in both (b) and (c).
(d) $T=0$ iPEPS results of the DQSF $\mathcal{S}^{++--}(\bm{k},E)$ calculated at $B=\qty{0.8}{T}$ with bond dimension $\chi=4$.}
\label{fig:multi-magnon}
\end{figure}

The iPEPS calculations of the dynamic quadrupolar structure factors (DQSFs) $\mathcal{S}^{++--}(\bm{k},E)$ and $\mathcal{S}^{--++}(\bm{k},E)$ at $T=0$ confirmed the presence of two-magnon bound states (see Supplemental Material~\cite{SM}), 
which form relatively narrow bands [Fig.~\ref{fig:multi-magnon}(d)]. The $\mathcal{S}^{++--}(\bm{k},E)$ branch contains degenerate minima at the $\Gamma$ and $K$ points, and the energy decreases as we lower the magnetic field. 
Figure \ref{fig:multi-magnon}(b) shows the energy evolution of the one- and two-magnon bound states at the $\Gamma$ point of the iPEPS calculation.
Remarkably, the two-magnon bound state in the UUD phase undergoes condensation before the 1-magnon BEC occurs [Fig.~\ref{fig:multi-magnon}(b)]. The critical field determined by the two-magnon BEC, marked by the red dot in Fig.~\ref{fig:multi-magnon}(b), agrees quantitatively with the critical field $B_\text{c1}$ derived from the order parameters (Fig.~\ref{fig:magnetization}). This correlation establishes two-magnon BEC transition at $B_\text{c1}$ as the origin of the SN-supersolid phase. 
Furthermore, DMRG calculations of the two-magnon BEC critical point [Fig.~\ref{fig:multi-magnon}(c)] also agree quantitatively with the $B_\text{c1}$ value in Fig.~\ref{fig:magnetization}, further supporting this conclusion.

%In \NiP{}, the two-magnon bound state in the FP phase was shown to go through a BEC transition that induces the SN phase below the saturation field~\cite{ShengJ2025_NiP}. A natural question is: Could the two-magnon bound state in the UUD phase also be responsible for the SN-supersolid phase below $B_\text{c1}$? 
%Before answering this question, we should keep in mind that the two-magnon bound states in the UUD phase are invisible to the INS measurements since they are located in the $\Delta S^z = \pm 2$ sectors, similar to their invisibility in the FP phase. This selection rule is due to the $U(1)$ symmetry of the model \eqref{eq:Ham} for \NiP{}; In the presence of $U(1)$-breaking anisotropies, the two-magnon bound states could show up in INS through hybridization with the 1-magnon bands~\cite{BaiX2021}. 
%Since the model \eqref{eq:Ham} has been proven to be highly accurate for \NiP{}, we can use iPEPS calculations to accurately track the evolution of two-magnon bound states as a function of the magnetic field.

%Figure~\ref{fig:multi-magnon}{\bf B} shows the iPEPS calculation of the 1-magnon and two-magnon states at the $\Gamma$ point. {\color{red} [xx check the iPEPS condensation point and $B_\text{c1}$ again. xx] }

%{\it Discussion.} 
The INS data at $B=\qty{1.72}{T}$ reveal additional weak-intensity modes near $0.1$ and \qty{0.4}{meV} that are not present in the iPEPS results [Fig.~\ref{fig:neutron}(c)]. This discrepancy is expected due to the first-order nature of the transition at $B_\text{c2}$ (Fig.~\ref{fig:magnetization}), which allows the UUD domains to  persist slightly above the critical field. 
Figure~\ref{fig:multi-magnon}(b), which tracks the field dependence of the magnetic excitations of the UUD state,  corroborates this explanation, as the extrapolation to \qty{1.72}{T} aligns with these additional modes (see Supplemental Material~\cite{SM}).
%By checking the INS data at \qty{1.6}{T}, we can indeed see that these modes are exactly the excitations of the domains that remain in the UUD phase even at \qty{1.72}{T} (see Fig.~?? in supplementary materials).

A key distinction exists in the quadrupole waves observed between the SN-supersolid and SN phases in this INS experiment at $T=\qty{60}{mK}$.
At $B=\qty{1.72}{T}$, the quadrupole waves originate from a long-range-ordered SN phase. However, at $B=\qty{0}{T}$, \qty{60}{mK} appears to be above the quadrupole transition temperature, suggesting that the low-energy narrow mode observed by INS represents ``paramagnon''-type quadrupole waves rather than those from a U(1)-broken phase. 
Nevertheless, the observation of this distinct mode and its agreement with the $T=0$ iPEPS calculations strongly indicate the presence of the SN-supersolid phase for $T \lesssim \qty{60}{mK}$.

%As a real material, there are always symmetry-allowed anisotropies on top of the nearest-neighbor-only model \eqref{eq:Ham}.
%Clearly, these anisotropies should be small enough to ensure that Eq.~\eqref{eq:Ham} reproduces the full INS spectra across all phases at a quantitative level, nor should they alter the nature of the SN-supersolid and SN phases. However, 

%{\color{red} [xx mention that the discussion of multipolar order is not completely new, particularly in f-electron systems~\cite{SantiniP2009_RMP} (also the Ir Nature paper). However, 1. most of them happen only at finite-T; 2. lack of clean/simple model in these systems, that complicates the identification of both the ground state and the multipolar exciations. 3. For actinide dioxides, multipole-multipole interations appear to be the key; similarly, typical discussion on quadrupolar order requires a large biquadratic term. These multipole-multipole interactions are shown to be not necessary in \NiP{} (conceptual advancement), providing new insights into some long-standing problems such as the multipolar order in AmO$_2$ without multipolar interactions. xx] }

Our study presents the first instance where the existence of SN orders is supported by a microscopic model that accurately reproduces the INS spectra across all fields.
The low-energy quadrupole waves observed in \NiP{} for phases adjacent to the UUD state offer compelling evidence for SN-supersolid and SN phases, whose common origin is found to be the condensation of two-magnon bound states. 
Remarkably, the TL model \eqref{eq:Ham} supports the SN phases in \NiP{} without invoking a finite quadrupole-quadrupole interaction typically considered essential for the SN order~\cite{TsunetsuguH2006,LauchliA2006,SantiniP2009_RMP}.

{\it Acknowledgments}--We thank Hai-Qing Lin, Qingming Zhang, Chao Cao, and Ziji Xiang for helpful discussions. 
This work was supported by the National Key Research and Development Program of China (Grants No.~2021YFA1400400, No.~2021ZD0301800, No.~2022YFA1402200, and No.~2024YFA1408303), the National Natural Science Foundation of China (Grants No.~12134020, No.~12034017, No.~12104255, No.~12322403, No.~12347107, No.~12374124, No.~12374146, No.~12374150, No.~12488201, and No.~12574157), the Strategic Priority Research Program of Chinese Academy of Sciences (Grants No.~XDB0500202 and No.~XDB33000000), the Youth Innovation Promotion Association of Chinese Academy of Sciences (Grant No.~2021004), the Key R\&D Program of Zhejiang Province, China (Grant No.~2021C01002), the Guangdong Provincial Quantum Science Strategic Initiative (Grants No.~GDZX240100 and No.~GDZX2401007), and the Fundamental Research Funds for the Central Universities (Grant No.~226-2024-00068). We also acknowledge the neutron beam time awarded by Materials and Life Science Experimental Facility of the Japan Proton Accelerator Research Complex (J-PARC) through Proposal No.~2024A0351.
A portion of this research used resources at the Spallation Neutron Source, a DOE Office of Science User Facility operated by the Oak Ridge National Laboratory, where the beam time was allocated to CNCS on Proposal No.~IPTS-29333.
%and Spallation Neutron Source, Oak Ridge National Laboratory (ORNL) through proposal No.~29333. 

{\it Data availability}--The data that support the findings of this article are openly available~\cite{das}.

\bibliography{refs}

\begin{thebibliography}{59}%
\makeatletter
\providecommand \@ifxundefined [1]{%
 \@ifx{#1\undefined}
}%
\providecommand \@ifnum [1]{%
 \ifnum #1\expandafter \@firstoftwo
 \else \expandafter \@secondoftwo
 \fi
}%
\providecommand \@ifx [1]{%
 \ifx #1\expandafter \@firstoftwo
 \else \expandafter \@secondoftwo
 \fi
}%
\providecommand \natexlab [1]{#1}%
\providecommand \enquote  [1]{``#1''}%
\providecommand \bibnamefont  [1]{#1}%
\providecommand \bibfnamefont [1]{#1}%
\providecommand \citenamefont [1]{#1}%
\providecommand \href@noop [0]{\@secondoftwo}%
\providecommand \href [0]{\begingroup \@sanitize@url \@href}%
\providecommand \@href[1]{\@@startlink{#1}\@@href}%
\providecommand \@@href[1]{\endgroup#1\@@endlink}%
\providecommand \@sanitize@url [0]{\catcode `\\12\catcode `\$12\catcode
  `\&12\catcode `\#12\catcode `\^12\catcode `\_12\catcode `\%12\relax}%
\providecommand \@@startlink[1]{}%
\providecommand \@@endlink[0]{}%
\providecommand \url  [0]{\begingroup\@sanitize@url \@url }%
\providecommand \@url [1]{\endgroup\@href {#1}{\urlprefix }}%
\providecommand \urlprefix  [0]{URL }%
\providecommand \Eprint [0]{\href }%
\providecommand \doibase [0]{https://doi.org/}%
\providecommand \selectlanguage [0]{\@gobble}%
\providecommand \bibinfo  [0]{\@secondoftwo}%
\providecommand \bibfield  [0]{\@secondoftwo}%
\providecommand \translation [1]{[#1]}%
\providecommand \BibitemOpen [0]{}%
\providecommand \bibitemStop [0]{}%
\providecommand \bibitemNoStop [0]{.\EOS\space}%
\providecommand \EOS [0]{\spacefactor3000\relax}%
\providecommand \BibitemShut  [1]{\csname bibitem#1\endcsname}%
\let\auto@bib@innerbib\@empty
%</preamble>
\bibitem [{\citenamefont {Blume}\ and\ \citenamefont
  {Hsieh}(1969)}]{BlumeM1969}%
  \BibitemOpen
  \bibfield  {author} {\bibinfo {author} {\bibfnamefont {M.}~\bibnamefont
  {Blume}}\ and\ \bibinfo {author} {\bibfnamefont {Y.~Y.}\ \bibnamefont
  {Hsieh}},\ }\bibfield  {title} {\bibinfo {title} {Biquadratic {{Exchange}}
  and {{Quadrupolar Ordering}}},\ }\href {https://doi.org/10.1063/1.1657616}
  {\bibfield  {journal} {\bibinfo  {journal} {J. Appl. Phys.}\ }\textbf
  {\bibinfo {volume} {40}},\ \bibinfo {pages} {1249} (\bibinfo {year}
  {1969})}\BibitemShut {NoStop}%
\bibitem [{\citenamefont {Matveev}(1973)}]{MatveevVM1973}%
  \BibitemOpen
  \bibfield  {author} {\bibinfo {author} {\bibfnamefont {V.~M.}\ \bibnamefont
  {Matveev}},\ }\bibfield  {title} {\bibinfo {title} {Quantum quadrupolar
  magnetism and phase transitions in the presence of biquadratic exchange},\
  }\href@noop {} {\bibfield  {journal} {\bibinfo  {journal} {Zh. Eksp. Teor.
  Fiz.}\ }\textbf {\bibinfo {volume} {65}},\ \bibinfo {pages} {1626} (\bibinfo
  {year} {1973})},\ \bibinfo {note} {[Sov. Phys. JETP \textbf{38}, 813
  (1974)]}\BibitemShut {NoStop}%
\bibitem [{\citenamefont {Andreev}\ and\ \citenamefont
  {Grishchuk}(1984)}]{AndreevAF1984}%
  \BibitemOpen
  \bibfield  {author} {\bibinfo {author} {\bibfnamefont {A.~F.}\ \bibnamefont
  {Andreev}}\ and\ \bibinfo {author} {\bibfnamefont {I.~A.}\ \bibnamefont
  {Grishchuk}},\ }\bibfield  {title} {\bibinfo {title} {Spin nematics},\
  }\href@noop {} {\bibfield  {journal} {\bibinfo  {journal} {Zh. Eksp. Teor.
  Fiz.}\ }\textbf {\bibinfo {volume} {87}},\ \bibinfo {pages} {467} (\bibinfo
  {year} {1984})},\ \bibinfo {note} {[Sov. Phys. JETP \textbf{60}, 267
  (1984)]}\BibitemShut {NoStop}%
\bibitem [{\citenamefont {Papanicolaou}(1988)}]{PapanicolaouN1988}%
  \BibitemOpen
  \bibfield  {author} {\bibinfo {author} {\bibfnamefont {N.}~\bibnamefont
  {Papanicolaou}},\ }\bibfield  {title} {\bibinfo {title} {Unusual phases in
  quantum spin-1 systems},\ }\href
  {https://doi.org/10.1016/0550-3213(88)90073-9} {\bibfield  {journal}
  {\bibinfo  {journal} {Nucl. Phys. B}\ }\textbf {\bibinfo {volume} {305}},\
  \bibinfo {pages} {367} (\bibinfo {year} {1988})}\BibitemShut {NoStop}%
\bibitem [{\citenamefont {Chubukov}(1990)}]{ChubukovAV1990}%
  \BibitemOpen
  \bibfield  {author} {\bibinfo {author} {\bibfnamefont {A.~V.}\ \bibnamefont
  {Chubukov}},\ }\bibfield  {title} {\bibinfo {title} {Fluctuations in spin
  nematics},\ }\href {https://doi.org/10.1088/0953-8984/2/6/018} {\bibfield
  {journal} {\bibinfo  {journal} {J. Phys.: Condens. Matter}\ }\textbf
  {\bibinfo {volume} {2}},\ \bibinfo {pages} {1593} (\bibinfo {year}
  {1990})}\BibitemShut {NoStop}%
\bibitem [{\citenamefont {Svistov}\ \emph {et~al.}(2011)\citenamefont
  {Svistov}, \citenamefont {Fujita}, \citenamefont {Yamaguchi}, \citenamefont
  {Kimura}, \citenamefont {Omura}, \citenamefont {Prokofiev}, \citenamefont
  {Smirnov}, \citenamefont {Honda},\ and\ \citenamefont
  {Hagiwara}}]{SvistovLE2011}%
  \BibitemOpen
  \bibfield  {author} {\bibinfo {author} {\bibfnamefont {L.~E.}\ \bibnamefont
  {Svistov}}, \bibinfo {author} {\bibfnamefont {T.}~\bibnamefont {Fujita}},
  \bibinfo {author} {\bibfnamefont {H.}~\bibnamefont {Yamaguchi}}, \bibinfo
  {author} {\bibfnamefont {S.}~\bibnamefont {Kimura}}, \bibinfo {author}
  {\bibfnamefont {K.}~\bibnamefont {Omura}}, \bibinfo {author} {\bibfnamefont
  {A.}~\bibnamefont {Prokofiev}}, \bibinfo {author} {\bibfnamefont {A.~I.}\
  \bibnamefont {Smirnov}}, \bibinfo {author} {\bibfnamefont {Z.}~\bibnamefont
  {Honda}},\ and\ \bibinfo {author} {\bibfnamefont {M.}~\bibnamefont
  {Hagiwara}},\ }\bibfield  {title} {\bibinfo {title} {New high magnetic field
  phase of the frustrated {{{\emph{S}}}}=1/2 chain compound
  {{LiCuVO}}{$_{4}$}},\ }\href {https://doi.org/10.1134/S0021364011010073}
  {\bibfield  {journal} {\bibinfo  {journal} {JETP Lett.}\ }\textbf {\bibinfo
  {volume} {93}},\ \bibinfo {pages} {21} (\bibinfo {year} {2011})}\BibitemShut
  {NoStop}%
\bibitem [{\citenamefont {Orlova}\ \emph {et~al.}(2017)\citenamefont {Orlova},
  \citenamefont {Green}, \citenamefont {Law}, \citenamefont {Gorbunov},
  \citenamefont {Chanda}, \citenamefont {Kr{\"a}mer}, \citenamefont
  {Horvati{\'c}}, \citenamefont {Kremer}, \citenamefont {Wosnitza},\ and\
  \citenamefont {Rikken}}]{OrlovaA2017}%
  \BibitemOpen
  \bibfield  {author} {\bibinfo {author} {\bibfnamefont {A.}~\bibnamefont
  {Orlova}}, \bibinfo {author} {\bibfnamefont {E.~L.}\ \bibnamefont {Green}},
  \bibinfo {author} {\bibfnamefont {J.~M.}\ \bibnamefont {Law}}, \bibinfo
  {author} {\bibfnamefont {D.~I.}\ \bibnamefont {Gorbunov}}, \bibinfo {author}
  {\bibfnamefont {G.}~\bibnamefont {Chanda}}, \bibinfo {author} {\bibfnamefont
  {S.}~\bibnamefont {Kr{\"a}mer}}, \bibinfo {author} {\bibfnamefont
  {M.}~\bibnamefont {Horvati{\'c}}}, \bibinfo {author} {\bibfnamefont {R.~K.}\
  \bibnamefont {Kremer}}, \bibinfo {author} {\bibfnamefont {J.}~\bibnamefont
  {Wosnitza}},\ and\ \bibinfo {author} {\bibfnamefont {G.~L. J.~A.}\
  \bibnamefont {Rikken}},\ }\bibfield  {title} {\bibinfo {title} {Nuclear
  {{Magnetic Resonance Signature}} of the {{Spin-Nematic Phase}} in
  {{LiCuVO}}{$_4$} at {{High Magnetic Fields}}},\ }\href
  {https://doi.org/10.1103/PhysRevLett.118.247201} {\bibfield  {journal}
  {\bibinfo  {journal} {Phys. Rev. Lett.}\ }\textbf {\bibinfo {volume} {118}},\
  \bibinfo {pages} {247201} (\bibinfo {year} {2017})}\BibitemShut {NoStop}%
\bibitem [{\citenamefont {Povarov}\ \emph {et~al.}(2019)\citenamefont
  {Povarov}, \citenamefont {Bhartiya}, \citenamefont {Yan},\ and\ \citenamefont
  {Zheludev}}]{PovarovKY2019}%
  \BibitemOpen
  \bibfield  {author} {\bibinfo {author} {\bibfnamefont {K.~{\relax Yu}.}\
  \bibnamefont {Povarov}}, \bibinfo {author} {\bibfnamefont {V.~K.}\
  \bibnamefont {Bhartiya}}, \bibinfo {author} {\bibfnamefont {Z.}~\bibnamefont
  {Yan}},\ and\ \bibinfo {author} {\bibfnamefont {A.}~\bibnamefont
  {Zheludev}},\ }\bibfield  {title} {\bibinfo {title} {Thermodynamics of a
  frustrated quantum magnet on a square lattice},\ }\href
  {https://doi.org/10.1103/PhysRevB.99.024413} {\bibfield  {journal} {\bibinfo
  {journal} {Phys. Rev. B}\ }\textbf {\bibinfo {volume} {99}},\ \bibinfo
  {pages} {024413} (\bibinfo {year} {2019})}\BibitemShut {NoStop}%
\bibitem [{\citenamefont {Skoulatos}\ \emph {et~al.}(2019)\citenamefont
  {Skoulatos}, \citenamefont {Rucker}, \citenamefont {Nilsen}, \citenamefont
  {Bertin}, \citenamefont {Pomjakushina}, \citenamefont {Ollivier},
  \citenamefont {Schneidewind}, \citenamefont {Georgii}, \citenamefont
  {Zaharko}, \citenamefont {Keller}, \citenamefont {R{\"u}egg}, \citenamefont
  {Pfleiderer}, \citenamefont {Schmidt}, \citenamefont {Shannon}, \citenamefont
  {Kriele}, \citenamefont {Senyshyn},\ and\ \citenamefont
  {Smerald}}]{SkoulatosM2019}%
  \BibitemOpen
  \bibfield  {author} {\bibinfo {author} {\bibfnamefont {M.}~\bibnamefont
  {Skoulatos}}, \bibinfo {author} {\bibfnamefont {F.}~\bibnamefont {Rucker}},
  \bibinfo {author} {\bibfnamefont {G.~J.}\ \bibnamefont {Nilsen}}, \bibinfo
  {author} {\bibfnamefont {A.}~\bibnamefont {Bertin}}, \bibinfo {author}
  {\bibfnamefont {E.}~\bibnamefont {Pomjakushina}}, \bibinfo {author}
  {\bibfnamefont {J.}~\bibnamefont {Ollivier}}, \bibinfo {author}
  {\bibfnamefont {A.}~\bibnamefont {Schneidewind}}, \bibinfo {author}
  {\bibfnamefont {R.}~\bibnamefont {Georgii}}, \bibinfo {author} {\bibfnamefont
  {O.}~\bibnamefont {Zaharko}}, \bibinfo {author} {\bibfnamefont
  {L.}~\bibnamefont {Keller}}, \bibinfo {author} {\bibfnamefont {{\relax
  Ch}.}~\bibnamefont {R{\"u}egg}}, \bibinfo {author} {\bibfnamefont
  {C.}~\bibnamefont {Pfleiderer}}, \bibinfo {author} {\bibfnamefont
  {B.}~\bibnamefont {Schmidt}}, \bibinfo {author} {\bibfnamefont
  {N.}~\bibnamefont {Shannon}}, \bibinfo {author} {\bibfnamefont
  {A.}~\bibnamefont {Kriele}}, \bibinfo {author} {\bibfnamefont
  {A.}~\bibnamefont {Senyshyn}},\ and\ \bibinfo {author} {\bibfnamefont
  {A.}~\bibnamefont {Smerald}},\ }\bibfield  {title} {\bibinfo {title}
  {Putative spin-nematic phase in {{BaCdVO}}({{PO}}{$_4$}){$_{2}$}},\ }\href
  {https://doi.org/10.1103/PhysRevB.100.014405} {\bibfield  {journal} {\bibinfo
   {journal} {Phys. Rev. B}\ }\textbf {\bibinfo {volume} {100}},\ \bibinfo
  {pages} {014405} (\bibinfo {year} {2019})}\BibitemShut {NoStop}%
\bibitem [{\citenamefont {Bhartiya}\ \emph {et~al.}(2019)\citenamefont
  {Bhartiya}, \citenamefont {Povarov}, \citenamefont {Blosser}, \citenamefont
  {Bettler}, \citenamefont {Yan}, \citenamefont {Gvasaliya}, \citenamefont
  {Raymond}, \citenamefont {Ressouche}, \citenamefont {Beauvois}, \citenamefont
  {Xu}, \citenamefont {Yokaichiya},\ and\ \citenamefont
  {Zheludev}}]{BhartiyaVK2019}%
  \BibitemOpen
  \bibfield  {author} {\bibinfo {author} {\bibfnamefont {V.~K.}\ \bibnamefont
  {Bhartiya}}, \bibinfo {author} {\bibfnamefont {K.~{\relax Yu}.}\ \bibnamefont
  {Povarov}}, \bibinfo {author} {\bibfnamefont {D.}~\bibnamefont {Blosser}},
  \bibinfo {author} {\bibfnamefont {S.}~\bibnamefont {Bettler}}, \bibinfo
  {author} {\bibfnamefont {Z.}~\bibnamefont {Yan}}, \bibinfo {author}
  {\bibfnamefont {S.}~\bibnamefont {Gvasaliya}}, \bibinfo {author}
  {\bibfnamefont {S.}~\bibnamefont {Raymond}}, \bibinfo {author} {\bibfnamefont
  {E.}~\bibnamefont {Ressouche}}, \bibinfo {author} {\bibfnamefont
  {K.}~\bibnamefont {Beauvois}}, \bibinfo {author} {\bibfnamefont
  {J.}~\bibnamefont {Xu}}, \bibinfo {author} {\bibfnamefont {F.}~\bibnamefont
  {Yokaichiya}},\ and\ \bibinfo {author} {\bibfnamefont {A.}~\bibnamefont
  {Zheludev}},\ }\bibfield  {title} {\bibinfo {title} {Presaturation phase with
  no dipolar order in a quantum ferro-antiferromagnet},\ }\href
  {https://doi.org/10.1103/PhysRevResearch.1.033078} {\bibfield  {journal}
  {\bibinfo  {journal} {Phys. Rev. Res.}\ }\textbf {\bibinfo {volume} {1}},\
  \bibinfo {pages} {033078} (\bibinfo {year} {2019})}\BibitemShut {NoStop}%
\bibitem [{\citenamefont {Ishikawa}\ \emph {et~al.}(2015)\citenamefont
  {Ishikawa}, \citenamefont {Yoshida}, \citenamefont {Nawa}, \citenamefont
  {Jeong}, \citenamefont {Kr{\"a}mer}, \citenamefont {Horvati{\'c}},
  \citenamefont {Berthier}, \citenamefont {Takigawa}, \citenamefont {Akaki},
  \citenamefont {Miyake}, \citenamefont {Tokunaga}, \citenamefont {Kindo},
  \citenamefont {Yamaura}, \citenamefont {Okamoto},\ and\ \citenamefont
  {Hiroi}}]{IshikawaH2015}%
  \BibitemOpen
  \bibfield  {author} {\bibinfo {author} {\bibfnamefont {H.}~\bibnamefont
  {Ishikawa}}, \bibinfo {author} {\bibfnamefont {M.}~\bibnamefont {Yoshida}},
  \bibinfo {author} {\bibfnamefont {K.}~\bibnamefont {Nawa}}, \bibinfo {author}
  {\bibfnamefont {M.}~\bibnamefont {Jeong}}, \bibinfo {author} {\bibfnamefont
  {S.}~\bibnamefont {Kr{\"a}mer}}, \bibinfo {author} {\bibfnamefont
  {M.}~\bibnamefont {Horvati{\'c}}}, \bibinfo {author} {\bibfnamefont
  {C.}~\bibnamefont {Berthier}}, \bibinfo {author} {\bibfnamefont
  {M.}~\bibnamefont {Takigawa}}, \bibinfo {author} {\bibfnamefont
  {M.}~\bibnamefont {Akaki}}, \bibinfo {author} {\bibfnamefont
  {A.}~\bibnamefont {Miyake}}, \bibinfo {author} {\bibfnamefont
  {M.}~\bibnamefont {Tokunaga}}, \bibinfo {author} {\bibfnamefont
  {K.}~\bibnamefont {Kindo}}, \bibinfo {author} {\bibfnamefont
  {J.}~\bibnamefont {Yamaura}}, \bibinfo {author} {\bibfnamefont
  {Y.}~\bibnamefont {Okamoto}},\ and\ \bibinfo {author} {\bibfnamefont
  {Z.}~\bibnamefont {Hiroi}},\ }\bibfield  {title} {\bibinfo {title}
  {One-{{Third Magnetization Plateau}} with a {{Preceding Novel Phase}} in
  {{Volborthite}}},\ }\href {https://doi.org/10.1103/PhysRevLett.114.227202}
  {\bibfield  {journal} {\bibinfo  {journal} {Phys. Rev. Lett.}\ }\textbf
  {\bibinfo {volume} {114}},\ \bibinfo {pages} {227202} (\bibinfo {year}
  {2015})}\BibitemShut {NoStop}%
\bibitem [{\citenamefont {Janson}\ \emph {et~al.}(2016)\citenamefont {Janson},
  \citenamefont {Furukawa}, \citenamefont {Momoi}, \citenamefont {Sindzingre},
  \citenamefont {Richter},\ and\ \citenamefont {Held}}]{JansonO2016}%
  \BibitemOpen
  \bibfield  {author} {\bibinfo {author} {\bibfnamefont {O.}~\bibnamefont
  {Janson}}, \bibinfo {author} {\bibfnamefont {S.}~\bibnamefont {Furukawa}},
  \bibinfo {author} {\bibfnamefont {T.}~\bibnamefont {Momoi}}, \bibinfo
  {author} {\bibfnamefont {P.}~\bibnamefont {Sindzingre}}, \bibinfo {author}
  {\bibfnamefont {J.}~\bibnamefont {Richter}},\ and\ \bibinfo {author}
  {\bibfnamefont {K.}~\bibnamefont {Held}},\ }\bibfield  {title} {\bibinfo
  {title} {Magnetic {{Behavior}} of {{Volborthite
  Cu}}{$_{3}$}{{V}}{$_{2}$}{{O}}{$_7$}({{OH}}){$_{2}\cdot$}{{2H}}{$_{2}$}{{O
  Determined}} by {{Coupled Trimers Rather}} than {{Frustrated Chains}}},\
  }\href {https://doi.org/10.1103/PhysRevLett.117.037206} {\bibfield  {journal}
  {\bibinfo  {journal} {Phys. Rev. Lett.}\ }\textbf {\bibinfo {volume} {117}},\
  \bibinfo {pages} {037206} (\bibinfo {year} {2016})}\BibitemShut {NoStop}%
\bibitem [{\citenamefont {Yoshida}\ \emph {et~al.}(2017)\citenamefont
  {Yoshida}, \citenamefont {Nawa}, \citenamefont {Ishikawa}, \citenamefont
  {Takigawa}, \citenamefont {Jeong}, \citenamefont {Kr{\"a}mer}, \citenamefont
  {Horvati{\'c}}, \citenamefont {Berthier}, \citenamefont {Matsui},
  \citenamefont {Goto}, \citenamefont {Kimura}, \citenamefont {Sasaki},
  \citenamefont {Yamaura}, \citenamefont {Yoshida}, \citenamefont {Okamoto},\
  and\ \citenamefont {Hiroi}}]{YoshidaM2017}%
  \BibitemOpen
  \bibfield  {author} {\bibinfo {author} {\bibfnamefont {M.}~\bibnamefont
  {Yoshida}}, \bibinfo {author} {\bibfnamefont {K.}~\bibnamefont {Nawa}},
  \bibinfo {author} {\bibfnamefont {H.}~\bibnamefont {Ishikawa}}, \bibinfo
  {author} {\bibfnamefont {M.}~\bibnamefont {Takigawa}}, \bibinfo {author}
  {\bibfnamefont {M.}~\bibnamefont {Jeong}}, \bibinfo {author} {\bibfnamefont
  {S.}~\bibnamefont {Kr{\"a}mer}}, \bibinfo {author} {\bibfnamefont
  {M.}~\bibnamefont {Horvati{\'c}}}, \bibinfo {author} {\bibfnamefont
  {C.}~\bibnamefont {Berthier}}, \bibinfo {author} {\bibfnamefont
  {K.}~\bibnamefont {Matsui}}, \bibinfo {author} {\bibfnamefont
  {T.}~\bibnamefont {Goto}}, \bibinfo {author} {\bibfnamefont {S.}~\bibnamefont
  {Kimura}}, \bibinfo {author} {\bibfnamefont {T.}~\bibnamefont {Sasaki}},
  \bibinfo {author} {\bibfnamefont {J.}~\bibnamefont {Yamaura}}, \bibinfo
  {author} {\bibfnamefont {H.}~\bibnamefont {Yoshida}}, \bibinfo {author}
  {\bibfnamefont {Y.}~\bibnamefont {Okamoto}},\ and\ \bibinfo {author}
  {\bibfnamefont {Z.}~\bibnamefont {Hiroi}},\ }\bibfield  {title} {\bibinfo
  {title} {Spin dynamics in the high-field phases of volborthite},\ }\href
  {https://doi.org/10.1103/PhysRevB.96.180413} {\bibfield  {journal} {\bibinfo
  {journal} {Phys. Rev. B}\ }\textbf {\bibinfo {volume} {96}},\ \bibinfo
  {pages} {180413} (\bibinfo {year} {2017})}\BibitemShut {NoStop}%
\bibitem [{\citenamefont {Kohama}\ \emph {et~al.}(2019)\citenamefont {Kohama},
  \citenamefont {Ishikawa}, \citenamefont {Matsuo}, \citenamefont {Kindo},
  \citenamefont {Shannon},\ and\ \citenamefont {Hiroi}}]{KohamaY2019}%
  \BibitemOpen
  \bibfield  {author} {\bibinfo {author} {\bibfnamefont {Y.}~\bibnamefont
  {Kohama}}, \bibinfo {author} {\bibfnamefont {H.}~\bibnamefont {Ishikawa}},
  \bibinfo {author} {\bibfnamefont {A.}~\bibnamefont {Matsuo}}, \bibinfo
  {author} {\bibfnamefont {K.}~\bibnamefont {Kindo}}, \bibinfo {author}
  {\bibfnamefont {N.}~\bibnamefont {Shannon}},\ and\ \bibinfo {author}
  {\bibfnamefont {Z.}~\bibnamefont {Hiroi}},\ }\bibfield  {title} {\bibinfo
  {title} {Possible observation of quantum spin-nematic phase in a frustrated
  magnet},\ }\href {https://doi.org/10.1073/pnas.1821969116} {\bibfield
  {journal} {\bibinfo  {journal} {Proc. Natl. Acad. Sci. U.S.A.}\ }\textbf
  {\bibinfo {volume} {116}},\ \bibinfo {pages} {10686} (\bibinfo {year}
  {2019})}\BibitemShut {NoStop}%
\bibitem [{\citenamefont {Kim}\ \emph {et~al.}(2024)\citenamefont {Kim},
  \citenamefont {Kim}, \citenamefont {Kwon}, \citenamefont {Kim}, \citenamefont
  {Kim}, \citenamefont {Ha}, \citenamefont {Kim}, \citenamefont {Lee},
  \citenamefont {Kim}, \citenamefont {Cho}, \citenamefont {Heo}, \citenamefont
  {Jang}, \citenamefont {Sahle}, \citenamefont {Longo}, \citenamefont
  {Strempfer}, \citenamefont {Fabbris}, \citenamefont {Choi}, \citenamefont
  {Haskel}, \citenamefont {Kim}, \citenamefont {Kim},\ and\ \citenamefont
  {Kim}}]{KimH2024}%
  \BibitemOpen
  \bibfield  {author} {\bibinfo {author} {\bibfnamefont {H.}~\bibnamefont
  {Kim}}, \bibinfo {author} {\bibfnamefont {J.-K.}\ \bibnamefont {Kim}},
  \bibinfo {author} {\bibfnamefont {J.}~\bibnamefont {Kwon}}, \bibinfo {author}
  {\bibfnamefont {J.}~\bibnamefont {Kim}}, \bibinfo {author} {\bibfnamefont
  {H.-W.~J.}\ \bibnamefont {Kim}}, \bibinfo {author} {\bibfnamefont
  {S.}~\bibnamefont {Ha}}, \bibinfo {author} {\bibfnamefont {K.}~\bibnamefont
  {Kim}}, \bibinfo {author} {\bibfnamefont {W.}~\bibnamefont {Lee}}, \bibinfo
  {author} {\bibfnamefont {J.}~\bibnamefont {Kim}}, \bibinfo {author}
  {\bibfnamefont {G.~Y.}\ \bibnamefont {Cho}}, \bibinfo {author} {\bibfnamefont
  {H.}~\bibnamefont {Heo}}, \bibinfo {author} {\bibfnamefont {J.}~\bibnamefont
  {Jang}}, \bibinfo {author} {\bibfnamefont {C.~J.}\ \bibnamefont {Sahle}},
  \bibinfo {author} {\bibfnamefont {A.}~\bibnamefont {Longo}}, \bibinfo
  {author} {\bibfnamefont {J.}~\bibnamefont {Strempfer}}, \bibinfo {author}
  {\bibfnamefont {G.}~\bibnamefont {Fabbris}}, \bibinfo {author} {\bibfnamefont
  {Y.}~\bibnamefont {Choi}}, \bibinfo {author} {\bibfnamefont {D.}~\bibnamefont
  {Haskel}}, \bibinfo {author} {\bibfnamefont {J.}~\bibnamefont {Kim}},
  \bibinfo {author} {\bibfnamefont {J.-W.}\ \bibnamefont {Kim}},\ and\ \bibinfo
  {author} {\bibfnamefont {B.~J.}\ \bibnamefont {Kim}},\ }\bibfield  {title}
  {\bibinfo {title} {Quantum spin nematic phase in a square-lattice iridate},\
  }\href {https://doi.org/10.1038/s41586-023-06829-4} {\bibfield  {journal}
  {\bibinfo  {journal} {Nature}\ }\textbf {\bibinfo {volume} {625}},\ \bibinfo
  {pages} {264} (\bibinfo {year} {2024})}\BibitemShut {NoStop}%
\bibitem [{\citenamefont {Wang}\ and\ \citenamefont
  {Batista}(2018)}]{WangZ2018_SSM}%
  \BibitemOpen
  \bibfield  {author} {\bibinfo {author} {\bibfnamefont {Z.}~\bibnamefont
  {Wang}}\ and\ \bibinfo {author} {\bibfnamefont {C.~D.}\ \bibnamefont
  {Batista}},\ }\bibfield  {title} {\bibinfo {title} {Dynamics and
  {{Instabilities}} of the {{Shastry-Sutherland Model}}},\ }\href
  {https://doi.org/10.1103/PhysRevLett.120.247201} {\bibfield  {journal}
  {\bibinfo  {journal} {Phys. Rev. Lett.}\ }\textbf {\bibinfo {volume} {120}},\
  \bibinfo {pages} {247201} (\bibinfo {year} {2018})}\BibitemShut {NoStop}%
\bibitem [{\citenamefont {Fogh}\ \emph {et~al.}(2024)\citenamefont {Fogh},
  \citenamefont {Nayak}, \citenamefont {Prokhnenko}, \citenamefont
  {Bartkowiak}, \citenamefont {Munakata}, \citenamefont {Soh}, \citenamefont
  {Turrini}, \citenamefont {Zayed}, \citenamefont {Pomjakushina}, \citenamefont
  {Kageyama}, \citenamefont {Nojiri}, \citenamefont {Kakurai}, \citenamefont
  {Normand}, \citenamefont {Mila},\ and\ \citenamefont
  {R{\o}nnow}}]{FoghE2024}%
  \BibitemOpen
  \bibfield  {author} {\bibinfo {author} {\bibfnamefont {E.}~\bibnamefont
  {Fogh}}, \bibinfo {author} {\bibfnamefont {M.}~\bibnamefont {Nayak}},
  \bibinfo {author} {\bibfnamefont {O.}~\bibnamefont {Prokhnenko}}, \bibinfo
  {author} {\bibfnamefont {M.}~\bibnamefont {Bartkowiak}}, \bibinfo {author}
  {\bibfnamefont {K.}~\bibnamefont {Munakata}}, \bibinfo {author}
  {\bibfnamefont {J.-R.}\ \bibnamefont {Soh}}, \bibinfo {author} {\bibfnamefont
  {A.~A.}\ \bibnamefont {Turrini}}, \bibinfo {author} {\bibfnamefont {M.~E.}\
  \bibnamefont {Zayed}}, \bibinfo {author} {\bibfnamefont {E.}~\bibnamefont
  {Pomjakushina}}, \bibinfo {author} {\bibfnamefont {H.}~\bibnamefont
  {Kageyama}}, \bibinfo {author} {\bibfnamefont {H.}~\bibnamefont {Nojiri}},
  \bibinfo {author} {\bibfnamefont {K.}~\bibnamefont {Kakurai}}, \bibinfo
  {author} {\bibfnamefont {B.}~\bibnamefont {Normand}}, \bibinfo {author}
  {\bibfnamefont {F.}~\bibnamefont {Mila}},\ and\ \bibinfo {author}
  {\bibfnamefont {H.~M.}\ \bibnamefont {R{\o}nnow}},\ }\bibfield  {title}
  {\bibinfo {title} {Field-induced bound-state condensation and spin-nematic
  phase in {{SrCu}}{$_2$}({{BO}}{$_3$}){$_2$} revealed by neutron scattering up
  to 25.9 {{T}}},\ }\href {https://doi.org/10.1038/s41467-023-44115-z}
  {\bibfield  {journal} {\bibinfo  {journal} {Nat. Commun.}\ }\textbf {\bibinfo
  {volume} {15}},\ \bibinfo {pages} {442} (\bibinfo {year} {2024})}\BibitemShut
  {NoStop}%
\bibitem [{\citenamefont {Liu}\ \emph {et~al.}(2025)\citenamefont {Liu},
  \citenamefont {Stone}, \citenamefont {Gao}, \citenamefont {Nakamura},
  \citenamefont {Kamazawa}, \citenamefont {Krajewska}, \citenamefont {Walker},
  \citenamefont {Cheng}, \citenamefont {Yu}, \citenamefont {Si}, \citenamefont
  {Dai},\ and\ \citenamefont {Lu}}]{LiuR2025}%
  \BibitemOpen
  \bibfield  {author} {\bibinfo {author} {\bibfnamefont {R.}~\bibnamefont
  {Liu}}, \bibinfo {author} {\bibfnamefont {M.~B.}\ \bibnamefont {Stone}},
  \bibinfo {author} {\bibfnamefont {S.}~\bibnamefont {Gao}}, \bibinfo {author}
  {\bibfnamefont {M.}~\bibnamefont {Nakamura}}, \bibinfo {author}
  {\bibfnamefont {K.}~\bibnamefont {Kamazawa}}, \bibinfo {author}
  {\bibfnamefont {A.}~\bibnamefont {Krajewska}}, \bibinfo {author}
  {\bibfnamefont {H.~C.}\ \bibnamefont {Walker}}, \bibinfo {author}
  {\bibfnamefont {P.}~\bibnamefont {Cheng}}, \bibinfo {author} {\bibfnamefont
  {R.}~\bibnamefont {Yu}}, \bibinfo {author} {\bibfnamefont {Q.}~\bibnamefont
  {Si}}, \bibinfo {author} {\bibfnamefont {P.}~\bibnamefont {Dai}},\ and\
  \bibinfo {author} {\bibfnamefont {X.}~\bibnamefont {Lu}},\ }\bibfield
  {title} {\bibinfo {title} {Spin correlations in the nematic quantum
  disordered state of {{FeSe}}},\ }\href
  {https://doi.org/10.1038/s41467-025-60071-2} {\bibfield  {journal} {\bibinfo
  {journal} {Nat. Commun.}\ }\textbf {\bibinfo {volume} {16}},\ \bibinfo
  {pages} {5212} (\bibinfo {year} {2025})}\BibitemShut {NoStop}%
\bibitem [{\citenamefont {Santini}\ \emph {et~al.}(2009)\citenamefont
  {Santini}, \citenamefont {Carretta}, \citenamefont {Amoretti}, \citenamefont
  {Caciuffo}, \citenamefont {Magnani},\ and\ \citenamefont
  {Lander}}]{SantiniP2009_RMP}%
  \BibitemOpen
  \bibfield  {author} {\bibinfo {author} {\bibfnamefont {P.}~\bibnamefont
  {Santini}}, \bibinfo {author} {\bibfnamefont {S.}~\bibnamefont {Carretta}},
  \bibinfo {author} {\bibfnamefont {G.}~\bibnamefont {Amoretti}}, \bibinfo
  {author} {\bibfnamefont {R.}~\bibnamefont {Caciuffo}}, \bibinfo {author}
  {\bibfnamefont {N.}~\bibnamefont {Magnani}},\ and\ \bibinfo {author}
  {\bibfnamefont {G.~H.}\ \bibnamefont {Lander}},\ }\bibfield  {title}
  {\bibinfo {title} {Multipolar interactions in {\emph{f}}-electron systems:
  {{The}} paradigm of actinide dioxides},\ }\href
  {https://doi.org/10.1103/RevModPhys.81.807} {\bibfield  {journal} {\bibinfo
  {journal} {Rev. Mod. Phys.}\ }\textbf {\bibinfo {volume} {81}},\ \bibinfo
  {pages} {807} (\bibinfo {year} {2009})}\BibitemShut {NoStop}%
\bibitem [{\citenamefont {Tsunetsugu}\ and\ \citenamefont
  {Arikawa}(2006)}]{TsunetsuguH2006}%
  \BibitemOpen
  \bibfield  {author} {\bibinfo {author} {\bibfnamefont {H.}~\bibnamefont
  {Tsunetsugu}}\ and\ \bibinfo {author} {\bibfnamefont {M.}~\bibnamefont
  {Arikawa}},\ }\bibfield  {title} {\bibinfo {title} {Spin {{Nematic Phase}} in
  {{{\emph{S}}}}=1 {{Triangular Antiferromagnets}}},\ }\href
  {https://doi.org/10.1143/JPSJ.75.083701} {\bibfield  {journal} {\bibinfo
  {journal} {J. Phys. Soc. Jpn.}\ }\textbf {\bibinfo {volume} {75}},\ \bibinfo
  {pages} {083701} (\bibinfo {year} {2006})}\BibitemShut {NoStop}%
\bibitem [{\citenamefont {L{\"a}uchli}\ \emph {et~al.}(2006)\citenamefont
  {L{\"a}uchli}, \citenamefont {Mila},\ and\ \citenamefont
  {Penc}}]{LauchliA2006}%
  \BibitemOpen
  \bibfield  {author} {\bibinfo {author} {\bibfnamefont {A.}~\bibnamefont
  {L{\"a}uchli}}, \bibinfo {author} {\bibfnamefont {F.}~\bibnamefont {Mila}},\
  and\ \bibinfo {author} {\bibfnamefont {K.}~\bibnamefont {Penc}},\ }\bibfield
  {title} {\bibinfo {title} {Quadrupolar {{Phases}} of the {{{\emph{S}}}}=1
  {{Bilinear-Biquadratic Heisenberg Model}} on the {{Triangular Lattice}}},\
  }\href {https://doi.org/10.1103/PhysRevLett.97.087205} {\bibfield  {journal}
  {\bibinfo  {journal} {Phys. Rev. Lett.}\ }\textbf {\bibinfo {volume} {97}},\
  \bibinfo {pages} {087205} (\bibinfo {year} {2006})}\BibitemShut {NoStop}%
\bibitem [{\citenamefont {Shindou}\ \emph {et~al.}(2013)\citenamefont
  {Shindou}, \citenamefont {Yunoki},\ and\ \citenamefont
  {Momoi}}]{ShindouR2013_nematic_DSF}%
  \BibitemOpen
  \bibfield  {author} {\bibinfo {author} {\bibfnamefont {R.}~\bibnamefont
  {Shindou}}, \bibinfo {author} {\bibfnamefont {S.}~\bibnamefont {Yunoki}},\
  and\ \bibinfo {author} {\bibfnamefont {T.}~\bibnamefont {Momoi}},\ }\bibfield
   {title} {\bibinfo {title} {Dynamical spin structure factors of quantum spin
  nematic states},\ }\href {https://doi.org/10.1103/PhysRevB.87.054429}
  {\bibfield  {journal} {\bibinfo  {journal} {Phys. Rev. B}\ }\textbf {\bibinfo
  {volume} {87}},\ \bibinfo {pages} {054429} (\bibinfo {year}
  {2013})}\BibitemShut {NoStop}%
\bibitem [{\citenamefont {Smerald}\ and\ \citenamefont
  {Shannon}(2013)}]{SmeraldA2013}%
  \BibitemOpen
  \bibfield  {author} {\bibinfo {author} {\bibfnamefont {A.}~\bibnamefont
  {Smerald}}\ and\ \bibinfo {author} {\bibfnamefont {N.}~\bibnamefont
  {Shannon}},\ }\bibfield  {title} {\bibinfo {title} {Theory of spin
  excitations in a quantum spin-nematic state},\ }\href
  {https://doi.org/10.1103/PhysRevB.88.184430} {\bibfield  {journal} {\bibinfo
  {journal} {Phys. Rev. B}\ }\textbf {\bibinfo {volume} {88}},\ \bibinfo
  {pages} {184430} (\bibinfo {year} {2013})}\BibitemShut {NoStop}%
\bibitem [{\citenamefont {Li}\ \emph {et~al.}(2021)\citenamefont {Li},
  \citenamefont {Huang}, \citenamefont {Brassington}, \citenamefont {Yue},
  \citenamefont {Chu}, \citenamefont {Guang}, \citenamefont {Zhou},
  \citenamefont {Gao}, \citenamefont {Feng}, \citenamefont {Cao}, \citenamefont
  {Choi}, \citenamefont {Sun}, \citenamefont {Li}, \citenamefont {Zhao},
  \citenamefont {Zhou},\ and\ \citenamefont {Sun}}]{LiN2021}%
  \BibitemOpen
  \bibfield  {author} {\bibinfo {author} {\bibfnamefont {N.}~\bibnamefont
  {Li}}, \bibinfo {author} {\bibfnamefont {Q.}~\bibnamefont {Huang}}, \bibinfo
  {author} {\bibfnamefont {A.}~\bibnamefont {Brassington}}, \bibinfo {author}
  {\bibfnamefont {X.~Y.}\ \bibnamefont {Yue}}, \bibinfo {author} {\bibfnamefont
  {W.~J.}\ \bibnamefont {Chu}}, \bibinfo {author} {\bibfnamefont {S.~K.}\
  \bibnamefont {Guang}}, \bibinfo {author} {\bibfnamefont {X.~H.}\ \bibnamefont
  {Zhou}}, \bibinfo {author} {\bibfnamefont {P.}~\bibnamefont {Gao}}, \bibinfo
  {author} {\bibfnamefont {E.~X.}\ \bibnamefont {Feng}}, \bibinfo {author}
  {\bibfnamefont {H.~B.}\ \bibnamefont {Cao}}, \bibinfo {author} {\bibfnamefont
  {E.~S.}\ \bibnamefont {Choi}}, \bibinfo {author} {\bibfnamefont
  {Y.}~\bibnamefont {Sun}}, \bibinfo {author} {\bibfnamefont {Q.~J.}\
  \bibnamefont {Li}}, \bibinfo {author} {\bibfnamefont {X.}~\bibnamefont
  {Zhao}}, \bibinfo {author} {\bibfnamefont {H.~D.}\ \bibnamefont {Zhou}},\
  and\ \bibinfo {author} {\bibfnamefont {X.~F.}\ \bibnamefont {Sun}},\
  }\bibfield  {title} {\bibinfo {title} {Quantum spin state transitions in the
  spin-1 equilateral triangular lattice antiferromagnet
  {{Na}}{$_{2}$}{{BaNi}}({{PO}}{$_4$}){$_{2}$}},\ }\href
  {https://doi.org/10.1103/PhysRevB.104.104403} {\bibfield  {journal} {\bibinfo
   {journal} {Phys. Rev. B}\ }\textbf {\bibinfo {volume} {104}},\ \bibinfo
  {pages} {104403} (\bibinfo {year} {2021})}\BibitemShut {NoStop}%
\bibitem [{\citenamefont {Ding}\ \emph {et~al.}(2021)\citenamefont {Ding},
  \citenamefont {Ma}, \citenamefont {Gong}, \citenamefont {Hu}, \citenamefont
  {Zhao}, \citenamefont {Li}, \citenamefont {Zheng}, \citenamefont {Zhang},
  \citenamefont {Yu}, \citenamefont {Zhang}, \citenamefont {Zhao},\ and\
  \citenamefont {Pan}}]{DingF2021}%
  \BibitemOpen
  \bibfield  {author} {\bibinfo {author} {\bibfnamefont {F.}~\bibnamefont
  {Ding}}, \bibinfo {author} {\bibfnamefont {Y.}~\bibnamefont {Ma}}, \bibinfo
  {author} {\bibfnamefont {X.}~\bibnamefont {Gong}}, \bibinfo {author}
  {\bibfnamefont {D.}~\bibnamefont {Hu}}, \bibinfo {author} {\bibfnamefont
  {J.}~\bibnamefont {Zhao}}, \bibinfo {author} {\bibfnamefont {L.}~\bibnamefont
  {Li}}, \bibinfo {author} {\bibfnamefont {H.}~\bibnamefont {Zheng}}, \bibinfo
  {author} {\bibfnamefont {Y.}~\bibnamefont {Zhang}}, \bibinfo {author}
  {\bibfnamefont {Y.}~\bibnamefont {Yu}}, \bibinfo {author} {\bibfnamefont
  {L.}~\bibnamefont {Zhang}}, \bibinfo {author} {\bibfnamefont
  {F.}~\bibnamefont {Zhao}},\ and\ \bibinfo {author} {\bibfnamefont
  {B.}~\bibnamefont {Pan}},\ }\bibfield  {title} {\bibinfo {title} {Structure
  and frustrated magnetism of the two-dimensional triangular lattice
  antiferromagnet
  {{Na}}{\textsubscript{2}}{{BaNi}}({{PO}}{\textsubscript{4}}){\textsubscript{2}}},\
  }\href {https://doi.org/10.1088/1674-1056/abff1d} {\bibfield  {journal}
  {\bibinfo  {journal} {Chin. Phys. B}\ }\textbf {\bibinfo {volume} {30}},\
  \bibinfo {pages} {117505} (\bibinfo {year} {2021})}\BibitemShut {NoStop}%
\bibitem [{\citenamefont {Zapf}\ \emph {et~al.}(2014)\citenamefont {Zapf},
  \citenamefont {Jaime},\ and\ \citenamefont {Batista}}]{ZapfV2014_RMP}%
  \BibitemOpen
  \bibfield  {author} {\bibinfo {author} {\bibfnamefont {V.}~\bibnamefont
  {Zapf}}, \bibinfo {author} {\bibfnamefont {M.}~\bibnamefont {Jaime}},\ and\
  \bibinfo {author} {\bibfnamefont {C.~D.}\ \bibnamefont {Batista}},\
  }\bibfield  {title} {\bibinfo {title} {Bose-{{Einstein}} condensation in
  quantum magnets},\ }\href {https://doi.org/10.1103/RevModPhys.86.563}
  {\bibfield  {journal} {\bibinfo  {journal} {Rev. Mod. Phys.}\ }\textbf
  {\bibinfo {volume} {86}},\ \bibinfo {pages} {563} (\bibinfo {year}
  {2014})}\BibitemShut {NoStop}%
\bibitem [{\citenamefont {Sheng}\ \emph
  {et~al.}(2025{\natexlab{a}})\citenamefont {Sheng}, \citenamefont {Mei},
  \citenamefont {Wang}, \citenamefont {Xu}, \citenamefont {Jiang},
  \citenamefont {Xu}, \citenamefont {Ge}, \citenamefont {Zhao}, \citenamefont
  {Li}, \citenamefont {Candini}, \citenamefont {Xi}, \citenamefont {Zhao},
  \citenamefont {Fu}, \citenamefont {Yang}, \citenamefont {Zhang},
  \citenamefont {Biasiol}, \citenamefont {Wang}, \citenamefont {Zhu},
  \citenamefont {Miao}, \citenamefont {Tong}, \citenamefont {Yu}, \citenamefont
  {Mole}, \citenamefont {Cui}, \citenamefont {Ma}, \citenamefont {Zhang},
  \citenamefont {Ouyang}, \citenamefont {Tong}, \citenamefont {Podlesnyak},
  \citenamefont {Wang}, \citenamefont {Ye}, \citenamefont {Yu}, \citenamefont
  {Yu}, \citenamefont {Wu},\ and\ \citenamefont {Wang}}]{ShengJ2025_NiP}%
  \BibitemOpen
  \bibfield  {author} {\bibinfo {author} {\bibfnamefont {J.}~\bibnamefont
  {Sheng}}, \bibinfo {author} {\bibfnamefont {J.-W.}\ \bibnamefont {Mei}},
  \bibinfo {author} {\bibfnamefont {L.}~\bibnamefont {Wang}}, \bibinfo {author}
  {\bibfnamefont {X.}~\bibnamefont {Xu}}, \bibinfo {author} {\bibfnamefont
  {W.}~\bibnamefont {Jiang}}, \bibinfo {author} {\bibfnamefont
  {L.}~\bibnamefont {Xu}}, \bibinfo {author} {\bibfnamefont {H.}~\bibnamefont
  {Ge}}, \bibinfo {author} {\bibfnamefont {N.}~\bibnamefont {Zhao}}, \bibinfo
  {author} {\bibfnamefont {T.}~\bibnamefont {Li}}, \bibinfo {author}
  {\bibfnamefont {A.}~\bibnamefont {Candini}}, \bibinfo {author} {\bibfnamefont
  {B.}~\bibnamefont {Xi}}, \bibinfo {author} {\bibfnamefont {J.}~\bibnamefont
  {Zhao}}, \bibinfo {author} {\bibfnamefont {Y.}~\bibnamefont {Fu}}, \bibinfo
  {author} {\bibfnamefont {J.}~\bibnamefont {Yang}}, \bibinfo {author}
  {\bibfnamefont {Y.}~\bibnamefont {Zhang}}, \bibinfo {author} {\bibfnamefont
  {G.}~\bibnamefont {Biasiol}}, \bibinfo {author} {\bibfnamefont
  {S.}~\bibnamefont {Wang}}, \bibinfo {author} {\bibfnamefont {J.}~\bibnamefont
  {Zhu}}, \bibinfo {author} {\bibfnamefont {P.}~\bibnamefont {Miao}}, \bibinfo
  {author} {\bibfnamefont {X.}~\bibnamefont {Tong}}, \bibinfo {author}
  {\bibfnamefont {D.}~\bibnamefont {Yu}}, \bibinfo {author} {\bibfnamefont
  {R.}~\bibnamefont {Mole}}, \bibinfo {author} {\bibfnamefont {Y.}~\bibnamefont
  {Cui}}, \bibinfo {author} {\bibfnamefont {L.}~\bibnamefont {Ma}}, \bibinfo
  {author} {\bibfnamefont {Z.}~\bibnamefont {Zhang}}, \bibinfo {author}
  {\bibfnamefont {Z.}~\bibnamefont {Ouyang}}, \bibinfo {author} {\bibfnamefont
  {W.}~\bibnamefont {Tong}}, \bibinfo {author} {\bibfnamefont {A.}~\bibnamefont
  {Podlesnyak}}, \bibinfo {author} {\bibfnamefont {L.}~\bibnamefont {Wang}},
  \bibinfo {author} {\bibfnamefont {F.}~\bibnamefont {Ye}}, \bibinfo {author}
  {\bibfnamefont {D.}~\bibnamefont {Yu}}, \bibinfo {author} {\bibfnamefont
  {W.}~\bibnamefont {Yu}}, \bibinfo {author} {\bibfnamefont {L.}~\bibnamefont
  {Wu}},\ and\ \bibinfo {author} {\bibfnamefont {Z.}~\bibnamefont {Wang}},\
  }\bibfield  {title} {\bibinfo {title} {Bose--{{Einstein}} condensation of a
  two-magnon bound state in a spin-1 triangular lattice},\ }\href
  {https://doi.org/10.1038/s41563-024-02071-z} {\bibfield  {journal} {\bibinfo
  {journal} {Nat. Mater.}\ }\textbf {\bibinfo {volume} {24}},\ \bibinfo {pages}
  {544} (\bibinfo {year} {2025}{\natexlab{a}})}\BibitemShut {NoStop}%
\bibitem [{SM()}]{SM}%
  \BibitemOpen
  \href@noop {} {}\bibinfo {note} {See Supplemental Material at [URL] for a
  description of the methods (material synthesis, neutron scattering, iPEPS,
  LSW, and GLSW) and additional neutron scattering data, which includes
  Refs.~\cite{NakajimaK2011,EhlersG2011,EhlersG2016,InamuraY2013,AzuahRT2009,VerstraeteF2004_arxiv,VerstraeteF2008_review,OrusR2009,CorbozP2014_tJ,LiaoHJ2017,NishinoT1996}.}\BibitemShut
  {Stop}%
\bibitem [{\citenamefont {Zhitomirsky}\ and\ \citenamefont
  {Chernyshev}(2013)}]{ZhitomirskyME2013_RMP}%
  \BibitemOpen
  \bibfield  {author} {\bibinfo {author} {\bibfnamefont {M.~E.}\ \bibnamefont
  {Zhitomirsky}}\ and\ \bibinfo {author} {\bibfnamefont {A.~L.}\ \bibnamefont
  {Chernyshev}},\ }\bibfield  {title} {\bibinfo {title} {Colloquium:
  {{Spontaneous}} magnon decays},\ }\href
  {https://doi.org/10.1103/RevModPhys.85.219} {\bibfield  {journal} {\bibinfo
  {journal} {Rev. Mod. Phys.}\ }\textbf {\bibinfo {volume} {85}},\ \bibinfo
  {pages} {219} (\bibinfo {year} {2013})}\BibitemShut {NoStop}%
\bibitem [{\citenamefont {Sheng}\ \emph
  {et~al.}(2025{\natexlab{b}})\citenamefont {Sheng}, \citenamefont {Wang},
  \citenamefont {Jiang}, \citenamefont {Ge}, \citenamefont {Zhao},
  \citenamefont {Li}, \citenamefont {Kofu}, \citenamefont {Yu}, \citenamefont
  {Zhu}, \citenamefont {Mei}, \citenamefont {Wang},\ and\ \citenamefont
  {Wu}}]{ShengJ2025_CoP}%
  \BibitemOpen
  \bibfield  {author} {\bibinfo {author} {\bibfnamefont {J.}~\bibnamefont
  {Sheng}}, \bibinfo {author} {\bibfnamefont {L.}~\bibnamefont {Wang}},
  \bibinfo {author} {\bibfnamefont {W.}~\bibnamefont {Jiang}}, \bibinfo
  {author} {\bibfnamefont {H.}~\bibnamefont {Ge}}, \bibinfo {author}
  {\bibfnamefont {N.}~\bibnamefont {Zhao}}, \bibinfo {author} {\bibfnamefont
  {T.}~\bibnamefont {Li}}, \bibinfo {author} {\bibfnamefont {M.}~\bibnamefont
  {Kofu}}, \bibinfo {author} {\bibfnamefont {D.}~\bibnamefont {Yu}}, \bibinfo
  {author} {\bibfnamefont {W.}~\bibnamefont {Zhu}}, \bibinfo {author}
  {\bibfnamefont {J.-W.}\ \bibnamefont {Mei}}, \bibinfo {author} {\bibfnamefont
  {Z.}~\bibnamefont {Wang}},\ and\ \bibinfo {author} {\bibfnamefont
  {L.}~\bibnamefont {Wu}},\ }\bibfield  {title} {\bibinfo {title} {Continuum of
  spin excitations in an ordered magnet},\ }\href
  {https://doi.org/10.1016/j.xinn.2024.100769} {\bibfield  {journal} {\bibinfo
  {journal} {The Innovation}\ }\textbf {\bibinfo {volume} {6}},\ \bibinfo
  {pages} {100769} (\bibinfo {year} {2025}{\natexlab{b}})}\BibitemShut
  {NoStop}%
\bibitem [{\citenamefont {Chi}\ \emph {et~al.}(2024)\citenamefont {Chi},
  \citenamefont {Hu}, \citenamefont {Liao},\ and\ \citenamefont
  {Xiang}}]{ChiR2024}%
  \BibitemOpen
  \bibfield  {author} {\bibinfo {author} {\bibfnamefont {R.}~\bibnamefont
  {Chi}}, \bibinfo {author} {\bibfnamefont {J.}~\bibnamefont {Hu}}, \bibinfo
  {author} {\bibfnamefont {H.-J.}\ \bibnamefont {Liao}},\ and\ \bibinfo
  {author} {\bibfnamefont {T.}~\bibnamefont {Xiang}},\ }\bibfield  {title}
  {\bibinfo {title} {Dynamical spectra of spin supersolid states in triangular
  antiferromagnets},\ }\href {https://doi.org/10.1103/PhysRevB.110.L180404}
  {\bibfield  {journal} {\bibinfo  {journal} {Phys. Rev. B}\ }\textbf {\bibinfo
  {volume} {110}},\ \bibinfo {pages} {L180404} (\bibinfo {year}
  {2024})}\BibitemShut {NoStop}%
\bibitem [{\citenamefont {Gao}\ \emph {et~al.}(2024)\citenamefont {Gao},
  \citenamefont {Zhang}, \citenamefont {Xiang}, \citenamefont {Yu},
  \citenamefont {Lu}, \citenamefont {Sun}, \citenamefont {Jin}, \citenamefont
  {Su},\ and\ \citenamefont {Li}}]{GaoY2024_CoP}%
  \BibitemOpen
  \bibfield  {author} {\bibinfo {author} {\bibfnamefont {Y.}~\bibnamefont
  {Gao}}, \bibinfo {author} {\bibfnamefont {C.}~\bibnamefont {Zhang}}, \bibinfo
  {author} {\bibfnamefont {J.}~\bibnamefont {Xiang}}, \bibinfo {author}
  {\bibfnamefont {D.}~\bibnamefont {Yu}}, \bibinfo {author} {\bibfnamefont
  {X.}~\bibnamefont {Lu}}, \bibinfo {author} {\bibfnamefont {P.}~\bibnamefont
  {Sun}}, \bibinfo {author} {\bibfnamefont {W.}~\bibnamefont {Jin}}, \bibinfo
  {author} {\bibfnamefont {G.}~\bibnamefont {Su}},\ and\ \bibinfo {author}
  {\bibfnamefont {W.}~\bibnamefont {Li}},\ }\bibfield  {title} {\bibinfo
  {title} {Double magnon-roton excitations in the triangular-lattice spin
  supersolid},\ }\href {https://doi.org/10.1103/PhysRevB.110.214408} {\bibfield
   {journal} {\bibinfo  {journal} {Phys. Rev. B}\ }\textbf {\bibinfo {volume}
  {110}},\ \bibinfo {pages} {214408} (\bibinfo {year} {2024})}\BibitemShut
  {NoStop}%
\bibitem [{\citenamefont {Vanderstraeten}\ \emph {et~al.}(2015)\citenamefont
  {Vanderstraeten}, \citenamefont {Mari{\"e}n}, \citenamefont {Verstraete},\
  and\ \citenamefont {Haegeman}}]{VanderstraetenL2015}%
  \BibitemOpen
  \bibfield  {author} {\bibinfo {author} {\bibfnamefont {L.}~\bibnamefont
  {Vanderstraeten}}, \bibinfo {author} {\bibfnamefont {M.}~\bibnamefont
  {Mari{\"e}n}}, \bibinfo {author} {\bibfnamefont {F.}~\bibnamefont
  {Verstraete}},\ and\ \bibinfo {author} {\bibfnamefont {J.}~\bibnamefont
  {Haegeman}},\ }\bibfield  {title} {\bibinfo {title} {Excitations and the
  tangent space of projected entangled-pair states},\ }\href
  {https://doi.org/10.1103/PhysRevB.92.201111} {\bibfield  {journal} {\bibinfo
  {journal} {Phys. Rev. B}\ }\textbf {\bibinfo {volume} {92}},\ \bibinfo
  {pages} {201111} (\bibinfo {year} {2015})}\BibitemShut {NoStop}%
\bibitem [{\citenamefont {Vanderstraeten}\ \emph {et~al.}(2019)\citenamefont
  {Vanderstraeten}, \citenamefont {Haegeman},\ and\ \citenamefont
  {Verstraete}}]{VanderstraetenL2019}%
  \BibitemOpen
  \bibfield  {author} {\bibinfo {author} {\bibfnamefont {L.}~\bibnamefont
  {Vanderstraeten}}, \bibinfo {author} {\bibfnamefont {J.}~\bibnamefont
  {Haegeman}},\ and\ \bibinfo {author} {\bibfnamefont {F.}~\bibnamefont
  {Verstraete}},\ }\bibfield  {title} {\bibinfo {title} {Simulating excitation
  spectra with projected entangled-pair states},\ }\href
  {https://doi.org/10.1103/PhysRevB.99.165121} {\bibfield  {journal} {\bibinfo
  {journal} {Phys. Rev. B}\ }\textbf {\bibinfo {volume} {99}},\ \bibinfo
  {pages} {165121} (\bibinfo {year} {2019})}\BibitemShut {NoStop}%
\bibitem [{\citenamefont {Ponsioen}\ and\ \citenamefont
  {Corboz}(2020)}]{PonsioenB2020}%
  \BibitemOpen
  \bibfield  {author} {\bibinfo {author} {\bibfnamefont {B.}~\bibnamefont
  {Ponsioen}}\ and\ \bibinfo {author} {\bibfnamefont {P.}~\bibnamefont
  {Corboz}},\ }\bibfield  {title} {\bibinfo {title} {Excitations with projected
  entangled pair states using the corner transfer matrix method},\ }\href
  {https://doi.org/10.1103/PhysRevB.101.195109} {\bibfield  {journal} {\bibinfo
   {journal} {Phys. Rev. B}\ }\textbf {\bibinfo {volume} {101}},\ \bibinfo
  {pages} {195109} (\bibinfo {year} {2020})}\BibitemShut {NoStop}%
\bibitem [{\citenamefont {Ponsioen}\ \emph {et~al.}(2022)\citenamefont
  {Ponsioen}, \citenamefont {Assaad},\ and\ \citenamefont
  {Corboz}}]{PonsioenB2022}%
  \BibitemOpen
  \bibfield  {author} {\bibinfo {author} {\bibfnamefont {B.}~\bibnamefont
  {Ponsioen}}, \bibinfo {author} {\bibfnamefont {F.~F.}\ \bibnamefont
  {Assaad}},\ and\ \bibinfo {author} {\bibfnamefont {P.}~\bibnamefont
  {Corboz}},\ }\bibfield  {title} {\bibinfo {title} {Automatic differentiation
  applied to excitations with projected entangled pair states},\ }\href
  {https://doi.org/10.21468/SciPostPhys.12.1.006} {\bibfield  {journal}
  {\bibinfo  {journal} {SciPost Phys.}\ }\textbf {\bibinfo {volume} {12}},\
  \bibinfo {pages} {006} (\bibinfo {year} {2022})}\BibitemShut {NoStop}%
\bibitem [{\citenamefont {Chi}\ \emph {et~al.}(2022)\citenamefont {Chi},
  \citenamefont {Liu}, \citenamefont {Wan}, \citenamefont {Liao},\ and\
  \citenamefont {Xiang}}]{ChiR2022}%
  \BibitemOpen
  \bibfield  {author} {\bibinfo {author} {\bibfnamefont {R.}~\bibnamefont
  {Chi}}, \bibinfo {author} {\bibfnamefont {Y.}~\bibnamefont {Liu}}, \bibinfo
  {author} {\bibfnamefont {Y.}~\bibnamefont {Wan}}, \bibinfo {author}
  {\bibfnamefont {H.-J.}\ \bibnamefont {Liao}},\ and\ \bibinfo {author}
  {\bibfnamefont {T.}~\bibnamefont {Xiang}},\ }\bibfield  {title} {\bibinfo
  {title} {Spin {{Excitation Spectra}} of {{Anisotropic Spin-1}}/2 {{Triangular
  Lattice Heisenberg Antiferromagnets}}},\ }\href
  {https://doi.org/10.1103/PhysRevLett.129.227201} {\bibfield  {journal}
  {\bibinfo  {journal} {Phys. Rev. Lett.}\ }\textbf {\bibinfo {volume} {129}},\
  \bibinfo {pages} {227201} (\bibinfo {year} {2022})}\BibitemShut {NoStop}%
\bibitem [{\citenamefont {{\"O}stlund}\ and\ \citenamefont
  {Rommer}(1995)}]{OstlundS1995}%
  \BibitemOpen
  \bibfield  {author} {\bibinfo {author} {\bibfnamefont {S.}~\bibnamefont
  {{\"O}stlund}}\ and\ \bibinfo {author} {\bibfnamefont {S.}~\bibnamefont
  {Rommer}},\ }\bibfield  {title} {\bibinfo {title} {Thermodynamic {{Limit}} of
  {{Density Matrix Renormalization}}},\ }\href
  {https://doi.org/10.1103/PhysRevLett.75.3537} {\bibfield  {journal} {\bibinfo
   {journal} {Phys. Rev. Lett.}\ }\textbf {\bibinfo {volume} {75}},\ \bibinfo
  {pages} {3537} (\bibinfo {year} {1995})}\BibitemShut {NoStop}%
\bibitem [{\citenamefont {Liao}\ \emph {et~al.}(2019)\citenamefont {Liao},
  \citenamefont {Liu}, \citenamefont {Wang},\ and\ \citenamefont
  {Xiang}}]{LiaoHJ2019}%
  \BibitemOpen
  \bibfield  {author} {\bibinfo {author} {\bibfnamefont {H.-J.}\ \bibnamefont
  {Liao}}, \bibinfo {author} {\bibfnamefont {J.-G.}\ \bibnamefont {Liu}},
  \bibinfo {author} {\bibfnamefont {L.}~\bibnamefont {Wang}},\ and\ \bibinfo
  {author} {\bibfnamefont {T.}~\bibnamefont {Xiang}},\ }\bibfield  {title}
  {\bibinfo {title} {Differentiable {{Programming Tensor Networks}}},\ }\href
  {https://doi.org/10.1103/PhysRevX.9.031041} {\bibfield  {journal} {\bibinfo
  {journal} {Phys. Rev. X}\ }\textbf {\bibinfo {volume} {9}},\ \bibinfo {pages}
  {031041} (\bibinfo {year} {2019})}\BibitemShut {NoStop}%
\bibitem [{\citenamefont {Xu}\ \emph {et~al.}(2025)\citenamefont {Xu},
  \citenamefont {Hasik}, \citenamefont {Ponsioen},\ and\ \citenamefont
  {Nevidomskyy}}]{XuY2025}%
  \BibitemOpen
  \bibfield  {author} {\bibinfo {author} {\bibfnamefont {Y.}~\bibnamefont
  {Xu}}, \bibinfo {author} {\bibfnamefont {J.}~\bibnamefont {Hasik}}, \bibinfo
  {author} {\bibfnamefont {B.}~\bibnamefont {Ponsioen}},\ and\ \bibinfo
  {author} {\bibfnamefont {A.~H.}\ \bibnamefont {Nevidomskyy}},\ }\bibfield
  {title} {\bibinfo {title} {Simulating spin dynamics of supersolid states in a
  quantum {{Ising}} magnet},\ }\href
  {https://doi.org/10.1103/PhysRevB.111.L060402} {\bibfield  {journal}
  {\bibinfo  {journal} {Phys. Rev. B}\ }\textbf {\bibinfo {volume} {111}},\
  \bibinfo {pages} {L060402} (\bibinfo {year} {2025})}\BibitemShut {NoStop}%
\bibitem [{\citenamefont {Holstein}\ and\ \citenamefont
  {Primakoff}(1940)}]{HolsteinT1940}%
  \BibitemOpen
  \bibfield  {author} {\bibinfo {author} {\bibfnamefont {T.}~\bibnamefont
  {Holstein}}\ and\ \bibinfo {author} {\bibfnamefont {H.}~\bibnamefont
  {Primakoff}},\ }\bibfield  {title} {\bibinfo {title} {Field {{Dependence}} of
  the {{Intrinsic Domain Magnetization}} of a {{Ferromagnet}}},\ }\href
  {https://doi.org/10.1103/PhysRev.58.1098} {\bibfield  {journal} {\bibinfo
  {journal} {Phys. Rev.}\ }\textbf {\bibinfo {volume} {58}},\ \bibinfo {pages}
  {1098} (\bibinfo {year} {1940})}\BibitemShut {NoStop}%
\bibitem [{\citenamefont {Batista}\ and\ \citenamefont
  {Ortiz}(2004)}]{BatistaCD2004_SUN}%
  \BibitemOpen
  \bibfield  {author} {\bibinfo {author} {\bibfnamefont {C.~D.}\ \bibnamefont
  {Batista}}\ and\ \bibinfo {author} {\bibfnamefont {G.}~\bibnamefont
  {Ortiz}},\ }\bibfield  {title} {\bibinfo {title} {Algebraic approach to
  interacting quantum systems},\ }\href
  {https://doi.org/10.1080/00018730310001642086} {\bibfield  {journal}
  {\bibinfo  {journal} {Adv. Phys.}\ }\textbf {\bibinfo {volume} {53}},\
  \bibinfo {pages} {1} (\bibinfo {year} {2004})}\BibitemShut {NoStop}%
\bibitem [{\citenamefont {Seifert}\ and\ \citenamefont
  {Savary}(2022)}]{SeifertUFP2022}%
  \BibitemOpen
  \bibfield  {author} {\bibinfo {author} {\bibfnamefont {U.~F.~P.}\
  \bibnamefont {Seifert}}\ and\ \bibinfo {author} {\bibfnamefont
  {L.}~\bibnamefont {Savary}},\ }\bibfield  {title} {\bibinfo {title} {Phase
  diagrams and excitations of anisotropic {{{\emph{S}}}}=1 quantum magnets on
  the triangular lattice},\ }\href
  {https://doi.org/10.1103/PhysRevB.106.195147} {\bibfield  {journal} {\bibinfo
   {journal} {Phys. Rev. B}\ }\textbf {\bibinfo {volume} {106}},\ \bibinfo
  {pages} {195147} (\bibinfo {year} {2022})}\BibitemShut {NoStop}%
\bibitem [{\citenamefont {Yamamoto}\ \emph {et~al.}(2014)\citenamefont
  {Yamamoto}, \citenamefont {Marmorini},\ and\ \citenamefont
  {Danshita}}]{YamamotoD2014}%
  \BibitemOpen
  \bibfield  {author} {\bibinfo {author} {\bibfnamefont {D.}~\bibnamefont
  {Yamamoto}}, \bibinfo {author} {\bibfnamefont {G.}~\bibnamefont
  {Marmorini}},\ and\ \bibinfo {author} {\bibfnamefont {I.}~\bibnamefont
  {Danshita}},\ }\bibfield  {title} {\bibinfo {title} {Quantum {{Phase
  Diagram}} of the {{Triangular-Lattice}} {{{\emph{XXZ}}}} {{Model}} in a
  {{Magnetic Field}}},\ }\href {https://doi.org/10.1103/PhysRevLett.112.127203}
  {\bibfield  {journal} {\bibinfo  {journal} {Phys. Rev. Lett.}\ }\textbf
  {\bibinfo {volume} {112}},\ \bibinfo {pages} {127203} (\bibinfo {year}
  {2014})}\BibitemShut {NoStop}%
\bibitem [{\citenamefont {Silberglitt}\ and\ \citenamefont
  {Torrance}(1970)}]{SilberglittR1970}%
  \BibitemOpen
  \bibfield  {author} {\bibinfo {author} {\bibfnamefont {R.}~\bibnamefont
  {Silberglitt}}\ and\ \bibinfo {author} {\bibfnamefont {J.~B.}\ \bibnamefont
  {Torrance}},\ }\bibfield  {title} {\bibinfo {title} {Effect of {{Single-Ion
  Anisotropy}} on {{Two-Spin-Wave Bound State}} in a {{Heisenberg
  Ferromagnet}}},\ }\href {https://doi.org/10.1103/PhysRevB.2.772} {\bibfield
  {journal} {\bibinfo  {journal} {Phys. Rev. B}\ }\textbf {\bibinfo {volume}
  {2}},\ \bibinfo {pages} {772} (\bibinfo {year} {1970})}\BibitemShut {NoStop}%
\bibitem [{\citenamefont {Oguchi}(1971)}]{OguchiT1971}%
  \BibitemOpen
  \bibfield  {author} {\bibinfo {author} {\bibfnamefont {T.}~\bibnamefont
  {Oguchi}},\ }\bibfield  {title} {\bibinfo {title} {Theory of {{Two-Magnon
  Bound States}} in the {{Heisenberg Ferro-}} and {{Antiferromagnet}}},\ }\href
  {https://doi.org/10.1143/JPSJ.31.394} {\bibfield  {journal} {\bibinfo
  {journal} {J. Phys. Soc. Jpn.}\ }\textbf {\bibinfo {volume} {31}},\ \bibinfo
  {pages} {394} (\bibinfo {year} {1971})}\BibitemShut {NoStop}%
\bibitem [{\citenamefont {Bai}\ \emph {et~al.}(2021)\citenamefont {Bai},
  \citenamefont {Zhang}, \citenamefont {Dun}, \citenamefont {Zhang},
  \citenamefont {Huang}, \citenamefont {Zhou}, \citenamefont {Stone},
  \citenamefont {Kolesnikov}, \citenamefont {Ye}, \citenamefont {Batista},\
  and\ \citenamefont {Mourigal}}]{BaiX2021}%
  \BibitemOpen
  \bibfield  {author} {\bibinfo {author} {\bibfnamefont {X.}~\bibnamefont
  {Bai}}, \bibinfo {author} {\bibfnamefont {S.-S.}\ \bibnamefont {Zhang}},
  \bibinfo {author} {\bibfnamefont {Z.}~\bibnamefont {Dun}}, \bibinfo {author}
  {\bibfnamefont {H.}~\bibnamefont {Zhang}}, \bibinfo {author} {\bibfnamefont
  {Q.}~\bibnamefont {Huang}}, \bibinfo {author} {\bibfnamefont
  {H.}~\bibnamefont {Zhou}}, \bibinfo {author} {\bibfnamefont {M.~B.}\
  \bibnamefont {Stone}}, \bibinfo {author} {\bibfnamefont {A.~I.}\ \bibnamefont
  {Kolesnikov}}, \bibinfo {author} {\bibfnamefont {F.}~\bibnamefont {Ye}},
  \bibinfo {author} {\bibfnamefont {C.~D.}\ \bibnamefont {Batista}},\ and\
  \bibinfo {author} {\bibfnamefont {M.}~\bibnamefont {Mourigal}},\ }\bibfield
  {title} {\bibinfo {title} {Hybridized quadrupolar excitations in the
  spin-anisotropic frustrated magnet {{FeI}}{$_{2}$}},\ }\href
  {https://doi.org/10.1038/s41567-020-01110-1} {\bibfield  {journal} {\bibinfo
  {journal} {Nat. Phys.}\ }\textbf {\bibinfo {volume} {17}},\ \bibinfo {pages}
  {467} (\bibinfo {year} {2021})}\BibitemShut {NoStop}%
\bibitem [{das()}]{das}%
  \BibitemOpen
  \href {https://doi.org/10.5061/dryad.zcrjdfnp9} {}\bibinfo {note} {J. Sheng
  {\it et al.}, DRYAD (2025), 10.5061/dryad.zcrjdfnp9}\BibitemShut {NoStop}%
\bibitem [{\citenamefont {Nakajima}\ \emph {et~al.}(2011)\citenamefont
  {Nakajima}, \citenamefont {{Ohira-Kawamura}}, \citenamefont {Kikuchi},
  \citenamefont {Nakamura}, \citenamefont {Kajimoto}, \citenamefont {Inamura},
  \citenamefont {Takahashi}, \citenamefont {Aizawa}, \citenamefont {Suzuya},
  \citenamefont {Shibata}, \citenamefont {Nakatani}, \citenamefont {Soyama},
  \citenamefont {Maruyama}, \citenamefont {Tanaka}, \citenamefont {Kambara},
  \citenamefont {Iwahashi}, \citenamefont {Itoh}, \citenamefont {Osakabe},
  \citenamefont {Wakimoto}, \citenamefont {Kakurai}, \citenamefont {Maekawa},
  \citenamefont {Harada}, \citenamefont {Oikawa}, \citenamefont {E.~Lechner},
  \citenamefont {Mezei},\ and\ \citenamefont {Arai}}]{NakajimaK2011}%
  \BibitemOpen
  \bibfield  {author} {\bibinfo {author} {\bibfnamefont {K.}~\bibnamefont
  {Nakajima}}, \bibinfo {author} {\bibfnamefont {S.}~\bibnamefont
  {{Ohira-Kawamura}}}, \bibinfo {author} {\bibfnamefont {T.}~\bibnamefont
  {Kikuchi}}, \bibinfo {author} {\bibfnamefont {M.}~\bibnamefont {Nakamura}},
  \bibinfo {author} {\bibfnamefont {R.}~\bibnamefont {Kajimoto}}, \bibinfo
  {author} {\bibfnamefont {Y.}~\bibnamefont {Inamura}}, \bibinfo {author}
  {\bibfnamefont {N.}~\bibnamefont {Takahashi}}, \bibinfo {author}
  {\bibfnamefont {K.}~\bibnamefont {Aizawa}}, \bibinfo {author} {\bibfnamefont
  {K.}~\bibnamefont {Suzuya}}, \bibinfo {author} {\bibfnamefont
  {K.}~\bibnamefont {Shibata}}, \bibinfo {author} {\bibfnamefont
  {T.}~\bibnamefont {Nakatani}}, \bibinfo {author} {\bibfnamefont
  {K.}~\bibnamefont {Soyama}}, \bibinfo {author} {\bibfnamefont
  {R.}~\bibnamefont {Maruyama}}, \bibinfo {author} {\bibfnamefont
  {H.}~\bibnamefont {Tanaka}}, \bibinfo {author} {\bibfnamefont
  {W.}~\bibnamefont {Kambara}}, \bibinfo {author} {\bibfnamefont
  {T.}~\bibnamefont {Iwahashi}}, \bibinfo {author} {\bibfnamefont
  {Y.}~\bibnamefont {Itoh}}, \bibinfo {author} {\bibfnamefont {T.}~\bibnamefont
  {Osakabe}}, \bibinfo {author} {\bibfnamefont {S.}~\bibnamefont {Wakimoto}},
  \bibinfo {author} {\bibfnamefont {K.}~\bibnamefont {Kakurai}}, \bibinfo
  {author} {\bibfnamefont {F.}~\bibnamefont {Maekawa}}, \bibinfo {author}
  {\bibfnamefont {M.}~\bibnamefont {Harada}}, \bibinfo {author} {\bibfnamefont
  {K.}~\bibnamefont {Oikawa}}, \bibinfo {author} {\bibfnamefont
  {R.}~\bibnamefont {E.~Lechner}}, \bibinfo {author} {\bibfnamefont
  {F.}~\bibnamefont {Mezei}},\ and\ \bibinfo {author} {\bibfnamefont
  {M.}~\bibnamefont {Arai}},\ }\bibfield  {title} {\bibinfo {title}
  {{{AMATERAS}}: {{A Cold-Neutron Disk Chopper Spectrometer}}},\ }\href
  {https://doi.org/10.1143/JPSJS.80SB.SB028} {\bibfield  {journal} {\bibinfo
  {journal} {J. Phys. Soc. Jpn.}\ }\textbf {\bibinfo {volume} {80}},\ \bibinfo
  {pages} {SB028} (\bibinfo {year} {2011})}\BibitemShut {NoStop}%
\bibitem [{\citenamefont {Ehlers}\ \emph {et~al.}(2011)\citenamefont {Ehlers},
  \citenamefont {Podlesnyak}, \citenamefont {Niedziela}, \citenamefont
  {Iverson},\ and\ \citenamefont {Sokol}}]{EhlersG2011}%
  \BibitemOpen
  \bibfield  {author} {\bibinfo {author} {\bibfnamefont {G.}~\bibnamefont
  {Ehlers}}, \bibinfo {author} {\bibfnamefont {A.~A.}\ \bibnamefont
  {Podlesnyak}}, \bibinfo {author} {\bibfnamefont {J.~L.}\ \bibnamefont
  {Niedziela}}, \bibinfo {author} {\bibfnamefont {E.~B.}\ \bibnamefont
  {Iverson}},\ and\ \bibinfo {author} {\bibfnamefont {P.~E.}\ \bibnamefont
  {Sokol}},\ }\bibfield  {title} {\bibinfo {title} {The new cold neutron
  chopper spectrometer at the {{Spallation Neutron Source}}: {{Design}} and
  performance},\ }\href {https://doi.org/10.1063/1.3626935} {\bibfield
  {journal} {\bibinfo  {journal} {Rev. Sci. Instrum.}\ }\textbf {\bibinfo
  {volume} {82}},\ \bibinfo {pages} {085108} (\bibinfo {year}
  {2011})}\BibitemShut {NoStop}%
\bibitem [{\citenamefont {Ehlers}\ \emph {et~al.}(2016)\citenamefont {Ehlers},
  \citenamefont {Podlesnyak},\ and\ \citenamefont {Kolesnikov}}]{EhlersG2016}%
  \BibitemOpen
  \bibfield  {author} {\bibinfo {author} {\bibfnamefont {G.}~\bibnamefont
  {Ehlers}}, \bibinfo {author} {\bibfnamefont {A.~A.}\ \bibnamefont
  {Podlesnyak}},\ and\ \bibinfo {author} {\bibfnamefont {A.~I.}\ \bibnamefont
  {Kolesnikov}},\ }\bibfield  {title} {\bibinfo {title} {The cold neutron
  chopper spectrometer at the {{Spallation Neutron Source}}---{{A}} review of
  the first 8 years of operation},\ }\href {https://doi.org/10.1063/1.4962024}
  {\bibfield  {journal} {\bibinfo  {journal} {Rev. Sci. Instrum.}\ }\textbf
  {\bibinfo {volume} {87}},\ \bibinfo {pages} {093902} (\bibinfo {year}
  {2016})}\BibitemShut {NoStop}%
\bibitem [{\citenamefont {Inamura}\ \emph {et~al.}(2013)\citenamefont
  {Inamura}, \citenamefont {Nakatani}, \citenamefont {Suzuki},\ and\
  \citenamefont {Otomo}}]{InamuraY2013}%
  \BibitemOpen
  \bibfield  {author} {\bibinfo {author} {\bibfnamefont {Y.}~\bibnamefont
  {Inamura}}, \bibinfo {author} {\bibfnamefont {T.}~\bibnamefont {Nakatani}},
  \bibinfo {author} {\bibfnamefont {J.}~\bibnamefont {Suzuki}},\ and\ \bibinfo
  {author} {\bibfnamefont {T.}~\bibnamefont {Otomo}},\ }\bibfield  {title}
  {\bibinfo {title} {Development {{Status}} of {{Software}} ``{{Utsusemi}}''
  for {{Chopper Spectrometers}} at {{MLF}}, {{J-PARC}}},\ }\href
  {https://doi.org/10.7566/JPSJS.82SA.SA031} {\bibfield  {journal} {\bibinfo
  {journal} {J. Phys. Soc. Jpn.}\ }\textbf {\bibinfo {volume} {82}},\ \bibinfo
  {pages} {SA031} (\bibinfo {year} {2013})}\BibitemShut {NoStop}%
\bibitem [{\citenamefont {Azuah}\ \emph {et~al.}(2009)\citenamefont {Azuah},
  \citenamefont {Kneller}, \citenamefont {Qiu}, \citenamefont
  {{Tregenna-Piggott}}, \citenamefont {Brown}, \citenamefont {Copley},\ and\
  \citenamefont {Dimeo}}]{AzuahRT2009}%
  \BibitemOpen
  \bibfield  {author} {\bibinfo {author} {\bibfnamefont {R.~T.}\ \bibnamefont
  {Azuah}}, \bibinfo {author} {\bibfnamefont {L.~R.}\ \bibnamefont {Kneller}},
  \bibinfo {author} {\bibfnamefont {Y.}~\bibnamefont {Qiu}}, \bibinfo {author}
  {\bibfnamefont {P.~L.~W.}\ \bibnamefont {{Tregenna-Piggott}}}, \bibinfo
  {author} {\bibfnamefont {C.~M.}\ \bibnamefont {Brown}}, \bibinfo {author}
  {\bibfnamefont {J.~R.~D.}\ \bibnamefont {Copley}},\ and\ \bibinfo {author}
  {\bibfnamefont {R.~M.}\ \bibnamefont {Dimeo}},\ }\bibfield  {title} {\bibinfo
  {title} {Dave: {{A}} compressive software suite for the reduction,
  visualization, and analysis of low energy neutron spectroscopic data},\
  }\href {https://doi.org/10.6028/jres.114.025} {\bibfield  {journal} {\bibinfo
   {journal} {J. Res. Natl. Inst. Stan. Technol.}\ }\textbf {\bibinfo {volume}
  {114}},\ \bibinfo {pages} {341} (\bibinfo {year} {2009})}\BibitemShut
  {NoStop}%
\bibitem [{\citenamefont {Verstraete}\ and\ \citenamefont
  {Cirac}(2004)}]{VerstraeteF2004_arxiv}%
  \BibitemOpen
  \bibfield  {author} {\bibinfo {author} {\bibfnamefont {F.}~\bibnamefont
  {Verstraete}}\ and\ \bibinfo {author} {\bibfnamefont {J.~I.}\ \bibnamefont
  {Cirac}},\ }\href {https://doi.org/10.48550/arXiv.cond-mat/0407066} {\bibinfo
  {title} {Renormalization algorithms for {{Quantum-Many Body Systems}} in two
  and higher dimensions}} (\bibinfo {year} {2004}),\ \Eprint
  {https://arxiv.org/abs/cond-mat/0407066} {arXiv:cond-mat/0407066}
  \BibitemShut {NoStop}%
\bibitem [{\citenamefont {Verstraete}\ \emph {et~al.}(2008)\citenamefont
  {Verstraete}, \citenamefont {Murg},\ and\ \citenamefont
  {Cirac}}]{VerstraeteF2008_review}%
  \BibitemOpen
  \bibfield  {author} {\bibinfo {author} {\bibfnamefont {F.}~\bibnamefont
  {Verstraete}}, \bibinfo {author} {\bibfnamefont {V.}~\bibnamefont {Murg}},\
  and\ \bibinfo {author} {\bibfnamefont {J.}~\bibnamefont {Cirac}},\ }\bibfield
   {title} {\bibinfo {title} {Matrix product states, projected entangled pair
  states, and variational renormalization group methods for quantum spin
  systems},\ }\href {https://doi.org/10.1080/14789940801912366} {\bibfield
  {journal} {\bibinfo  {journal} {Adv. Phys.}\ }\textbf {\bibinfo {volume}
  {57}},\ \bibinfo {pages} {143} (\bibinfo {year} {2008})}\BibitemShut
  {NoStop}%
\bibitem [{\citenamefont {Or{\'u}s}\ and\ \citenamefont
  {Vidal}(2009)}]{OrusR2009}%
  \BibitemOpen
  \bibfield  {author} {\bibinfo {author} {\bibfnamefont {R.}~\bibnamefont
  {Or{\'u}s}}\ and\ \bibinfo {author} {\bibfnamefont {G.}~\bibnamefont
  {Vidal}},\ }\bibfield  {title} {\bibinfo {title} {Simulation of
  two-dimensional quantum systems on an infinite lattice revisited: {{Corner}}
  transfer matrix for tensor contraction},\ }\href
  {https://doi.org/10.1103/PhysRevB.80.094403} {\bibfield  {journal} {\bibinfo
  {journal} {Phys. Rev. B}\ }\textbf {\bibinfo {volume} {80}},\ \bibinfo
  {pages} {094403} (\bibinfo {year} {2009})}\BibitemShut {NoStop}%
\bibitem [{\citenamefont {Corboz}\ \emph {et~al.}(2014)\citenamefont {Corboz},
  \citenamefont {Rice},\ and\ \citenamefont {Troyer}}]{CorbozP2014_tJ}%
  \BibitemOpen
  \bibfield  {author} {\bibinfo {author} {\bibfnamefont {P.}~\bibnamefont
  {Corboz}}, \bibinfo {author} {\bibfnamefont {T.~M.}\ \bibnamefont {Rice}},\
  and\ \bibinfo {author} {\bibfnamefont {M.}~\bibnamefont {Troyer}},\
  }\bibfield  {title} {\bibinfo {title} {Competing {{States}} in the
  {\emph{t}}-{{{\emph{J}}}} {{Model}}: {{Uniform}} {\emph{d}}-{{Wave State}}
  versus {{Stripe State}}},\ }\href
  {https://doi.org/10.1103/PhysRevLett.113.046402} {\bibfield  {journal}
  {\bibinfo  {journal} {Phys. Rev. Lett.}\ }\textbf {\bibinfo {volume} {113}},\
  \bibinfo {pages} {046402} (\bibinfo {year} {2014})}\BibitemShut {NoStop}%
\bibitem [{\citenamefont {Liao}\ \emph {et~al.}(2017)\citenamefont {Liao},
  \citenamefont {Xie}, \citenamefont {Chen}, \citenamefont {Liu}, \citenamefont
  {Xie}, \citenamefont {Huang}, \citenamefont {Normand},\ and\ \citenamefont
  {Xiang}}]{LiaoHJ2017}%
  \BibitemOpen
  \bibfield  {author} {\bibinfo {author} {\bibfnamefont {H.~J.}\ \bibnamefont
  {Liao}}, \bibinfo {author} {\bibfnamefont {Z.~Y.}\ \bibnamefont {Xie}},
  \bibinfo {author} {\bibfnamefont {J.}~\bibnamefont {Chen}}, \bibinfo {author}
  {\bibfnamefont {Z.~Y.}\ \bibnamefont {Liu}}, \bibinfo {author} {\bibfnamefont
  {H.~D.}\ \bibnamefont {Xie}}, \bibinfo {author} {\bibfnamefont {R.~Z.}\
  \bibnamefont {Huang}}, \bibinfo {author} {\bibfnamefont {B.}~\bibnamefont
  {Normand}},\ and\ \bibinfo {author} {\bibfnamefont {T.}~\bibnamefont
  {Xiang}},\ }\bibfield  {title} {\bibinfo {title} {Gapless {{Spin-Liquid
  Ground State}} in the {{{\emph{S}}}}=1/2 {{Kagome Antiferromagnet}}},\ }\href
  {https://doi.org/10.1103/PhysRevLett.118.137202} {\bibfield  {journal}
  {\bibinfo  {journal} {Phys. Rev. Lett.}\ }\textbf {\bibinfo {volume} {118}},\
  \bibinfo {pages} {137202} (\bibinfo {year} {2017})}\BibitemShut {NoStop}%
\bibitem [{\citenamefont {Nishino}\ and\ \citenamefont
  {Okunishi}(1996)}]{NishinoT1996}%
  \BibitemOpen
  \bibfield  {author} {\bibinfo {author} {\bibfnamefont {T.}~\bibnamefont
  {Nishino}}\ and\ \bibinfo {author} {\bibfnamefont {K.}~\bibnamefont
  {Okunishi}},\ }\bibfield  {title} {\bibinfo {title} {Corner {{Transfer Matrix
  Renormalization Group Method}}},\ }\href
  {https://doi.org/10.1143/JPSJ.65.891} {\bibfield  {journal} {\bibinfo
  {journal} {J. Phys. Soc. Jpn.}\ }\textbf {\bibinfo {volume} {65}},\ \bibinfo
  {pages} {891} (\bibinfo {year} {1996})}\BibitemShut {NoStop}%
\end{thebibliography}%

\onecolumngrid
\appendix

\setcounter{figure}{0} 
\renewcommand{\thefigure}{S\arabic{figure}} 
\setcounter{equation}{0} 
\renewcommand{\theequation}{S\arabic{equation}}

\newpage
\begin{center}
{\bf ---Supplemental Material---}
\end{center}

\section{Materials Synthesis}
Single crystals of \NiP{} used in this study were grown using the flux method ~\cite{DingF2021,LiN2021}. The starting materials, Na$_2$CO$_3$, BaCO$_3$, NiO, (NH$_4$)$_2$HPO$_4$, and NaCl, were thoroughly mixed in a molar ratio of 1:1:1:2:5. The mixture was placed in an alumina crucible with a lid and heated to \qty{950}{\degreeCelsius} for 20 hours, followed by slow cooling to \qty{750}{\degreeCelsius} at a rate of \qty{1.3}{\degreeCelsius/\hour}. After cooling, the residual NaCl was removed by soaking the product in water, and yellow single crystals with a layered hexagonal shape [see Fig.~\ref{crystal}(a)] were mechanically separated from the bulk.

\section{Neutron Scattering}
\begin{figure}[tbp!]
\includegraphics[width=0.6\textwidth]{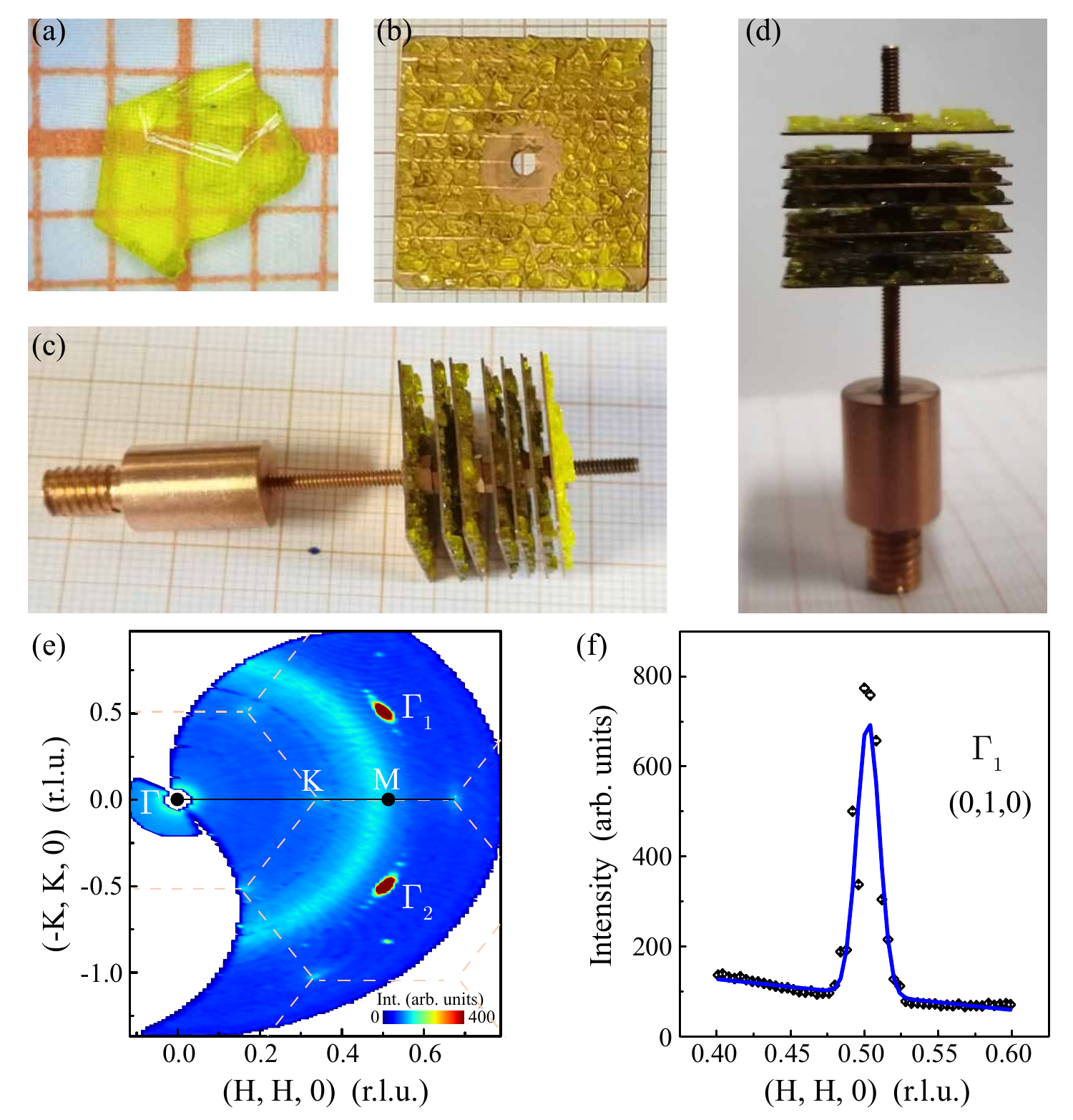}
\caption{(a) Picture of a \NiP{} single crystal. Each square in the picture represents $\qty{1}{mm}\times\qty{1}{mm}$. (b) Co-alignment of single crystals on the oxygen-free copper plate. (c)-(d) Assembly pictures of the co-aligned samples for neutron scattering experiments. (e) (H,K,L=0) scattering plane collected at $T=\qty{60}{mK}$ and $B=\qty{0}{T}$. (f) A line cut of the diffraction pattern along the (H,H,0) direction around (0,1,0) peak.  The blue line represents a Gaussian fit to the experimental data, yielding a Full Width at Half Maximum (FWHM) of $\qty{0.0365(2)}{\angstrom^{-1}}$, indicating good co-alignment of the sample.}
\label{crystal}
\end{figure}

The inelastic neutron scattering (INS) experiments were performed using the cold-neutron disk chopper spectrometer AMATERAS (BL14 beamline) at the Materials and Life Science Experimental Facility (MLF), J-PARC, with a fixed incident energy of $E_i=\qty{1.482}{meV}$ and an energy resolution of approximately $\qty{0.021}{meV}$ ~\cite{NakajimaK2011}. Additional measurements were carried out at the time-of-flight Cold Neutron Chopper Spectrometer (CNCS) at the Spallation Neutron Source, Oak Ridge National Laboratory, using a fixed incident energy of $E_i=\qty{1.55}{meV}$ and an energy resolution of approximately $\qty{0.033}{meV}$ ~\cite{EhlersG2011, EhlersG2016}. For these experiments, around 400 single crystals of \NiP{} were co-aligned in the (HK0) scattering plane on oxygen-free copper plates, with a total mass of approximately 2 grams, as shown in Figs.~\ref{crystal}(b)--(d). The samples were cooled using a dilution refrigerator insert within a \qty{7}{T} magnet on both spectrometers, with the magnetic field applied along the $c$-axis. INS data were collected at a base temperature of around $\qty{60}{mK}$ under various magnetic fields and processed using the freely available Utsusemi~\cite{InamuraY2013} and Dave~\cite{AzuahRT2009} software tools.

%\newpage

%\section{Supplementary Text}

\section{iPEPS ansatz for ground state and spin excitation spectra calculations}

The infinite projected entangled pair states (iPEPS) ansatz, a type of tensor network state, has been widely applied in studying two-dimensional strongly correlated systems~\cite{VerstraeteF2004_arxiv,VerstraeteF2008_review,OrusR2009,CorbozP2014_tJ,LiaoHJ2017,LiaoHJ2019}. In this research, we employed this ansatz to calculate the ground state properties and spin excitation spectra of the spin-1 XXZ model. The ground state $|0\rangle$ is represented by an infinite tensor network on a modified square lattice, transformed from the original triangular lattice by merging three sites on a triangle:
\begin{equation}
|0\rangle =
\begin{array}{l}
\begin{tikzpicture}[every node/.style={scale=1},scale=0.45]
          \trianglelattice{1.0}
    
          \triangular{0}{0}{1.04}
          \triangular{0}{-1.7321}{1.04}
          \triangular{1.5}{0.5*1.7321}{1.04}
          \triangular{1.5}{-0.5*1.7321}{1.04}
          \triangular{-1.5}{-0.5*1.7321}{1.04}
          \triangular{-1.5}{0.5*1.7321}{1.04}
          \triangular{3}{0}{1.04}
          \triangular{3.0}{-1.0*1.7321}{1.04}
    
          \draw[gray,thick] (0,2.05)--(0,-2.3);
          \draw[gray,thick] (-1.5,2.)--(-1.5,-2.3);
          \draw[gray,thick] (1.5,2.)--(1.5,-2.3);
          \draw[gray,thick] (3,2.05)--(3,-2.3);
    
          \blackline{-2}{1/3*1.7321}{0.1}{2.65}
          \blackline{0}{0}{2.4}{3.85}
          \blackline{1.5}{-0.5*1.7321}{2.6}{2.6}
          \blackline{3.0}{-1.0*1.7321}{1}{1}
    
          \Tensor{0}{0}{0.15}{0.2}
          \Tensor{0}{-1.7321}{0.15}{0.2}
          \Tensor{1.5}{0.5*1.7321}{0.15}{0.2}
          \Tensor{1.5}{-0.5*1.7321}{0.15}{0.2}
          \Tensor{-1.5}{-0.5*1.7321}{0.15}{0.2}
          \Tensor{-1.5}{0.5*1.7321}{0.15}{0.2}
          \Tensor{3}{0}{0.15}{0.2}
          \Tensor{3.0}{-1.0*1.7321}{0.15}{0.2}
          \Tensor{3.0}{1.7321}{0.15}{0.2}
          \Tensor{0}{1.7321}{0.15}{0.2}
\end{tikzpicture}
\end{array}
=
\begin{array}{l}
\begin{tikzpicture}[baseline,every node/.style={scale=1},scale=0.4]
        \draw[step=1 cm,gray] (-2.4,-2.4) grid (2.4,2.4);
        \Tensor{-1}{0}{0.15}{0.2}
        \Tensor{-1}{1}{0.15}{0.2}
        \Tensor{-1}{2}{0.15}{0.2}
        \Tensor{-1}{-1}{0.15}{0.2}
        \Tensor{-1}{-2}{0.15}{0.2}
        \Tensor{-2}{0}{0.15}{0.2}
        \Tensor{-2}{1}{0.15}{0.2}
        \Tensor{-2}{2}{0.15}{0.2}
        \Tensor{-2}{-1}{0.15}{0.2}
        \Tensor{-2}{-2}{0.15}{0.2}
        \CenterA{0}{0}{0.15}{0.2}
        \Tensor{0}{1}{0.15}{0.2}
        \Tensor{0}{2}{0.15}{0.2}
        \Tensor{0}{-1}{0.15}{0.2}
        \Tensor{0}{-2}{0.15}{0.2}
        \Tensor{1}{0}{0.15}{0.2}
        \Tensor{1}{1}{0.15}{0.2}
        \Tensor{1}{2}{0.15}{0.2}
        \Tensor{1}{-1}{0.15}{0.2}
        \Tensor{1}{-2}{0.15}{0.2}
        \Tensor{2}{0}{0.15}{0.2}
        \Tensor{2}{1}{0.15}{0.2}
        \Tensor{2}{2}{0.15}{0.2}
        \Tensor{2}{-1}{0.15}{0.2}
        \Tensor{2}{-2}{0.15}{0.2}
\end{tikzpicture}.
\end{array}
\end{equation}
The tensor network is built using translationally invariant tensors $A$. Each $A$ is a rank-5 tensor that contains one physical index of dimension $d=27$ and four virtual indices. The dimension of the virtual indices, known as the bond dimension $\chi$, determines the ansatz's capacity for entanglement, thereby controlling computational accuracy. The energy and the expectation values of local observables were calculated approximately using the corner-transfer-matrix renormalization group (CTMRG) method~\cite{NishinoT1996,OrusR2009,CorbozP2014_tJ}. To obtain the tensor $A$, we employed variational optimization of the ground state energy using automatic differentiation~\cite{LiaoHJ2019}. 

%Fig.~\ref{fig:magnetization} shows the ground state magnetization per site as a function of the magnetic field, calculated using iPEPS with bond dimensions $D=4$ and $5$. We find that iPEPS with $D=4$ provides sufficient precision for characterizing the four phases of the ground state, and all excited state calculations in the main text are performed with $D=4$.

%\begin{figure}[tbp!]
%\includegraphics[width=0.5\textwidth]{Fig_SM1.pdf}
%\caption{Field dependence of the ground state magnetization calculated by iPEPS with bond dimensions $D=4$ and $5$. The dashed lines mark the estimated phase transition points $B_\text{c1}=0.06$T and $B_\text{c2}=1.68$T.}
%\label{fig:magnetization}
%\end{figure}

For spin excitation spectra calculations, we employed the single-mode approximation~\cite{OstlundS1995,VanderstraetenL2015,PonsioenB2020} to construct the excited states. We started by replacing the tensor $A$ at position $\bm{r}$ in the ground state $|0\rangle$ with a new tensor $B$, then performed a Fourier transformation to get an excited state with momentum $\bm{k}$:
% ===========================================================================$
\begin{equation}
|\Phi_{\bm{k}}(B)\rangle = \sum_{\bm{r}} e^{\iu \bm{k}\cdot \bm{r}} |\Phi_{\bm{r}}(B)\rangle = \sum_{\bm{r}} e^{\iu \bm{k}\cdot \bm{r}}
\begin{array}{l}
\begin{tikzpicture}[every node/.style={scale=1},scale=0.5]
    \draw[step=1 cm,gray] (-2.4,-2.4) grid (2.4,2.4);
    \Tensor{-1}{0}{0.15}{0.2}
    \Tensor{-1}{1}{0.15}{0.2}
    \Tensor{-1}{2}{0.15}{0.2}
    \Tensor{-1}{-1}{0.15}{0.2}
    \Tensor{-1}{-2}{0.15}{0.2}
    \Tensor{-2}{0}{0.15}{0.2}
    \Tensor{-2}{1}{0.15}{0.2}
    \Tensor{-2}{2}{0.15}{0.2}
    \Tensor{-2}{-1}{0.15}{0.2}
    \Tensor{-2}{-2}{0.15}{0.2}
    \CenterB{0}{0}{0.15}{0.2}
    \Tensor{0}{1}{0.15}{0.2}
    \Tensor{0}{2}{0.15}{0.2}
    \Tensor{0}{-1}{0.15}{0.2}
    \Tensor{0}{-2}{0.15}{0.2}
    \Tensor{1}{0}{0.15}{0.2}
    \Tensor{1}{1}{0.15}{0.2}
    \Tensor{1}{2}{0.15}{0.2}
    \Tensor{1}{-1}{0.15}{0.2}
    \Tensor{1}{-2}{0.15}{0.2}
    \Tensor{2}{0}{0.15}{0.2}
    \Tensor{2}{1}{0.15}{0.2}
    \Tensor{2}{2}{0.15}{0.2}
    \Tensor{2}{-1}{0.15}{0.2}
    \Tensor{2}{-2}{0.15}{0.2}
\end{tikzpicture}
\end{array}.
\end{equation}

\begin{figure}[b]
\includegraphics[width=0.4\textwidth]{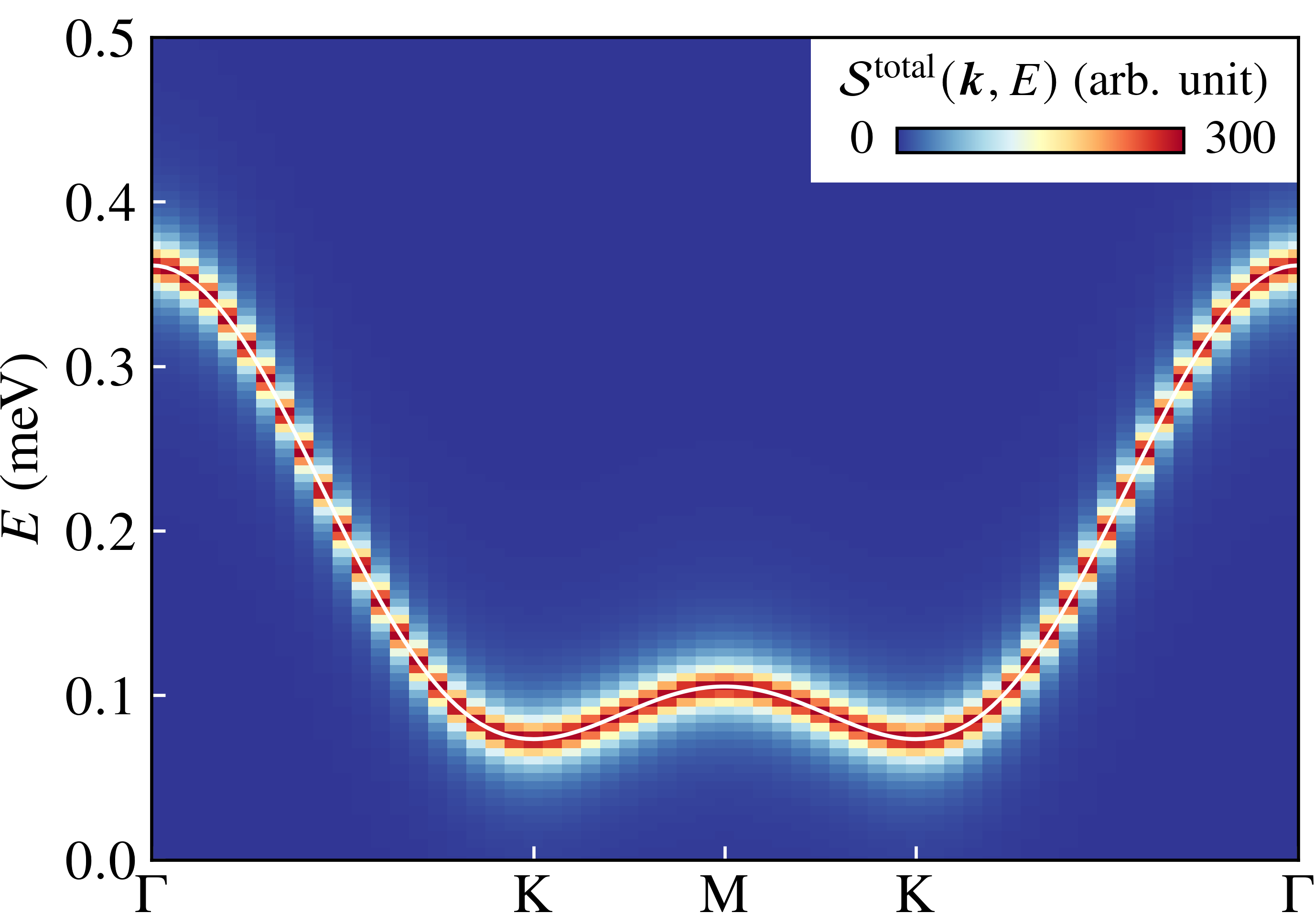}
\caption{$T=0$ iPEPS results of the spin excitation spectra at $B=\qty{2}{T}$ with bond dimension $\chi=4$. The white solid line shows the exact solution from Ref.~\cite{ShengJ2025_NiP}.}
\label{fig:fp}
\end{figure} 

To ensure orthogonality between the excited states and the ground state, we confined $\Phi_{\bm{r}}(B)$ within the tangent space of the ground state, i.e., $\langle \Phi_{\bm{r}}(B)|\Psi(A)\rangle=0$. After choosing a basis $|\Phi_{\bm{k}}(\tilde B_m)\rangle$ within the tangent space, we then calculated the Hamiltonian matrix and the norm matrix 
\begin{align}
H^\mathrm{eff}_{mn,\bm{k}} &= \langle\Phi_{\bm{k}} (\tilde B_m) |\mathcal{H}| \Phi_{\bm{k}} (\tilde B_n), \rangle\\
N^\mathrm{eff}_{mn,\bm{k}} &= \langle\Phi_{\bm{k}} (\tilde B_m)| \Phi_{\bm{k}} (\tilde B_n) \rangle,
\end{align} 
and solved the generalized eigen-equation 
\begin{equation} 
  \sum_n H^\mathrm{eff}_{mn,\bm{k}} v_{np} = \sum_n E_p N^\mathrm{eff}_{mn,\bm{k}} v_{np},
\end{equation}
to get the excited eigenstates $|\Phi_{\bm{k}}(B_m)\rangle$ and corresponding energy $E_m$. Automatic differentiation techniques~\cite{LiaoHJ2019,ChiR2022,PonsioenB2022} were used to accelerate Hamiltonian matrix calculations.

The DSSF at $T=0$ is given by the formula
\begin{equation}
 \mathcal{S}^{\alpha \alpha}(\bm{k},E) = \sum_{m} |\langle \Phi_{\bm{k}}(B_m) | S^{\alpha}_{\bm{k}}  | 0 \rangle|^2 \delta(E - E_m + E_0),
\end{equation}
where the delta function was replaced by Lorentzian function with coefficient $\eta$ to account for the instrumental resolution.

\begin{figure}[tbp!]
\includegraphics[width=0.9\textwidth]{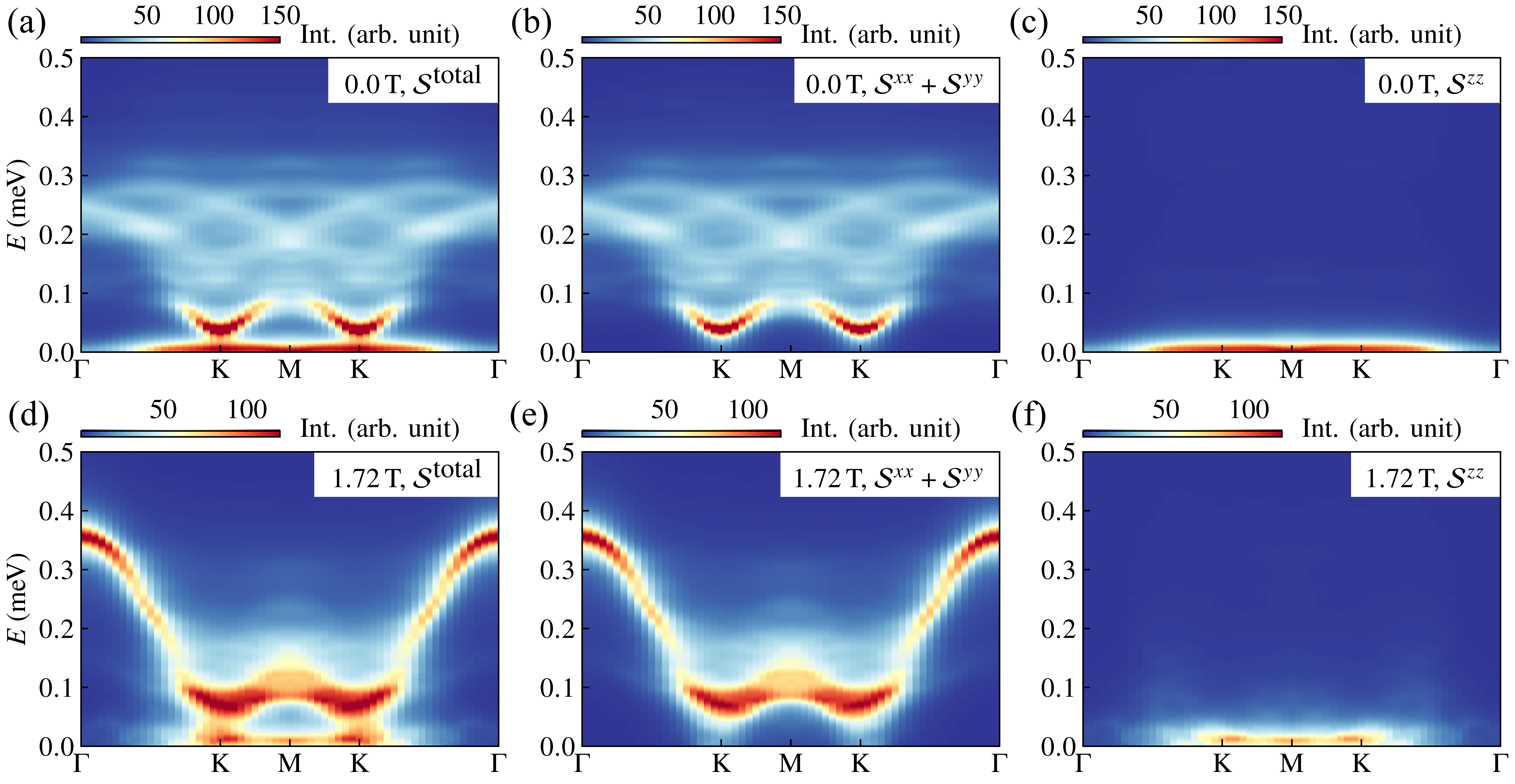}
\caption{$T=0$ iPEPS results of the spin excitation spectra at $B=\qty{0}{T}$ and \qty{1.72}{T}  with bond dimension $\chi=4$.
The component of the DSSF is indicated on the top right of each panel.
%Left column: total weights $\mathcal{S}^{\text{total}}=\mathcal{S}^{xx}+\mathcal{S}^{yy}+\mathcal{S}^{zz}$. Middle column: transverse fluctuation weights $\mathcal{S}^{xx}+\mathcal{S}^{yy}$. Right column: longitudinal fluctuation weights $\mathcal{S}^{zz}$.
}
\label{fig:sn}
\end{figure}

Figure~\ref{fig:fp} shows the iPEPS results of the DSSF in the FP phase, which is found to agree quantitativley with the exact solution. 
%We observe perfect consistency with the linear spin wave theory result, which is exact in the FP phase due to the lack of spin fluctuations. 
Figure~\ref{fig:sn} shows the DSSF in SN-supersolid and SN phases.
By separating the intensity of DSSF into different components, it is clear that the quadrupole wave (narrow low-energy mode) only appear in the $\mathcal{S}^{zz}(\bm{k},E)$ component.
%including contributions from transverse fluctuation modes and longitudinal fluctuation modes. We find that in these two SN phases, the narrow modes near zero energy both belong to longitudinal fluctuation modes.

\begin{figure}[tbp!]
\includegraphics[width=0.8\textwidth]{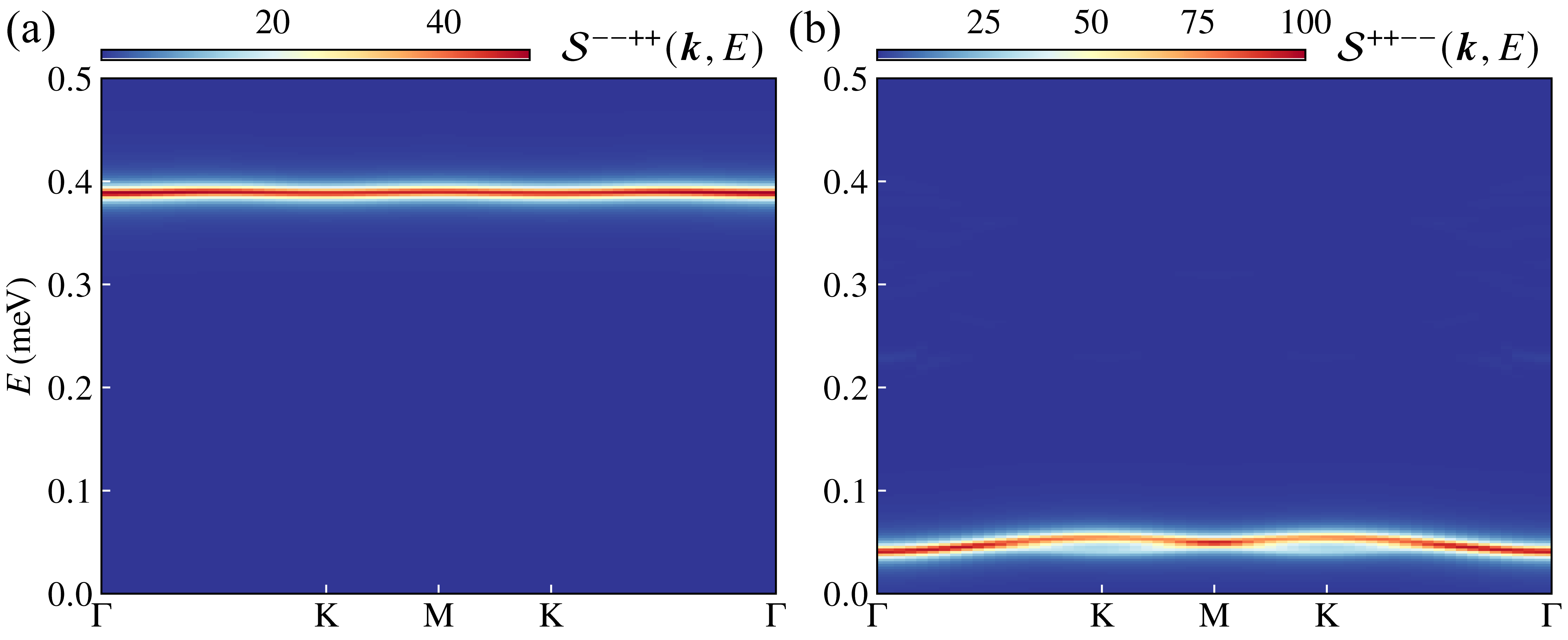}
\caption{$T=0$ iPEPS results of the DQSF at $B=\qty{0.2}{T}$ with bond dimension $\chi=4$.}
\label{fig:twomagnon}
\end{figure} 

The 2-magnon bound states in the UUD phase are invisible in the DSSF due to the $U(1)$ symmetry.
To reveal its existence, we calculated the dynamic quadrupolar structure factors (DQSF):
\begin{align}
\mathcal{S}^{--++}(\bm{k},E) &= \sum_{m} |\langle \Phi_{\bm{k}}(B_m) | S^{++}_{\bm{k}}  | 0 \rangle|^2 \delta(E - E_m + E_0),\\
\mathcal{S}^{++--}(\bm{k},E) &= \sum_{m} |\langle \Phi_{\bm{k}}(B_m) | S^{--}_{\bm{k}}  | 0 \rangle|^2 \delta(E - E_m + E_0),
\end{align}
where $S^{++}_{\bm{k}}\equiv\sum_{i}e^{\iu \bm{k}\cdot \bm{r}_i} (S^+_i)^2$ and $S^{--}_{\bm{k}} \equiv \sum_{i}e^{\iu \bm{k}\cdot \bm{r}_i} (S^-_i)^2$. 
%$\mathcal{S}^{--++}$ and $\mathcal{S}^{++--}$ correspond to spectral weights of $\Delta S^z=2$ and $-2$, respectively. 
Figure~\ref{fig:twomagnon} shows the results at $B=\qty{0.2}{T}$ in the UUD phase, where  two bound state modes are identified. %modes corresponding to $\Delta S^z=\pm2$, both with minimal dispersion and lowest energies at the $\Gamma$ point.

\section{Linear spin wave theory for the low-energy branches}

The Linear spin wave (LSW) theory was applied to the effective spin-1/2 model in the main text. We first used the site-factorized $SU(2)$ coherent state representation to minimize the classical energy of the effective model, which leads to four distinct phases:
\begin{enumerate}
\item SN-supersolid, with 3 sublattices (ordering wavevector at K): $\vec{s}_1/s=\vec{s}_2/s=(\cos \theta_1, 0, \sin \theta_1)$ and $\vec{s}_3/s=(\cos \theta_2, 0, -\sin \theta_2)$, up to a global rotation in the $xy$ plane;
\item UUD, with 3 sublattices (ordering wavevector at K): $\vec{s}_1/s=\vec{s}_2/s=(0,0,1)$, and $\vec{s}_3/s=(0,0,-1)$;
\item SN, with a single sublattice (ordering wavevector at $\Gamma$): $\vec{s}_1/s=(\cos \theta_1, 0, \sin \theta_1)$, up to a global rotation in the $xy$ plane;
\item FP, with a single sublattice (ordering wavevector at $\Gamma$): $\vec{s}_1/s=(0,0,1)$.
\end{enumerate}
Here, $s=1/2$ is the magnitude of the effective spins, and \{$\theta_1$, $\theta_2$\} are parameters that minimize the classical energy of the effective model.

The connection between these semiclassical phases of the effective model and those of the original model was established in Ref.~\cite{ShengJ2025_NiP}. Since the mapping between the spin operators is $s_i^-=P_0 (S_i^- S_i^-/2)P_0$, it is clear that any in-plane dipolar moment of $s=1/2$ is translated to in-plane quadrupolar moment of the original $S=1$ model.

Based on these semiclassical states, we carried out the standard $T=0$ LSW calculations for the SN-supersolid and SN phases with Holstein-Primakoff bosons. As discussed in the main text, only $\mathcal{S}_\text{eff}^{zz}(\bm{k},E)$ contributes to the DSSF of the original $S=1$ model.

\section{Generalized linear spin wave calculations}
The generalized linear spin wave (GLSW) theory has been widely used for systems where the magnitude of spin $S>1/2$, and sometimes also termed flavor wave theory in the literature~\cite{MatveevVM1973,PapanicolaouN1988,BatistaCD2004_SUN,TsunetsuguH2006,LauchliA2006,BaiX2021,SeifertUFP2022}. Here we followed the standard approach and used the $SU(3)$ coherent state representation for the original $S=1$ model. We note that this semiclassical approximation is not adequate to produce all the phases that appear under a magnetic field, where only the UUD and FP phases survive. 

We started by introducing the $SU(3)$ Schwinger bosons in the fundamental representation:
\begin{equation}
S_i^\alpha = \bm{b}_i^{(0) \dagger} L^\alpha \bm{b}_i^{(0)},
\end{equation}
where
\begin{equation}
\bm{b}_i^{(0)} \equiv \left( b_{ix}^{(0)} \quad b_{iy}^{(0)} \quad b_{iz}^{(0)} \right)^T,
\end{equation}
with the constraint $\sum_\alpha b_{i,\alpha}^{(0)\dagger} b_{i,\alpha}^{(0)} = 1$.
The matrices $L^\alpha$ are the spin-1 Pauli matrices in the quadrupolar representation:
\begin{equation}
L^x \equiv 
\begin{pmatrix}
0 & 0 & 0\\
0 & 0 & -\iu\\
0 & \iu & 0
\end{pmatrix},\quad
L^y \equiv 
\begin{pmatrix}
0 & 0 & \iu \\
0 & 0 & 0 \\
-\iu & 0 & 0
\end{pmatrix},\quad
L^z \equiv 
\begin{pmatrix}
0 & -\iu & 0 \\
\iu & 0 & 0 \\
0 & 0 & 0
\end{pmatrix}.
\end{equation}

The quadrupolar basis $|\alpha \rangle \equiv b_{i,\alpha}^{(0)\dagger} | 0 \rangle$ was adopted in this work:
\begin{equation}
|x\rangle \equiv \iu \frac{|1 \rangle - |-1\rangle}{\sqrt{2}},\quad |y\rangle \equiv \frac{|1 \rangle + |-1\rangle}{\sqrt{2}}, \quad |z\rangle = -\iu |0\rangle.
\end{equation}

The UUD phase contains 3 sublattices in the minimal magnetic unit cell described by a superlattice basis \{$\bm{A}_1 = 2\bm{a}_1 + \bm{a}_2$, $\bm{A}_2=\bm{a}_1 + 2\bm{a}_2$\}, where \{$\bm{a}_1$, $\bm{a}_2$\} is the basis of the TL.
Accordingly, the reciprocal space basis of the superlattice is
\begin{equation}
\begin{split}
\bm{B}_1 &= \frac{2}{3} \bm{b}_1 - \frac{1}{3} \bm{b}_2, \\
\bm{B}_2 &= - \frac{1}{3} \bm{b}_1 + \frac{2}{3} \bm{b}_2,
\end{split}
\end{equation}
where \{$\bm{b}_1$, $\bm{b}_2$\} is the reciprocal space basis of the TL.

The wave function of each sublattice in the UUD phase is: 
\begin{equation}
|Z_1^\text{(GS)}\rangle = |Z_2^\text{(GS)}\rangle = |1 \rangle = \frac{-\iu |x\rangle + |y\rangle}{\sqrt{2}}, \quad |Z_3^\text{(GS)}\rangle = |-1 \rangle = \frac{\iu |x\rangle + |y\rangle}{\sqrt{2}}.
\end{equation}

To describe excitations on top of this spin configuration,  we performed an $SU(3)$ rotation of the basis:
\begin{align}
\left(b_{i0}^\dagger \quad b_{i1}^\dagger \quad b_{i2}^\dagger \right)
&= \left( b_{ix}^{(0)\dagger} \quad b_{iy}^{(0)\dagger} \quad b_{iz}^{(0)\dagger} \right) R_i, \\
\tilde{L}_i^\alpha &= R_i^\dagger L^\alpha R_i,
\end{align}
where
\begin{equation}
R_1 = R_2 = 
\begin{pmatrix}
-\frac{\iu}{\sqrt{2}} & 0 & -\frac{1}{\sqrt{2}} \\
\frac{1}{\sqrt{2}} & 0 & \frac{\iu}{\sqrt{2}} \\
0 & -1 & 0
\end{pmatrix},\quad
R_3 =
\begin{pmatrix}
\frac{\iu}{\sqrt{2}} & 0 & -\frac{1}{\sqrt{2}} \\
\frac{1}{\sqrt{2}} & 0 & -\frac{\iu}{\sqrt{2}} \\
0 & -1 & 0
\end{pmatrix}.
\end{equation}

The GLSW calculation of the $S=1$ Hamiltonian followed the standard procedure, where the spin operators were replaced by Schwinger bosons. Then, the $b_{i0}$ boson was condensed:
\begin{equation}\label{eq:condense}
b_{i0} = b_{i0}^\dagger = \sqrt{1-b_{i1}^\dagger b_{i1} - b_{i2}^\dagger b_{i2}}.
\end{equation}

The square roots were Taylor expanded, and the Hamiltonian was kept to quadratic order, then further Fourier transformed into momentum space:
\begin{equation}\label{eq:Ham_GLSW}
\mathcal{H} \approx \sum_{\tilde{\bm{k}}}^\prime \Psi_{\tilde{\bm{k}}}^\dagger H_\text{GLSW} (\tilde{\bm{k}}) \Psi_{\tilde{\bm{k}}} + C,
\end{equation}
where $C$ is a constant. The prime on sum denotes that only half of the folded Brillouin zone is counted, and
\begin{equation}
\Psi_{\tilde{\bm{k}}} \equiv \left(
b_{\tilde{\bm{k}},1,\bm{d}_1},\,
b_{\tilde{\bm{k}},1,\bm{d}_2},\, 
b_{\tilde{\bm{k}},1,\bm{d}_3},\,
b_{\tilde{\bm{k}},2,\bm{d}_1},\,
b_{\tilde{\bm{k}},2,\bm{d}_2},\, 
b_{\tilde{\bm{k}},2,\bm{d}_3},\,
b_{-\tilde{\bm{k}},1,\bm{d}_1}^\dagger,\,
b_{-\tilde{\bm{k}},1,\bm{d}_2}^\dagger,\, 
b_{-\tilde{\bm{k}},1,\bm{d}_3}^\dagger,\,
b_{-\tilde{\bm{k}},2,\bm{d}_1}^\dagger,\,
b_{-\tilde{\bm{k}},2,\bm{d}_2}^\dagger,\, 
b_{-\tilde{\bm{k}},2,\bm{d}_3}^\dagger
\right)^T.
\end{equation}

Here, the Fourier transformation is defined as
\begin{equation}
\begin{split}
b_{i1} & \equiv b_{\tilde{\bm{r}}+\bm{d},1} = \sqrt{\frac{3}{N}} \sum_{\tilde{\bm{k}}} e^{\iu \tilde{\bm{k}} \cdot \tilde{\bm{r}}} b_{\tilde{\bm{k}},1,\bm{d}},\\
b_{i2} & \equiv b_{\tilde{\bm{r}}+\bm{d},2} = \sqrt{\frac{3}{N}} \sum_{\tilde{\bm{k}}} e^{\iu \tilde{\bm{k}} \cdot \tilde{\bm{r}}} b_{\tilde{\bm{k}},2,\bm{d}},
\end{split}
\end{equation}
where $N$ is the total number of lattice sites, $\tilde{\bm{r}}=m\bm{A}_1 + n\bm{A}_2$ denotes the coordinates of the superlattice ($m$, $n$ are integers), and $\bm{d}=\{\bm{d}_1,\,\bm{d}_2, \, \bm{d}_3\}$ are coordinates of the 3 sublattices:
\begin{equation}
\bm{d}_1 = 0, \quad \bm{d}_2 = \bm{a}_1, \quad \bm{d}_3 = \bm{a}_1 + \bm{a}_2.
\end{equation}
For each momentum $\bm{k}$, there is a unique decomposition 
\begin{equation}
\bm{k} = \tilde{\bm{k}} + \bm{K},
\end{equation}
where $\bm{K}=m \bm{B}_1 + n \bm{B}_2$ ($m$, $n$ are integers) denotes the coordinates of the reciprocal lattice, and $\tilde{\bm{k}}=\sum_i \tilde{k}_i \bm{B}_i$ where $\tilde{k}_i \in [0,1)$.

As indicated from Eq.~\eqref{eq:Ham_GLSW}, only the $\tilde{\bm{k}}$ part of a given momentum $\bm{k}$ is relevant in the dispersion relation, which can be obtained by the Bogoliubov transformation:
\begin{align}
\Psi_{\tilde{\bm{k}}} &= V_{\tilde{\bm{k}}} \tilde{\Psi}_{\tilde{\bm{k}}},\\
V_{\tilde{\bm{k}}}^\dagger H_\text{GLSW}(\tilde{\bm{k}}) V_{\tilde{\bm{k}}} &= \text{diag} \{ \omega_{\tilde{\bm{k}},6}, \ldots , \omega_{\tilde{\bm{k}},1}, \omega_{-\tilde{\bm{k}},1}, \ldots , \omega_{-\tilde{\bm{k}},6} \}.
\end{align}

The $T=0$ dynamic spin structure factor (DSSF) is defined as
\begin{equation}
\mathcal{S}^{ab}(\bm{k}, E) = 2 \pi \sum_{\nu \neq 0} \langle 0 | S_{\bm{k}}^a | \nu \rangle \langle \nu | S_{-\bm{k}}^b | 0 \rangle \delta \left( E + E_0 - E_\nu  \right),
\end{equation}
where $|\nu \rangle$ enumerates the excited states. $E_0$ and $E_\nu$ are the energies of the ground state and the excited states, respectively.

To compute the DSSF within the GLSW scheme, we expanded the spin operators by the Schwinger bosons:
\begin{align}
S_{\bm{k}}^\alpha &=  \frac{1}{\sqrt{N}} \sum_{\bm{r}} e^{-\iu \bm{k} \cdot \bm{r}} \bm{b}_{\bm{r}}^\dagger \tilde{L}_{\bm{r}}^\alpha \bm{b}_{\bm{r}} \nonumber \\
& = \frac{\sqrt{N}}{3} \delta_{\tilde{\bm{k}},0} \sum_{\bm{d}} e^{-\iu \bm{k} \cdot \bm{d}} \left( \tilde{L}_{\bm{d}}^\alpha \right)_{11} \nonumber \\
& \quad + \frac{1}{\sqrt{3}} \sum_{\bm{d}} \sum_{a=2}^3 e^{-\iu \bm{k} \cdot \bm{d}} \left[ \left( \tilde{L}_{\bm{d}}^\alpha \right)_{1a} b_{\tilde{\bm{k}},a,\bm{d}} + \left( \tilde{L}_{\bm{d}}^\alpha \right)_{a1} b_{-\tilde{\bm{k}},a,\bm{d}}^\dagger \right] \nonumber \\
& \quad + \frac{1}{\sqrt{N}} \sum_{\bm{d}} \sum_{a,b=2}^3 e^{-\iu \bm{k} \cdot \bm{d}} \left[ \left( \tilde{L}_{\bm{d}}^\alpha \right)_{ab} - \left( \tilde{L}_{\bm{d}}^\alpha \right)_{11} \delta_{ab} \right] \sum_{\tilde{\bm{q}}} b_{\tilde{\bm{q}},a,\bm{d}}^\dagger b_{\widetilde{\bm{k}+\bm{q}},b,\bm{d}} + \ldots ,
\end{align}

The DSSF including both the 1-boson contribution and the 2-boson continuum is:
\begin{equation}\label{eq:DSSF_GLSW}
\begin{split}
\mathcal{S}^{ab}(\bm{k},E) &= 2 \pi \sum_{j=1}^6 A_j^a (\bm{k}) \left[ A_j^b (\bm{k}) \right]^* \delta \left( E- \omega_{\tilde{\bm{k}},j} \right) \\
&\quad + \pi \int \frac{\mathrm{d} \tilde{\bm{q}}}{\mathcal{A}_\text{BZ}} \sum_{j_1,j_2=1}^6 A_{j_1j_2}^a (\bm{k},\tilde{\bm{q}}) \left[ A_{j_1j_2}^b (\bm{k},\tilde{\bm{q}})\right]^* \delta \left( E - \omega_{-\tilde{\bm{q}},j_1} - \omega_{\widetilde{\bm{k}+\bm{q}},j_2} \right),
\end{split}
\end{equation}
where $\mathcal{A}_\text{BZ}$ is the area of the folded Brillouin zone, and 
\begin{subequations}
\begin{align}
A_j^a(\bm{k}) &\equiv \frac{1}{\sqrt{3}} \sum_{i=1}^3 e^{-\iu \bm{k} \cdot \bm{d}_i} \left[ 
\left( \tilde{L}_{\bm{d}_i}^a \right)_{12} \left( V_{\tilde{\bm{k}}} \right)_{i,7-j} + 
\left( \tilde{L}_{\bm{d}_i}^a \right)_{13} \left( V_{\tilde{\bm{k}}} \right)_{i+3,7-j} +
\left( \tilde{L}_{\bm{d}_i}^a \right)_{21} \left( V_{\tilde{\bm{k}}} \right)_{i+6,7-j} +
\left( \tilde{L}_{\bm{d}_i}^a \right)_{31} \left( V_{\tilde{\bm{k}}} \right)_{i+9,7-j} \right], \\
A_{j_1j_2}^a (\bm{k},\tilde{\bm{q}}) &\equiv \frac{1}{\sqrt{3}} \sum_{i=1}^3 e^{-\iu \bm{k} \cdot \bm{d}_i} \nonumber \\
&\quad \quad \cdot \bigg\{ \left( \left( \tilde{L}_{\bm{d}_i}^a \right)_{22} - \left( \tilde{L}_{\bm{d}_i}^a \right)_{11} \right) \cdot 
\left[ 
\left( V_{-\tilde{\bm{q}}} \right)_{i,7-j_1} 
\left( V_{\widetilde{\bm{k}+\bm{q}}} \right)_{i+6,7-j_2}
+
\left( V_{-\tilde{\bm{q}}} \right)_{i+6,7-j_1}
\left( V_{\widetilde{\bm{k}+\bm{q}}} \right)_{i,7-j_2}
\right] \nonumber \\
&\quad \quad \,\, + \left( \left( \tilde{L}_{\bm{d}_i}^a \right)_{33} - \left( \tilde{L}_{\bm{d}_i}^a \right)_{11} \right) \cdot 
\left[ 
\left( V_{-\tilde{\bm{q}}} \right)_{i+3,7-j_1} 
\left( V_{\widetilde{\bm{k}+\bm{q}}} \right)_{i+9,7-j_2}
+
\left( V_{-\tilde{\bm{q}}} \right)_{i+9,7-j_1}
\left( V_{\widetilde{\bm{k}+\bm{q}}} \right)_{i+3,7-j_2}
\right] \nonumber \\
&\quad \quad \,\, + \left( \tilde{L}_{\bm{d}_i}^a \right)_{23} \cdot \left[ 
\left( V_{-\tilde{\bm{q}}} \right)_{i+3,7-j_1} 
\left( V_{\widetilde{\bm{k}+\bm{q}}} \right)_{i+6,7-j_2}
+
\left( V_{-\tilde{\bm{q}}} \right)_{i+6,7-j_1}
\left( V_{\widetilde{\bm{k}+\bm{q}}} \right)_{i+3,7-j_2}
\right] \nonumber \\
& \quad \quad \,\, + \left( \tilde{L}_{\bm{d}_i}^a \right)_{32} \cdot \left[ 
\left( V_{-\tilde{\bm{q}}} \right)_{i,7-j_1} 
\left( V_{\widetilde{\bm{k}+\bm{q}}} \right)_{i+9,7-j_2}
+
\left( V_{-\tilde{\bm{q}}} \right)_{i+9,7-j_1}
\left( V_{\widetilde{\bm{k}+\bm{q}}} \right)_{i,7-j_2}
\right] \bigg\}.
\end{align}
\end{subequations}

%Figure shows the DSSF at $B=\qty{0.8}{T}$ calculated according to Eq.~\eqref{eq:DSSF_GLSW}.

% \newpage
\section{Additional spin excitation spectra}

\begin{figure}[tbp!]
\includegraphics[width=0.8\textwidth]{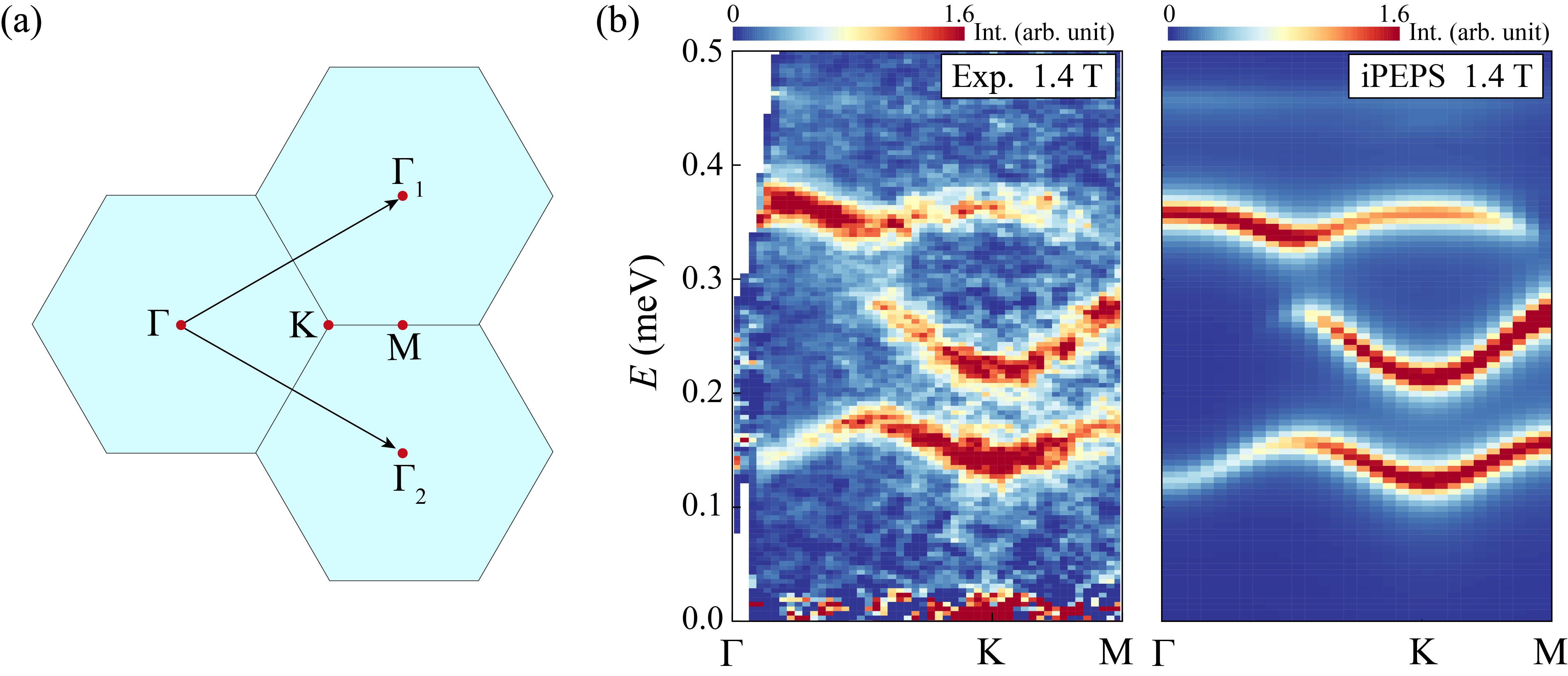}
\caption{(a) Schematics of the Brillouin zones with high-symmetry points $\Gamma=(0,0,0)$, K=(1/3,\,1/3,\,0), and M=(1/2,\,1/2,\,0). (b) Spin excitation spectra of \NiP{} at $B=\qty{1.4}{T}$. Left: Background subtracted INS results at \qty{60}{mK}, where INS data collected at \qty{60}{mK} and \qty{4}{T} was used as the background for subtraction. Right: iPEPS results at $T=0$ with bond dimension $\chi=4$.}
\label{fig:B1p4T_SW}
\end{figure}

Figure~\ref{fig:B1p4T_SW} shows the spin excitation spectra at \qty{1.4}{T}. Similarly to the results in the main text, we again see quantitative agreement between the INS and iPEPS results.

\begin{figure}[tbp!]
\includegraphics[width=0.6\textwidth]{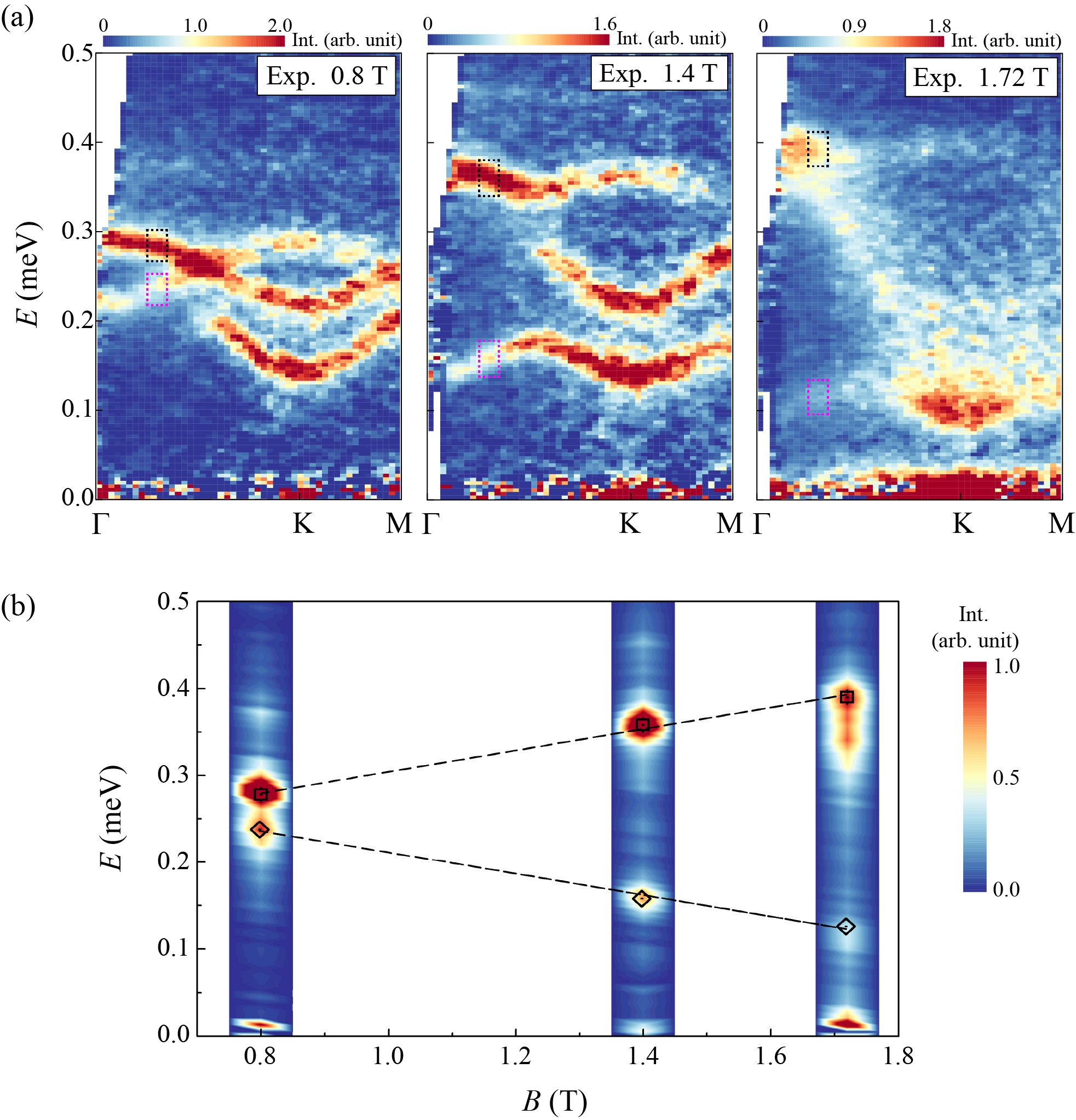}
\caption{(a) Background-subtracted INS spectra of \NiP{} at different magnetic fields measured at \qty{60}{mK}, where the dashed boxes indicate the spin wave excitations characteristic of the UUD phase near $\bm{k}=(0.1,\,0.1,\,0)$. (b) Background-subtracted INS spectra of \NiP{} at $\bm{k}=(0.1,\,0.1,\,0)$ [dashed boxes in (a)], where the dashed lines are guides to the eye. }
\label{fig:extrapolation}
\end{figure}

As we discussed in the main text, due to the first-order nature of $B_\text{c2}$, the INS data at $B=\qty{1.72}{T}$ have additional contributions from residual UUD domains. In Fig.~\ref{fig:extrapolation}(a) we compare the INS data between the UUD and the SN phases side by side. By tracking the INS spectra at a specific wavevector $\bm{k}=(0.1,\, 0.1,\, 0)$ [dashed boxes in Fig.~\ref{fig:extrapolation}(a)], Fig.~\ref{fig:extrapolation}(b) confirms that the extra modes in the INS data at $\qty{1.72}{T}$ are indeed consistent with those extrapolated from the UUD phase.

\begin{figure}[tbp!]
\includegraphics[width=0.9\textwidth]{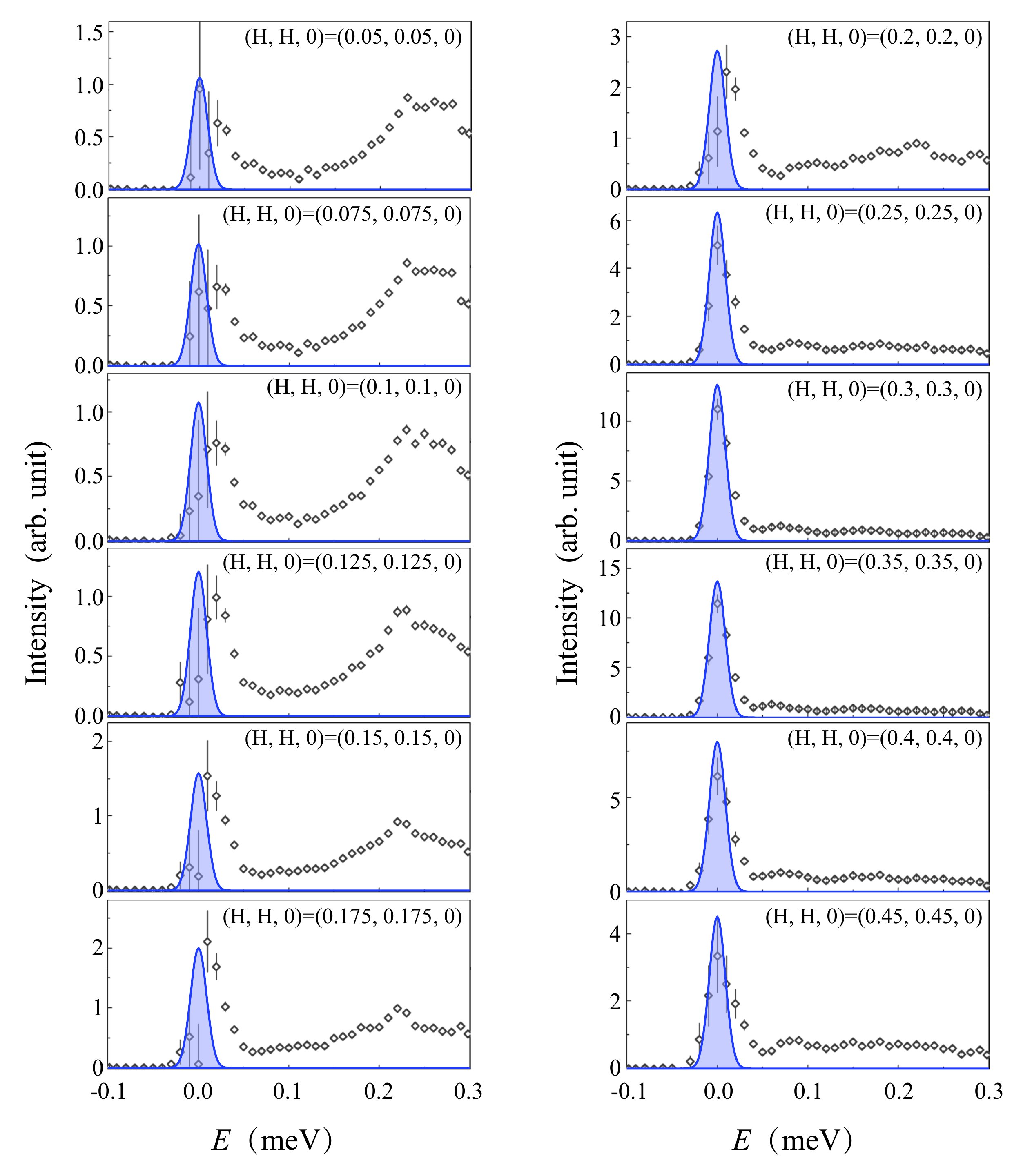}
\caption{Background-subtracted INS intensities at different wave vectors measured at $T=\qty{60}{mK}$ with $B=\qty{0}{T}$. The blue solid lines are the Gaussian fits to the residual incoherent scattering near the elastic line, with a fixed FWHM of \qty{0.021}{meV} that corresponds to the energy resolution of the spectrometer. The dark blue shaded areas represent the fitted residual incoherent scattering. The error bars are obtained via uncertainty propagation in subtracting the background at \qty{4}{T} and \qty{60}{mK}. }
\label{B0T}
\end{figure} 

\begin{figure}[tbp!]
\includegraphics[width=0.9\textwidth]{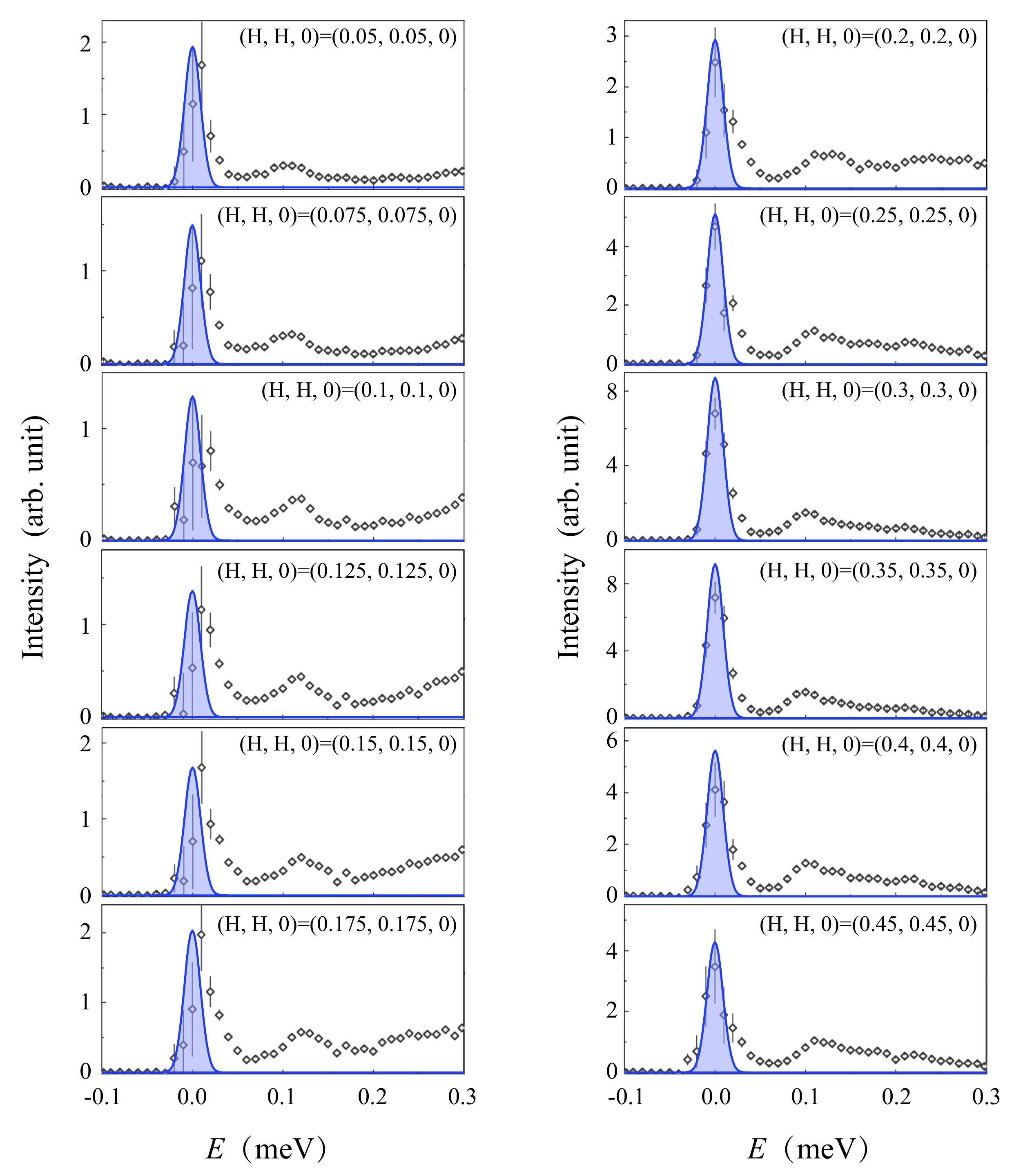}
\caption{Background-subtracted INS intensities at different wave vectors measured at $T=\qty{60}{mK}$ with $B=\qty{1.72}{T}$. The blue solid lines are the Gaussian fits to the residual incoherent scattering near the elastic line, with a fixed FWHM of \qty{0.021}{meV} that corresponds to the energy resolution of the spectrometer.  The dark blue shaded areas represent the fitted residual incoherent scattering. The error bars are obtained via uncertainty propagation in subtracting the background at \qty{4}{T} and \qty{60}{mK}.}
\label{B1p72T}
\end{figure}

Figures~\ref{B0T} and \ref{B1p72T} show the background-subtracted INS intensities at different wave vectors measured at $T=\qty{60}{mK}$, with $B=\qty{0}{T}$ and $B=\qty{1.72}{T}$, respectively. Here, data collected at \qty{60}{mK} and \qty{4}{T} was used as the background.
This subtraction effectively removed most of the low-energy incoherent scattering background, with a small amount of residual incoherent scattering remaining near the elastic line. 

To further isolate the quadrupole signal, we fitted the residual incoherent component using a Gaussian function with a fixed full width at half maximum (FWHM) of \qty{0.021}{meV} that corresponds to the instrument’s energy resolution. These fits are shown as the dark blue shaded areas in Figs.~\ref{B0T} and \ref{B1p72T}.

\begin{figure}[tbp!]
\includegraphics[width=0.9\textwidth]{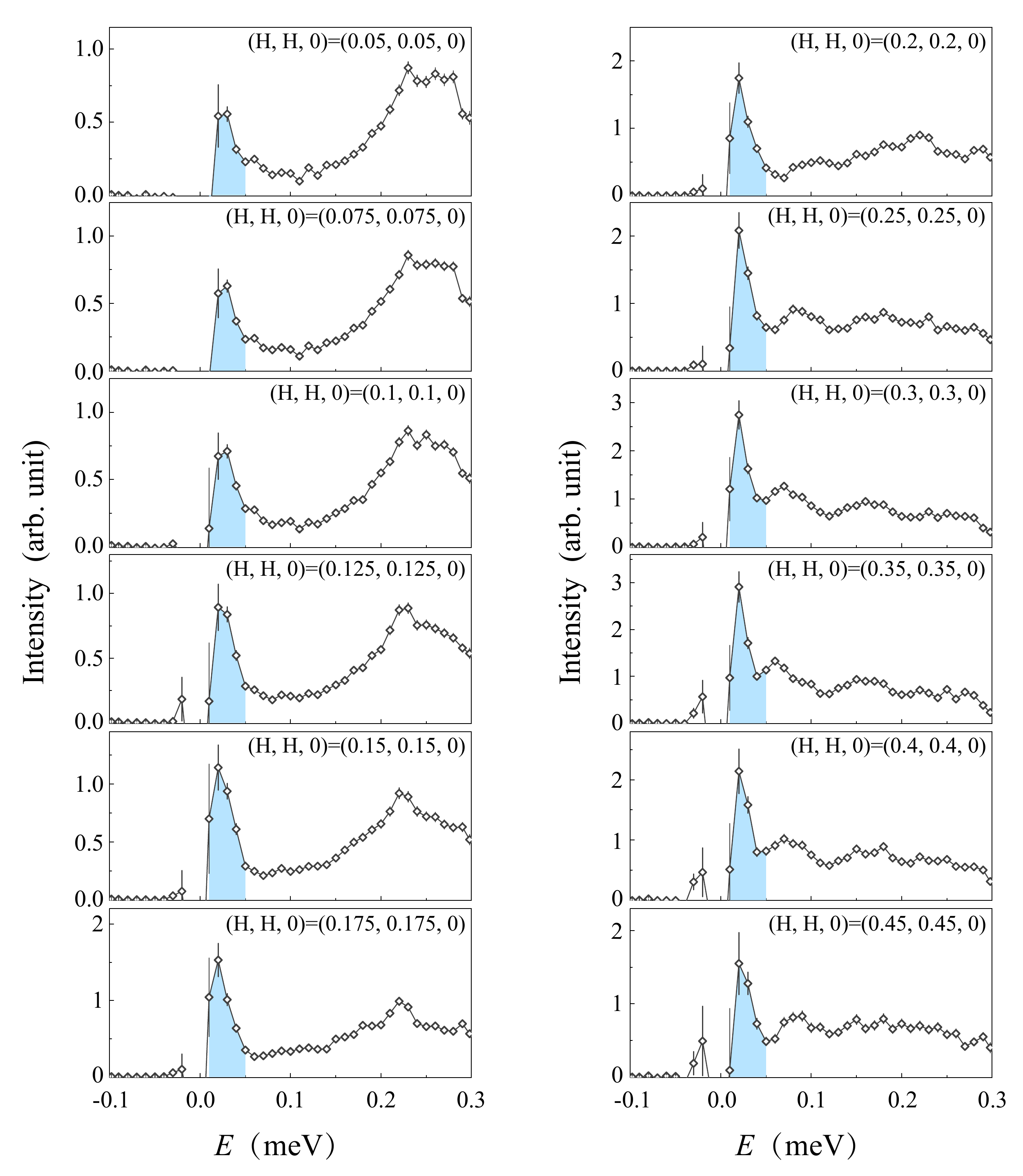}
\caption{INS intensity derived from Fig.~\ref{B0T} after subtracting the fitted residual incoherent scattering (dark blue shaded areas in Fig.~\ref{B0T}). The light blue areas in the 0.01--\qty{0.05}{meV} range indicate spectral weight that mainly arises from the quadrupole wave contribution. The error bars are identical to those in Fig.~\ref{B0T}.}
\label{B0T_subBG}
\end{figure}

\begin{figure}[tbp!]
\includegraphics[width=0.9\textwidth]{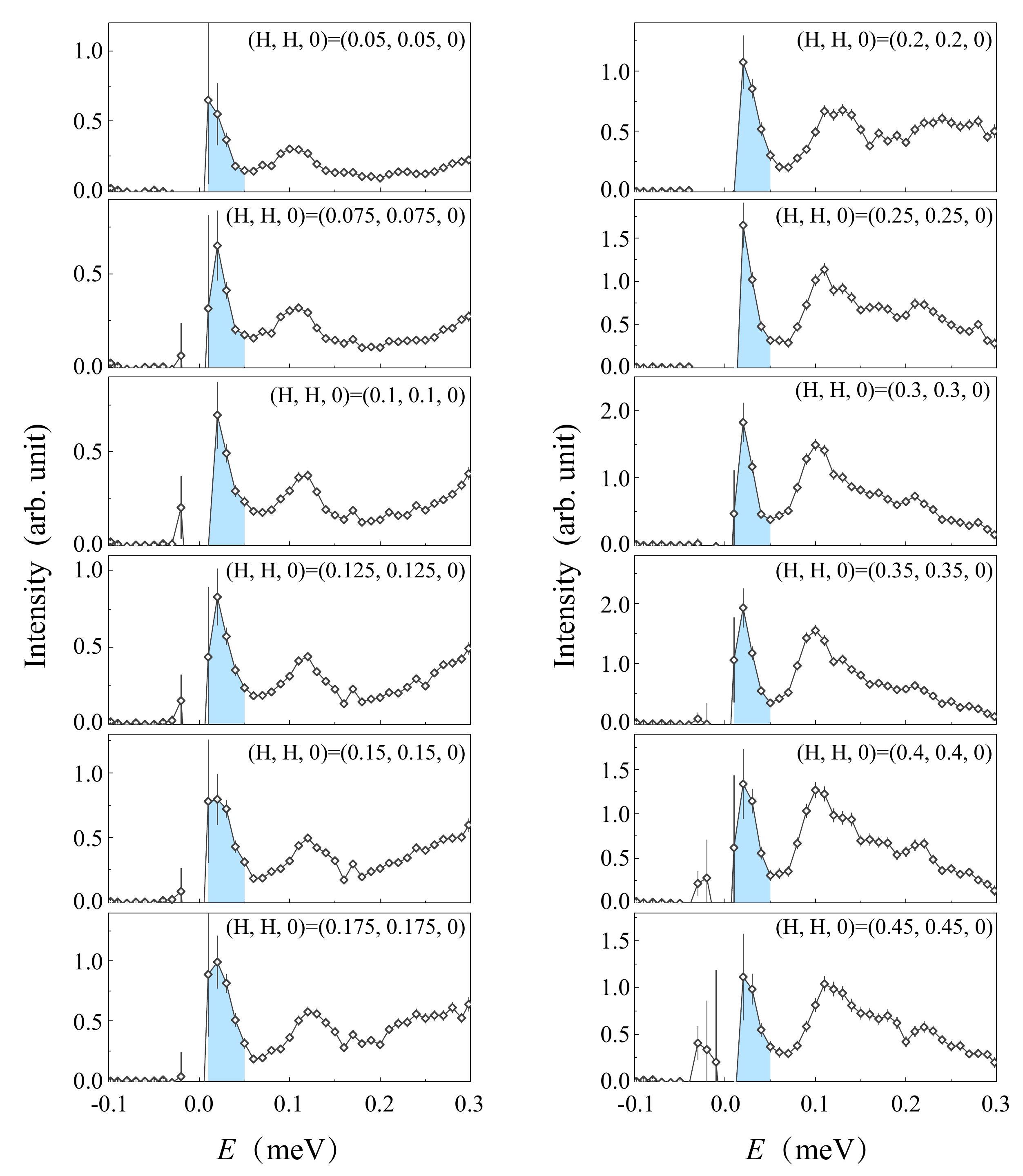}
\caption{INS intensity derived from Fig.~\ref{B1p72T} after subtracting the fitted residual incoherent scattering (dark blue shaded areas in Fig.~\ref{B1p72T}). The light blue areas in the 0.01--\qty{0.05}{meV} range indicate spectral weight that mainly arises from the quadrupole wave contribution. The error bars are identical to those in Fig.~\ref{B1p72T}.}
\label{B1p72T_subBG}
\end{figure}

After subtracting the fits to these incoherent backgrounds, no significant residual signal remains in the negative energy range. In contrast, a remaining spectral weight can be observed in the positive energy range (0.01 to \qty{0.05}{meV}), highlighted by the light blue areas in Figs.~\ref{B0T_subBG} and \ref{B1p72T_subBG}. This remaining spectral weight is consistent with the theoretical prediction of a narrow low-energy mode, i.e., the quadrupole wave.

\begin{figure}[tbp!]
\includegraphics[width=0.9\textwidth]{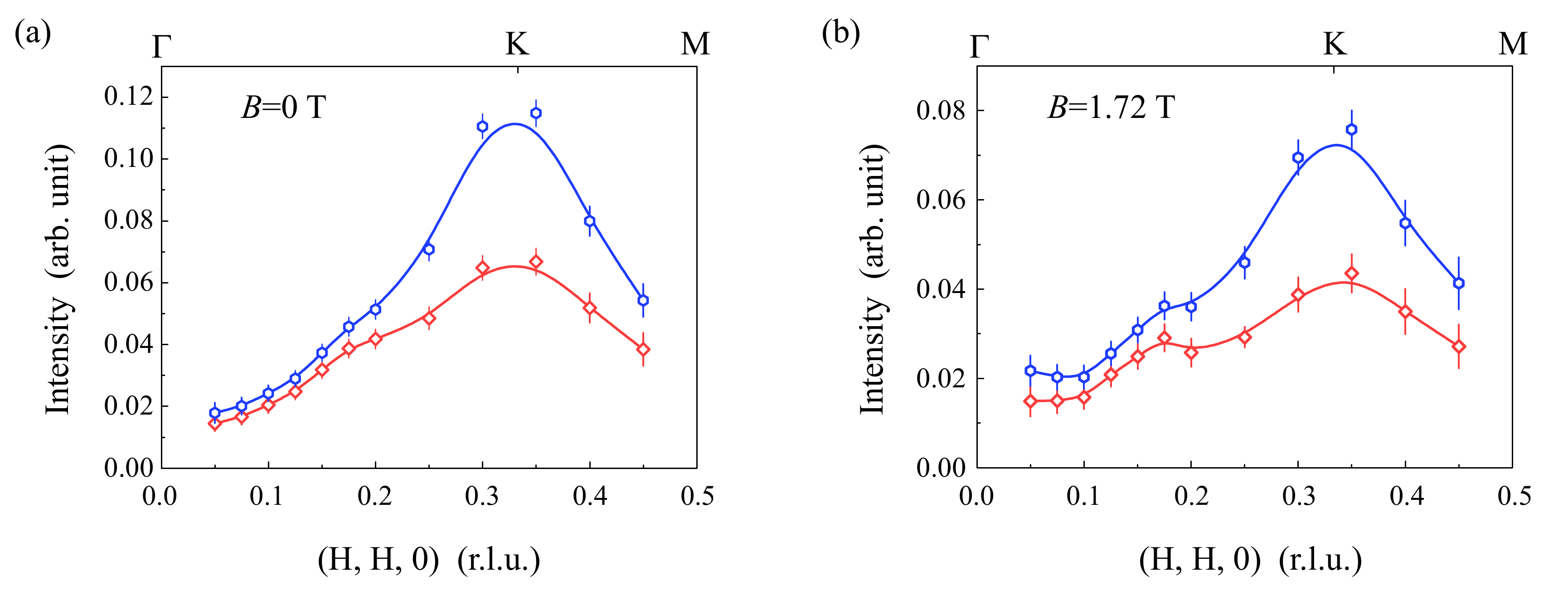}
\caption{Momentum dependence of the quadrupole wave intensity at $T=\qty{60}{mK}$ with (a) $B=\qty{0}{T}$ and (b) $B=\qty{1.72}{T}$. Red symbols: integrating the intensity of Figs. S8 and S10 in the energy range [0.01, 0.05] meV; Blue symbols: integrating the intensity of Figs. S7 and S9 in the energy range [0.01, 0.05] meV. The error bars are obtained via uncertainty propagation in accumulating the data in [0.01, 0.05] meV.}
\label{QW}
\end{figure}

Finally, integrating this positive energy signal (0.01 to \qty{0.05}{meV}) at each $\bm{k}$ gives us an estimate of the experimental $\bm{k}$-dependence of the quadrupole wave intensity, presented in Fig.~\ref{QW}. We note that the overall intensities for $B=\qty{0}{T}$ and $B=\qty{1.72}{T}$ are comparable, despite the fact that at $B=\qty{0}{T}$ the transition temperature into the SN-supersolid phase is likely below \qty{60}{mK} where the measurements were performed. We speculate that the transition temperature could be very close to \qty{60}{mK} that accounts for the short-range order and the paramagnon type of quadrupole waves at \qty{0}{T}.

%The resulting intensity profile shows a clear increase with $Q$, in good agreement with the theoretically calculated dynamical structure factor, thus providing spectroscopic evidence for the quadrupole wave mode.

\end{document}